\documentclass[10pt,a4paper,twoside]{article}

\usepackage{a4wide}

\usepackage[utf8]{inputenc} 
\usepackage[TS1,T1]{fontenc}

\usepackage{amssymb, amsmath,subeqnarray}

\usepackage{mathrsfs}

\usepackage{bbold}
\DeclareFontFamily{U}{bbold}{}
\DeclareFontShape{U}{bbold}{m}{n}{<-5.5> bbold5 <5.5-7.5> bbold7 <7.5-> bbold10}{}

\makeatletter

\@addtoreset{equation}{section}
\makeatother

\usepackage{graphicx}

\usepackage[english]{babel}

\newcommand{\be}{\begin{equation}}
\newcommand{\ee}{\end{equation}}
\newcommand{\bea}{\begin{eqnarray}}
\newcommand{\eea}{\end{eqnarray}}

\newcommand{\Aff}{A}
\newcommand{\Avkappa}{\overline{\kappa}}
\newcommand{\Avkappaeq}{\overline{\kappa}_\text{\mdseries eq}}
\newcommand{\betaeff}{{\beta_\star}}

\newcommand{\C}{{\cal C}}

\newcommand{\dexch}{d_\text{\mdseries exch}}
\newcommand{\deltaexch}{\delta_\text{\mdseries exch}}

\newcommand{\deltaint}{\delta_\text{\mdseries int}}
\newcommand{\Deltaexch}{\Delta_\text{\mdseries exch}}
\newcommand{\Deltaexchexcess}{\Delta_\text{\mdseries exch}^{\text{\mdseries excs},\beta_0}}
\newcommand{\Deltaexchexcessgen}{\Delta_\text{\mdseries exch}^{\text{\mdseries excs},\{F_i^0\}}}

\newcommand{\dint}{d_\text{\mdseries int}}
\newcommand{\En}{{\cal E}}
\newcommand{\Escale}{\Delta e}
\newcommand{\Esp}[1]{\langle#1\rangle}

\newcommand{\Espeq}[1]{\langle#1\rangle_\text{\mdseries eq}}
\newcommand{\Espst}[1]{\langle#1\rangle_\text{\mdseries st}}

\newcommand{\F}{\mathbb{F}}
\newcommand{\Fth}{{\cal F}}

\newcommand{\Heat}{{\cal Q}}

\newcommand{\Heatd}{{\cal Q}^\text{d}}
\newcommand{\Heatm}{q}
\newcommand{\Hist}{{{\cal H}ist}}

\newcommand{\jinst}{j}
\newcommand{\Jcumav}{J}
\newcommand{\Jcum}{{\cal J}}
\newcommand{\Jcumd}{{\cal J}^d}
\newcommand{\kB}{k_{\scriptscriptstyle B}}

\newcommand{\limTH}{\lim_\text{\mdseries TH}}
\newcommand{\M}{\mathbb{M}}
\newcommand{\N}{{\cal N}}

\newcommand{\Oobs}{{\cal O}}

\newcommand{\Prob}{P}
\newcommand{\Probcan}{P_\text{\mdseries can}}
\newcommand{\Probeq}{P_\text{\mdseries eq}}
\newcommand{\Probmc}{P_\text{\mdseries mc}}
\newcommand{\Probst}{P_\text{\mdseries st}}
\newcommand{\Probdist}{\Pi}
\newcommand{\sigexch}{\sigma_\text{\mdseries exch}}
\newcommand{\sigint}{\sigma_\text{\mdseries int}}

\newcommand{\syst}{{\cal S}}

\newcommand{\SB}{S^{\scriptscriptstyle B}}
\newcommand{\SG}{S^{\scriptscriptstyle SG}}
\newcommand{\STH}{S^{\scriptscriptstyle TH}}

\newcommand{\TimeRev}{\mathbb{T}}
 
\newcommand{\Trans}{\mathbb{W}}
\newcommand{\Udisc}{{\cal T}}

\newcommand{\UevG}{\widehat{\mathbb{U}}}

\title{\textbf{
 Thermal Contact I : Symmetries ruled by Exchange Entropy Variations}
}
\author{
F. Cornu\\ Laboratoire de Physique Théorique,  UMR 8627 du CNRS\\
Université Paris-Sud, Bât. 210
\\ F-91405 Orsay, France
 \\ \vspace{3mm}\\
 M. Bauer\\
 Institut de Physique Théorique de Saclay\footnote{CEA/DSM/IPhT, Unité de recherche associée au CNRS}, CEA Saclay
\\ F-91191 Gif-sur-Yvette Cedex, France
}

\date{May 2, 2015}

\begin{document}
\maketitle

\vspace{0.25cm}
\begin{itshape}

Most results contained in the  joined papers arXiv:1302.4538  and arXiv:1302.4540 put on cond-mat.stat-mech on February 19 2013  have now appeared in print as 

\vspace{0.15cm}
- [a]  \textnormal{Thermal Contact through a Diathermal Wall: A Solvable Toy Model},
\hfill\break\indent\indent
 F. Cornu and M. Bauer, J. Stat. Mech. (2013)  P10009

\vspace{0.15cm}
- [b] Part of \textnormal{Affinity and Fluctuations in a Mesoscopic Noria},
\hfill\break\indent\indent
 M. Bauer and F. Cornu,   J. Stat. Phys.   {\bf 155} (2014) 703, [arXiv:1402.2422]

\vspace{0.15cm}
- [c] \textnormal{ Local detailed balance : a microscopic derivation},
\hfill\break\indent\indent
M. Bauer and F. Cornu, J. Phys. A: Math. Theor. {\bf 48} (2015) 015008, [arXiv:1412.8179]

\vspace{0.25cm}

The present revised version takes into account the minor corrections made in the published articles but the bibliography is that of the first version (not updated). In reference [c] the denomination \textnormal{local detailed balance} has been used in place of \textnormal{modified detailed balance} in order to fit the  terminology that seems to be  most commonly used nowadays.

\end{itshape}
\vspace{0.5cm}

\begin{abstract}

  Thermal contact  is the archetype of non-equilibrium
  processes driven by constant non-equilibrium constraints  enforced by
  reservoirs  exchanging conserved microscopic quantities. 
    At a mesoscopic scale only the energies of the macroscopic bodies are accessible together with  the configurations of the contact system.  We consider a class of models where the contact system, as well as macroscopic bodies, have a finite number of possible configurations. The global system with only discrete degrees of freedom has no
  microscopic Hamiltonian dynamics, but  it is shown that, if the microscopic 
  dynamics   is assumed to be deterministic and ergodic  and to conserve energy according to some specific pattern, and if the mesoscopic evolution of the global system is approximated by a Markov process as closely as possible, then  the
mesoscopic  transition rates obey three constraints. In the limit where macroscopic bodies can be considered as reservoirs at thermodynamic equilibrium (but with different intensive parameters) the mesoscopic transition rates turn into transition rates for the contact system and the third constraint becomes modified detailed balance (MDB) ; the latter 
   is generically expressed in terms of the microscopic exchange entropy variation, namely the
  opposite of the variation of the thermodynamic entropy of the reservoir involved in
  a given microscopic jump of the contact system configuration. 
    We investigate the generic
  statistical properties for measurable  quantities that  arise from the MDB
  constraint.  For a finite-time evolution after the system prepared in an equilibrium state 
  has been set in contactwith thermostats at different temperatures,
   we derive a detailed fluctuation relation for the excess
  exchange entropy variation and an associated integral fluctuation relation. In
  the non-equilibrium stationary state (long-time limit), the proper
  mathematical definition of a large deviation function is introduced together
  with alternative definitions, and fluctuation relations are derived. The
  fluctuation relation for the exchange entropy variation is merely a particular
  case of the Lebowitz-Spohn fluctuation relation for the action functional
  \cite{LebowitzSpohn1999}. The generalization to systems exchanging  energy, volume and matter with several reservoirs, with a possible conservative external force acting on the contact system, is given explicitly.
 In the case of several independent  macroscopic currents, the infinite time limit of any odd cumulant per unit
  time of exchanged quantities is expressed in terms of a series involving
  higher even cumulants and powers of the thermodynamic forces associated to the
   currents. Every relation can be seen as a generalized Einstein-Green-Kubo relation
  valid far from equilibrium.  It entails a relation between
  the $n$th-order non-linear response coefficients of any odd cumulant per unit
  time in the vicinity of equilibrium and a finite number of lower-order
  non-linear response coefficients of even cumulants per unit time. The latter
  relations, already known in the literature, can be seen as another kind of generalizations of the standard
  Einstein-Green-Kubo formula pertaining to the first-order coefficient.


 \vskip 0.25cm 
{\it Corresponding author ~:} 
CORNU Françoise,
E-mail: Francoise.Cornu@u-psud.fr

\end{abstract}

{\small{\tableofcontents}}

\section{Introduction}

\subsection{Issues at stake}

Thermal contact between two energy reservoirs is one of the first issues
addressed by early thermodynamics and its second principle about entropy
variation. It has been revisited in the context of the description of transport
phenomena either from the statistical approach by the Boltzmann equation and its
BBGKY hierarchy generalization or from the phenomenological
thermodynamics of irreversible processes \cite{Onsager1931A, Onsager1931B,
  Callen1960} and the local equilibrium phenomenological approach for
inhomogeneous continuous media \cite{DeGroot1952,DeGrootMazur1962}. Until recently there have
been repeated attempts to formulate thermodynamics for non-equilibrium steady
states \cite{OonoPaniconi1998,SasaTasaki2006}.  Nowadays the statistical
description for the fluctuations of heat exchanges between a small system and
two thermal baths is part of the so-called `` stochastic thermodynamics'' of
small systems incorporating the effects of fluctuations
\cite{Seifert2008,Seifert2012}.  In the last two decades the latter domain has
been the subject of an increasingly intense activity, from the theoretical as
well as from the experimental point of view with very fast technological
improvements \cite{BustamanteETAL2005,Ritort2006}. A recent
extended review about stochastic thermodynamics is to be found in Ref.\cite{Seifert2012}, and its formulation in  the specific class of Markov jump processes  in continuous time is reviewed  in Refs.\cite{HarrisSchutz2007}.
(For brief  introductions see also  for instance Refs.\cite{Mallick2009,VanDenBroeck2013Varenna}.)

First steps in out-of-equilibrium statistical mechanics have been the study of
the linear (static or dynamic) response to an external constraint that drives
the system out of its equilibrium state. This vast topic began with the Einstein
fluctuation-dissipation relation for Brownian motion (see collected translated
papers in \cite{Einstein1956}) and is still active (for a review see
\cite{MarconiETAL2008}). Nowadays, the question of linear response in the
vicinity of a non-equilibrium steady state is under investigation (see for
instance \cite{ChetriteFalkovichGawedzki2008,BaiesiMaesWynants2009PRL,BaiesiMaesWynants2009,BaiesiETAL2010,
  VerleyChetriteLacoste2011}), but the subject is beyond the scope of the
present work.

Non-perturbative approaches for systems far from equilibrium have followed
various ways.  One trend has been the search for a unifying variational
principle based on a large deviation point of view, which would generalize the
entropy-probability relation introduced by Einstein in his theory of
thermodynamic equilibrium fluctuations \cite{Einstein1910} in order to explain
the opalescence phenomenon.  Indeed, principles of equilibrium statistical
mechanics can be seen as resulting from the maximization of the Shannon entropy
under various constraints about macroscopic observables in the framework of
information theory \cite{Jaynes1957A,Jaynes1957B}. From this point of view,
statistical mechanics, which allows one to retrieve the laws of thermodynamics
while describing mesoscopic fluctuations, may be interpreted as an example of a
large deviation theory, which provides the probabilities for an observable to
deviate from its most probable value \cite{Ellis1985}. In order to go beyond the
phenomenological irreversible thermodynamics, Oono \cite{Oono1989} has promoted
the idea of applying the large deviation theory to  non-equilibrium
statistical physics seen as a statistical mechanics along the time axis. In this
direction efforts have been devoted to the study of the maximization of the
Shannon entropy for the system histories under some constraints, such as fixed
values for the macroscopic out-of-equilibrium currents in a  non-equilibrium
steady state (see \cite{FilyokovKarpox1967Translated,Monthus2011} and references
therein).

In the same spirit a second path, the quest for some relevant physical
quantities that would obey some universal principle, is the subject of the
fluctuation relations either at finite time (transient regimes) or in the
infinite-time limit (stationary regimes). The active topic of fluctuation
relations was initiated by works about a symmetry of the large deviation
function of some entropy production rate in the non-equilibrium stationary
states of some chaotic dynamical systems \cite{EvansCohenMoriss1993,
  EvansSearles1994,GallavottiCohen1995PRL, GallavottiCohen1995JStatPhys}, and
about the Jarzynski identity for the finite-time cumulative work when a
Hamiltonian system is driven out of an initial equilibrium state by the
variation of some external parameter \cite{Jarzynski1997PRL}.  The first results
pertained to classical systems but fluctuation relations in quantum systems have
also been investigated (see for instance the reviews
\cite{EspositoHarbolaMukamel2009,Hanggi&al2011}).  A short list for successive
steps on the narrower pathway of fluctuation relations for systems with
Markovian stochastic dynamics is given by the following references
\cite{Jarzynski1997III, Kurchan1998, Crooks1998, Crooks1999, LebowitzSpohn1999,
  Maes1999, Seifert2005, Seifert2008, EspositoVanDenBroeck2010PREa}.  Among
these results, in the more specific case of systems with stochastic evolutions
described by a master equation for the probability of the system configurations,
the milestone is the fluctuation relation obeyed by the dimensionless action
functional introduced by Lebowitz and Spohn \cite{LebowitzSpohn1999}.

In the even more restricted case where the system exchanges conserved
microscopic quantities, such as energy quanta or particles or elementary
volumes, with infinite size reservoirs which stay in their equilibrium states
during the considered evolution of the finite system, the transition rates must
obey a relation, which will be referred to as the Modified Detailed Balance
(MDB)\footnote{The denomination ``generalized'' detailed balance is also to be
  found in the literature.}. When the transition rates obey the MDB, the action
functional introduced by Lebowitz and Spohn coincides with the variation of the
sum of the reservoirs thermodynamic entropies. Then, apart from the fluctuation
relation obeyed by the latter entropy variation, the known results about large
deviation functions which are the most relevant for the topic of the present
paper deal with the following quantities in finite size systems : the heat
current that goes through the finite system that sets up a thermal contact
between two thermostats \cite{Derrida2007, BodineauDerrida2007} (also first
addressed in the context of classical or quantum Hamiltonian dynamics in
Ref.\cite{JarzynskiWojcik2004}); more generally the various macroscopic
currents through a system in a non-equilibrium stationary state sustained by
various kinds of reservoirs \cite{AndrieuxGaspard2007JStatPhys,
  AndrieuxGaspard2007Bruxelles}. 

Apart from general approaches, a third trend in the quest for some
non-perturbative statistical theory of out-of-equilibrium phenomena has been the
search for solvable models which could give some hints in the comprehension of
these phenomena in the absence of any theoretical framework. We are interested
in a class of models where the system has a finite number of possible
configurations, the heat exchanges are described as changes in the populations
of its energy levels, and the configurations stochastically evolve under a
master equation with transition rates bound to obey the MDB relation arising
from the existence of an underlying ergodic deterministic microscopic dynamics
which conserves energy. In paper II \cite{CornuBauerB} we perform explicit analytical calculations
for the very simple case where the thermal contact system is reduced to two
spins, each of which is flipped by a single thermostat, the two thermostats
being at different temperatures. This model allows the description of a thin diathermal interface between two macroscopic bodies as a collection of independent spin pairs.  

\subsection{Main results valid beyond the thermal contact example}
\label{Outline}

In the present section we point out the main ideas and generic results of our study  when  we consider a class of models that generalize the case of thermal contact in the sense that the contact system $\syst$ can exchange not only energy but also matter or some portion of occupied volume with several macroscopic bodies according to some specific conservative exchange pattern where one and only one macroscopic body is involved in a microscopic jump of the whole system (see section \ref{GeneraliztionSeveralCurrents}). 
For the sake of pedagogy,  we express the main ideas and generic results in the language of the thermal contact example. The general arguments and results in the more general situation are more detailed in section \ref{GeneraliztionSeveralCurrents}.

In the whole paper we put emphasis on the \textit{exchange entropy  variation} $\Deltaexch S$  of the system\footnote{\label{exchangename} Exchange entropy variation is an abbreviation for ``variation of entropy due to exchanges'' of energy  with a thermal bath, or more generally,  of some measurable conserved quantities with various reservoirs.}, which is opposite to the well-defined entropy variation of the reservoirs described in the thermodynamic limit, because it seems to be the crucial quantity to consider in order to answer the  key-point question of identifying the relevant measurable quantities which are to obey non-trivial universal properties (beyond long-time decorrelations and the subsequent existence of large deviation functions). We stress that the term ``measurable quantities'' excludes the stationary probability distribution of the system configurations, which is very hard to determine from experiments. This point of view slightly differs from the common picture where the focus is put on the variation of the system Shannon-Gibbs entropy which explicitly  involves the
probability distribution of the system configurations. The latter picture has arisen from the seminal works by Crooks \cite{Crooks1998, Crooks1999, Crooks2000} and has led to the notion of entropy production along a stochastic trajectory of the system microscopic configurations  (see \cite{Seifert2005,Seifert2012} for a recent formulation). In the whole paper Boltzmann constant $\kB$ is set to 1 and $S$ denotes dimensionless entropies.

We point out that  Appendix \ref{LDRemarks} contains the proper mathematical definition of the large deviation function, and some variants for the models at stake, together with demonstrations of properties about large deviation functions which are not usually exhibited in the  literature of the physicists community.

\subsubsection{Constraints from ergodicity and energy conservation pattern}

We consider models for thermal contact where the energy exchanges between several
very large systems with indices $a$'s are realized through interactions with a
system $\syst$ which has a finite number of configurations, namely which has discrete
variables (also called degrees of freedom) in finite number, and whose configurations can changed only thanks to interactions with the large systems.  
More precisely, the domain ${\cal D}$ occupied by the so-called small system $\syst$
can be divided in several disjoint areas ${\cal D}_a$, such that the value of every degree of freedom of $\syst$ that sits inside  ${\cal D}_a$ can change only thanks to  an energy exchange with  a macroscopic body ${\cal B}_a$.  The small system has an internal energy which is
bounded, since its configuration space is finite.  When each large part with
index $a$ is described only at a macroscopic level, i.e. only by its
energy $E_a$, we have to resort to a statistical description, where the
knowledge of the system is specified by the following data : the configuration $\C$ of the small system and the energies  of the large parts. Such a set of data
defines what we call a mesoscopic configuration of the total system in the
sequel.  Then the stochastic evolution is determined by transition rates between
these mesoscopic configurations. The key points are the following.

First we have to answer the question : which constraints must be obeyed by the
choice for the transition rates ?   Indeed, one cannot define a Hamiltonian dynamics for discrete variables. However,  when the microscopic dynamics is deterministic, ergodic and conserves the global energy, then in a given energy level all ergodic microscopic
dynamics have the same period, and the probability of a mesoscopic configuration  $(\C,\{E_a\})$ calculated as a time average over a period of  microscopic dynamics coincides with  the microcanonical distribution.  Moreover,  the chosen
 interaction pattern entails that, during a period of 
microscopic dynamics, the number of deterministic jumps that occur from one mesoscopic
configuration of the global system to another one is equal to the number of
jumps in the opposite sense (subsection \ref{Ergodicity} and Appendix \ref{CoarseGrainedProperty}).

As a consequence, if the  mesoscopic dynamics is approximated by a Markov process according to the prescription derived in Appendix  \ref{Markovapprox}, then the corresponding transition rates between mesoscopic configurations  $(\C,\{E_a\})$ must satisfy three constraints (subsection \ref{MarkovianApproximationConseq}). In particular these transition rates  obey the  microcanonical detailed balance \eqref{ratioWmc}. (For a comparison the derivation of the microcanonical detailed balance for mesoscopic variables defined from  continuous microscopic variables evolving according to  a Hamiltonian dynamics invariant under time reversal is presented in Appendix \ref{TimeReversal}). The interaction pattern further transforms the microcanonical detailed  balance into the reduced expression \eqref{ergodic} which only involves   the  variation of the Boltzmann entropy of the  large part ${\cal B}_a$  that exchanges energy in the transition. Eventually, the three constraints for the transition rates  between mesoscopic configurations  $(\C,\{E_a\})$ ensure that  in the infinite-time limit the stochastic evolution  does lead to a unique stationary state which coincides with  the microcanonical probability distribution $\Probmc(\C,\{E_a\})$.

In the limit where the sizes of the large parts go to
infinity before the time evolution of the system is considered (subsection \ref{TransientSec}), the global
system does not reach equilibrium but, in a time window where the variations of the macroscopic energy  of every  large part divided by its number of degrees of freedom are negligible with respect to some microscopic energy scale, the large parts can be described as if they were in  the corresponding thermodynamic
equilibrium state. The \textit{thermodynamic entropy}  of a macroscopic body at equilibrium characterized by its internal energy $U_a$  and its  number of degrees of freedom $\N_a$ (and possibly other  extensive quantities such as volume which we omit) is an extensive quantity :
$\STH_a(U_a, \N_a)=\N_a s_a(U_a/\N_a)$
where $s_a(\epsilon)$ is a differentiable function of the mean energy per constituent $\epsilon$. 
In equilibrium statistical mechanics, for an isolated system $s_a(\epsilon)$ can be calculated  (on principle) in the microcanonical ensemble from  the \textit{Boltzmann entropy} 
$\SB(E_a,\N_a)\equiv\ln\Omega(E_a,\N_a)$,
where  Boltzmann constant $\kB$ has been set equal to 1 and 
$\Omega(E_a,\N_a)$ 
is the configuration number when the system is characterized by the  constant global extensive quantities $E_a$ and $\N_a$. The relation reads
$s_a(\epsilon)=\limTH \SB(E_a,\N_a)/\N_a$,
where  the thermodynamic limit  $\limTH$ corresponds to the coupled limits
$\N_a\to +\infty$ and $E_a\to+\infty$ with
$\lim E_a/\N_a=\epsilon$.   
In the case of thermal contact
the large parts behave as thermostats, each of which is characterized by the intensive thermodynamic parameter $\beta_a$ conjugated with the extensive macroscopic quantity $U_a$  through 
$ds_a(\epsilon)=\beta_a d\epsilon$. When Boltzmann constant is set equal to 1, $\beta_a$ is the inverse temperature of thermostat $a$. Then the  constraints for the transition rates between mesoscopic configurations $(\C, \{E_a\})$ of the global system become three conditions
which must be obeyed by the transition rates between the small system configurations $\C$  in the corresponding transient
regime. Let $(\C'\vert \Trans \vert \C)$ denote the transition rate from
configuration $\C$ to configuration $\C'$ in the transient regime. The three
conditions are the following ones. First, the set of transition rates must be
such that any configuration $\C$ can be reached by a succession of jumps with
non-zero transition rates from any configuration $\C'$, namely in the network
representation of the stochastic evolution 
\be
\label{Mirreducible0}
\quad\textit{the graph $G$ associated with the transition rates is connected.}\quad
\ee
In other words the Markov matrix $\M$ defined from the transition rates by \eqref{defMarkovM} must  be irreducible and the property \eqref{Mirreducible0} will be referred  to as the \textit{irreducibility} condition. Second, the transition rates must obey
 the \textit{microscopic reversibility} condition 
 for any couple of configurations $(\C,\C')$, namely
\be
\label{MicroRevCond}
(\C'\vert \Trans \vert \C)\not=0 \qquad \Leftrightarrow \qquad (\C\vert \Trans
\vert \C')\not=0.
\ee 
Third, the transition rates have to satisfy  the \textit{modified detailed balance}  which takes the form \eqref{ergodicThermo} in the case of pure thermal contact. According to the derivation,
the modified detailed balance  reads quite generally 
\be
\label{MDBexch0}
\frac{(\C'\vert \Trans \vert C)}{(\C\vert \Trans \vert C')}=e^{-\deltaexch S(\C'\leftarrow\, \C)},
\ee
where the exchange entropy variation $\deltaexch S(\C'\leftarrow\, \C)$ associated with a jump  of the small system from a microscopic configuration  $\C$ to another one $\C'$  is defined  as the opposite of the infinitesimal variation of the   thermodynamic entropy of the reservoir ${\cal B}_a$ that  causes the jump of configuration from $\C$ to $\C'$ by  exchanging energy and/or volume and/or matter with the constituents of the small system. The explicit expression for $\deltaexch S(\C'\leftarrow\, \C)$ in terms of configuration observables is given in \eqref{ExplicitDeltaSexchGen}; it includes the case where a conservative external force acts on some global coordinate of the contact system.
 In the case of thermal contact, namely energy transfers, $\deltaexch S(\C'\leftarrow\, \C)=\beta_a\delta\Heatm_a(\C'\leftarrow\, \C)$ where 
$\delta\Heatm_a(\C'\leftarrow\, \C)$ is the opposite of the heat amount received by  heat source ${\cal B}_a$, which may be associated with its heat capacity at fixed volume or its heat capacity at fixed pressure (when the volume of the interface and that of the heat reservoirs both vary by opposite amounts). 
The modified detailed balance has been used in various specific forms, for instance, for heat currents between two heat sources or particle currents between two particle reservoirs  or for coupled exchanges of energy and particles in molecular motor models (see among others \cite{AndrieuxGaspard2006PRE, BodineauDerrida2007,  SchmiedlSpeckSeifert2007,LiepeltLipowsky2007}). 
We stress that 
$\deltaexch S(\C'\leftarrow\, \C)$ is defined unambiguously and involves  experimentally measurable quantities (in the thermal contact case the  inverse temperatures of the energy reservoirs $\beta_a$'s as well as the heat amounts received by the system from the reservoirs).

\subsubsection{Various entropy variations inherent to Markovian dynamics}

In section  \ref{Entropysection} we stress some consequences of the fact that
the configuration probability obeys a master equation. First, even if the
transition rates do not obey the MDB, the irreducibility of the Markov matrix implies the uniqueness of the stationary state and the role played by the \textit{relative entropy with respect to the stationary solution}, 
$S_\text{\mdseries rel}[\Prob(t)\vert \Probst]$ is recalled (subsection \ref{EvolutionSec}).
We introduce the generic currents \eqref{defjinst0}-\eqref{defjinstdeltaK} for a microscopic variation when the system goes out of a given configuration $\C$; the averages of such currents  show off either in the  time derivative of the average of an observable  or in the  flow of an exchange quantity (subsection \ref{MicroCurrentSec}). In order to make the comparison with the thermodynamics of irreversible processes, we recall how the time-derivative of the \textit{Shannon-Gibbs entropy }
$\SG[\Prob(t)]$ of the system is split into two contributions, an exchange (or external) part 
$\dexch S/dt$ arising from exchanges with the external reservoirs, hereafter called  the \textit{exchange entropy flow},
 and an internal (or irreversible) part $\dint \SG/dt$  due to the internal irreversible processes in the system, also called the \textit{entropy production rate} (subsection \ref{EvolSSGSec}). The notation is chosen so as to emphasize that neither the exchange entropy flow $\dexch S/dt$ nor the entropy production rate are time derivatives of some function.

  In the stationary state, the time derivative $d \SG[\Probst]/dt$ vanishes and the entropy production rate becomes opposite to the exchange entropy flow. Henceforth, in the case where there are only two energy reservoirs,  the stationary entropy production rate can be rewritten as in phenomenological irreversible thermodynamics, namely as
\be
\label{defFth}
\left.\frac{\dint \SG}{dt}\right\vert_\text{st}=-\left.\frac{\dexch S}{dt}\right\vert_\text{st}
=\Fth\Jcumav,
\ee 
 where $\Jcumav$ is the mean heat current  that goes through the system from  heat bath 2 to heat bath 1 and $\Fth=\beta_1-\beta_2$ is the associated ``thermodynamic force''. In the transient regime, when the MDB is satisfied the expressions for $\dexch S/dt$ and $\dint \SG/dt$ coincide with the splitting of $d\SG/dt$ into two functionals $\sigexch[\Prob(t)]$ and $\sigint[\Prob(t)]$ \cite{Schnakenberg1976,LebowitzSpohn1999}. As noticed in \cite{EspositoVanDenBroeck2010PREa} for the more generic case where the transition rates are time-dependent, $\sigint[\Prob(t)]$ appears as the sum of two positive terms. In the  case of time-homogeneous Markov evolution,  the 
 first positive term is  the opposite of the time-derivative of the relative entropy of $\Prob(t)$ with respect to $\Probst$ and we rewrite the second positive term as the average  of the microscopic current associated with some stationary affinity variation when the system jumps out  of a configuration $\delta \Aff[\Probst]$, average which is calculated with respect to the probability $\Prob(t)$. Then the
 entropy production rate can be written as the sum of the following two positive contributions,
 \be
 \label{decompSProductionRate}
 \frac{\dint \SG}{dt}\underset{MDB}{=}-\frac{d S_\text{\mdseries rel}[\Prob(t)\vert \Probst]}{dt}+ \Esp{\jinst_{\delta \Aff[\Probst]}}_t.
 \ee
The current $\jinst_{\delta \Aff[\Probst]}(\C)$  proves to be positive even before averaging over the configurations (see \eqref{jinspositiv}). For the sake of completeness we recall  how, in the graph theory where the master equation is represented by a network, a non equilibrium stationary state is characterized by the affinities and  probability currents associated with a restricted number of cycles defined from the graph built with the transition rates. When the graph is a pure cycle, there is a  single affinity which has a simple probabilistic interpretation given in paper II.

\subsubsection{MDB and non-perturbative symmetries in transient regimes}

In the form \eqref{MDBexch0} where it involves the microscopic exchange entropy variation 
$\deltaexch S(\C'\leftarrow\, \C)$ associated with a  jump from configuration $\C$ to  configuration $\C'$,  the MDB entails time-reversal symmetries for finite time intervals at more and more mesoscopic levels as follows.
The ratio \eqref{historypropertySexch} of the probabilities for a microscopic history $\Hist$ and the time-reversed one is determined by the \textit{cumulative exchange entropy variation  along a history}, $\Deltaexch S[\Hist]$, defined in  \eqref{DefSexch}.
Then, the ratio \eqref{TimeRevPQ1PQ2CfC0} between, on the one hand, the probability for all evolutions with fixed initial and final configurations $\C_0$ and $\C_f$ and given heat amounts $\Heat_1$ and $\Heat_2$ received from the thermostats and, on the other hand, the probability for all backward evolutions with  exchanged initial and final configurations  and opposite heat amounts $-\Heat_1$ and $-\Heat_2$  is determined only by the exchange entropy variation
\be
\label{def1}
\Deltaexch S(\Heat_1,\Heat_2)= \beta_1 \Heat_1+\beta_2 \Heat_2.
\ee
The ratio of probabilities does not  explicitly  depend on the initial and final configurations $\C_0$ and $\C_f$. There is only an implicit dependence on these configurations through the energy conservation rule that the  quantities $\Heat_1$ and $\Heat_2$ for a given history must satisfy, $\Heat_1+\Heat_2=\En(\C_f)-\En(\C_0)$.

When the system is prepared in an equilibrium state at the inverse temperature $\beta_0$ and suddenly put in thermal contact at the initial time of measurements with  two heat baths at the inverse temperatures $\beta_1$ and $\beta_2$, the symmetry  \eqref{TimeRevPQ1PQ2CfC0} together with the specific form of the equilibrium canonical distribution lead one  to consider
 the following  measurable quantity : the \textit{excess exchange entropy variation} $\Deltaexchexcess S(\Heat_1,\Heat_2)$ defined as the difference between the exchange entropy variation under the non-equilibrium external constraint $\beta_1\not=\beta_2$ and that under the equilibrium condition $\beta_1=\beta_2\equiv\beta_0$, namely
 \be
 \label{def2}
\Deltaexchexcess S(\Heat_1,\Heat_2)=\Deltaexch S(\Heat_1,\Heat_2)-\beta_0(\Heat_1+\Heat_2).
\ee 
$\Deltaexchexcess S(\Heat_1,\Heat_2)$ obeys  the symmetry relation at any finite time, or finite-time ``detailed fluctuation relation'', 
\be
\label{DFRDeltaexchexcess0}
\frac{\Prob_{\Probcan^{\beta_0}}\left(\Deltaexchexcess S\right)}{\Prob_{\Probcan^{\beta_0}}\left(-\Deltaexchexcess S\right)}=e^{-\Deltaexchexcess S}.
\ee
The latter relation itself entails the identity, or finite-time ``integral fluctuation relation''
\be
\label{IFRDeltaexchexcess0}
\Esp{e^{\Deltaexchexcess S(t)}}_{\Probcan^{\beta_0}}=1
\ee
in the spirit of Jarzynski identity \cite{Jarzynski1997PRL}.
The integral fluctuation relation 
\eqref{IFRDeltaexchexcess0} 
entails through Jensen's inequality $\Esp{e^x}\geq e^{\Esp{x}}$ that, the mean heat amounts that are exchanged during the time interval $t$ from the initial time where the thermal contact is set on between the system  initially at equilibrium at inverse temperature $\beta_0$ and the two heat sources must satisfy 
\be
\label{ClausiusLawRetrieved} 
(\beta_0-\beta_1) Q_1(t)+(\beta_0-\beta_2) Q_2(t)=-\Esp{\Deltaexchexcess S}\geq 0,
\ee
where $Q_a(t)$ denotes the expectation value of $\Heat_a$ when the experiment is repeated a large number of times, $Q_a(t)=\Esp{\Heat_a(t)}$.
To our knowledge the two relations \eqref{DFRDeltaexchexcess0} and \eqref{IFRDeltaexchexcess0}  have not appeared explicitly in the literature, though the calculations involved in the derivation of the present finite-time fluctuation relations are analogous to those that lead to finite-time detailed fluctuation relations for protocols where the system is in thermal contact with only one heat bath and is driven out of equilibrium by a time-dependent external parameter (see the argument first exhibited by Crooks \cite{Crooks1999} for work fluctuations and then Seifert \cite{Seifert2005} for the entropy production along a stochastic trajectory (see also  the review \cite{Seifert2008}). 
The latter class of protocols is very different as for the physical mechanisms that they involve : the changes in energy level populations are caused by energy exchanges with only one thermal bath and the system is driven out of equilibrium by the time dependence of the energy levels enforced by external time-dependent forces  \cite{Crooks1999Thesis}. Moreover, in Jarzynski-like protocols the system evolves from an initial equilibrium state and measurements are performed until work ceases to be provided to the system, which then relaxes to another equilibrium state, whereas in Hatano-Sasa-like protocols  the system evolves from an initial non-equilibrium steady state to another one (and then housekeeping heats and excess heats are introduced as in the steady state thermodynamics introduced by Oono and Paniconi \cite{OonoPaniconi1998}). In the  finite-time protocol considered here the system starts in an equilibrium state and at time $t$ it has not yet  reached the steady state controlled by $\beta_1$ and $\beta_2$. The present protocol does not either involve the comparison of forward and backward  evolutions corresponding to  two different series of experiments.
Moreover, the integral fluctuation relation \eqref{IFRDeltaexchexcess0}  differs from the Hatano-Sasa relation \cite{HatanoSasa2001} in the sense that the quantity to be averaged over repeated experiments does not involve the probability distribution of the system.

In a time window sufficient long so that  $\beta_1$ times the maximal possible energy variation of the system that sets up thermal contact is negligible with respect to $(\beta_1-\beta_2) \Esp{\Heat_2}$, the definition \eqref{def1}  can be replaced by its typical value, $\Deltaexch S(\Heat_1,\Heat_2)\simeq  -(\beta_1- \beta_2) \Heat_2$. Moreover, if  the time window is also such that $\beta_0$ times  the maximal possible energy variation of the system is also negligible with respect to $(\beta_1- \beta_2) \Esp{\Heat_2}$,  the definition \eqref{def2} can be replaced by the approximation $\Deltaexchexcess S(\Heat_1,\Heat_2)\simeq\Deltaexch S(\Heat_1,\Heat_2)$. 
Then the relations \eqref{DFRDeltaexchexcess0} and \eqref{IFRDeltaexchexcess0} are compatible with the finite-time equalities settled in \cite{JarzynskiWojcik2004} for a heat transfer $\Heat_2$ between two finite bodies initially prepared at different inverse temperatures $\beta_1$ and $\beta_2$ and whose microscopic Hamiltonian dynamics involves a negligible interaction turned on at time $0$ and switched off at time $t$ (while the temperatures of both bodies may vary).
We also notice that in the considered time window the inequality \eqref{ClausiusLawRetrieved} can be approximated by $(\beta_1-\beta_2)Q_2(t)\geq 0$. We retrieve the result derived from thermodynamics principles :  on the average heat flows from the hotter heat bath to the colder one, in the absence of work given to the system that ensures contact between them.

We notice that if the configuration probability distribution in a stationary state of the system
happens to be the canonical distribution at an effective  inverse temperature $\beta_{\star}(\beta_1,\beta_2)$, as it is the case in the solvable model considered in paper II, there exist similar detailed and integral fluctuation relations for a protocol where  the system is initially prepared in a non-equilibrium stationary state  with  two heat baths at the inverse temperatures $\beta_1^0$ and $\beta_2^0$ and suddenly put at the initial time of measurements in thermal contact  with  two heat baths at the inverse temperatures $\beta_1$ and $\beta_2$. Then
$\Deltaexchexcess S(\Heat_1,\Heat_2)$ is to be replaced by 
$\Delta_\text{\mdseries exch}^{\text{\mdseries excs},\beta_\star^0}S(\Heat_1,\Heat_2)$ where $\beta_\star^0$ is  the effective inverse temperature $\beta_\star^0(\beta_1^0,\beta_2^0)$.

\subsubsection{MDB and fluctuation relations in the stationary regime}

For a system with a finite number of configurations,
when the Markov stochastic matrix for the continous-time  evolution of the
configuration probabilities is irreducible, (see definition
\eqref{Mirreducible0}), the Perron-Frobenius theorem can be applied: the system
has a single stationary state, and it is such that every configuration has a non-vanishing weight (see for instance Ref. \cite{CoxMiller1965}).  Moreover the system reaches  its stationary state in a exponentially-short time \cite{vanKampen1992}.
Then the symmetry relation \eqref{TimeRevPQ1PQ2CfC0} enforced by the MDB at finite time leads to the existence of  lower and upper bounds for the ratio between  the finite-time probability  to measure heat amounts $\Heat_1$ and $\Heat_2$ and the corresponding probability for the opposite values, when the system is in its stationary state (see \eqref{InequalityHeats}).
Similar bounds are exhibited in 
Ref.\cite{BodineauDerrida2007}.

The Markovian property of the evolution  implies that the cumulative heats $\Heat_1$ and $\Heat_2$, and subsequently the cumulative exchange entropy variation $\Deltaexch S$, all grow linearly with time in the long-time limit and that there exist large deviation functions for each of them \cite{LewisRussell1997, Varadhan2008}. The proper mathematical definition of a large deviation function is recalled in Appendix \ref{MathDefSec}, and other alternative definitions when exchanged quantities are discrete are introduced in Appendix \ref{MathDefAltSec}.

The finite-time  inequalities \eqref{InequalityHeats} entail  a symmetry in the long-time limit, according to the general results derived in Appendix \ref{IneqLargeDeviation} :
  the dimensionless  exchange entropy variation $\Delta_\text{exch}S$ obeys the fluctuation relation 
 \be
\label{FTSexch0}
f_{\Deltaexch S}(\Jcum)-f_{\Deltaexch S}(-\Jcum)=-\Jcum,
\ee
where $\Jcum$ denotes the values taken by the  cumulative current $\Delta_\text{exch}S(t)/t$.
The fluctuation relation  for $\Delta_\text{exch}S$ is a special case of the more general fluctuation relation for the action functional introduced by Lebowitz and Spohn 
 \cite{LebowitzSpohn1999} ; indeed when the transition rates obey the MDB, the action functional for a given history becomes equal to the entropy variation of the reservoirs, namely to the opposite of the exchange entropy variation.

 For a  system with a finite number of configurations, $\Heat_1+\Heat_2$ is bounded. Then, according to the results of Appendix  \ref{JointLargeDeviation}, the large deviation function for $\Heat_2$ is equal to that for $-\Heat_1$, while 
the fluctuation relation \eqref{FTSexch0} for $f_{\Deltaexch S}$ also entails   a fluctuation relation for the  large deviation function $f_{\Heat_2}$ for the cumulative heat $\Heat_2$, because the difference between  $\Deltaexch S$ and 
 $-(\beta_1-\beta_2)\Heat_2$ is finite at any time $t$. The fluctuation relation can be written in the generic form
 \be
 \label{FRJcumAffIntro}
 f_{\Heat_2}(\Jcum)-f_{\Heat_2}(-\Jcum)=\Fth\Jcum,
  \ee
  where $\Jcum$ denotes the values taken by the  cumulative current $\Heat_2(t)/t$ and
 $\Fth$ is the thermodynamic force conjugated to the mean instantaneous heat current from heat bath $2$ in the stationary state $J\equiv\Espst{\jinst_2}$ through the expression \eqref{defFth} of the exchange entropy flow in the stationary state.  
 Indeed  on the one hand $\lim_{t\to+\infty} \Esp{\Deltaexch S}/t=-(\beta_1-\beta_2)\lim_{t\to+\infty} \Esp{\Heat_2(t)}/t$ and on the other hand $\lim_{t\to+\infty} \Esp{\Deltaexch S}/t=\dexch S/dt\vert_\text{st}$ and $\lim_{t\to+\infty} \Esp{\Heat_2(t)}/t=\Espst{\jinst_2}$ ; then comparison of $-\dexch S/dt\vert_\text{st}=(\beta_1-\beta_2)\Espst{\jinst_2}$ with definition \eqref{defFth} leads to identify the coefficient $(\beta_1-\beta_2)$ that arises in the fluctuation relation  with the thermodynamic force $\Fth$.
 
\clearpage
\subsubsection{MDB and generalized Einstein-Green-Kubo relations}

Quite generally, when there exists a large deviation function for the cumulative
current $\Jcum_t\equiv X_t/t$ associated with the cumulative quantity $X_t$, the
generic expression of the linear response in the non-equilibrium steady state far from
equilibrium reads
\be
\label{genericlinearresponse0}
\frac{\partial \Jcumav}{\partial \Fth}=
\left.\frac{\partial^2 f(\Jcum;\Fth)}{\partial \Fth\partial \Jcum}\right\vert_{\Jcum=\Jcumav}
\times \lim_{t\to+\infty} \frac{\Espst{X_t^2}-\Espst{X_t}^2}{t}.
\ee 
It relates on the one hand,
the coefficient $\partial \Jcumav/\partial \Fth$ of the linear response of the
heat current $\Jcumav$ to a variation of the thermodynamic force $\Fth$ and, on
the other hand, the infinite-time limit of the variance per unit time of the
cumulative quantity $X_t$ in the non-equilibrium steady state, with a coefficient
whose expression depends on the system.

 In the limit where the thermodynamic force $\Fth$ vanishes, 
 $\partial \Jcumav/\partial \Fth$  tends to the linear-response  coefficient near equilibrium, namely the  kinetic Onsager coefficient,   
 $L\equiv \partial \Jcumav/\partial \Fth\vert_{\Fth=0}$. If the system obeys the fluctuation relation \eqref{FRJcumAffIntro}, with $X$ in place of $\Heat_2$,  then the coefficient in the identity \eqref{genericlinearresponse0} 
 takes the universal value $\frac{1}{2}$ and the identity becomes the fluctuation-dissipation relation (also referred to as the Einstein-Green-Kubo relation) between the kinetic Onsager coefficient  and the infinite-time limit of the variance per unit time of $X_t$ at equilibrium. In the case of thermal contact, in the limit where the difference $\beta_1-\beta_2$ between the inverse temperatures vanishes, the ratio between the stationary heat current that goes through the system from  heat bath 2 to heat bath 1 and the difference $\beta_1-\beta_2$ becomes equal to $\frac{1}{2}$ times the  variance per unit time of  the heat amount exchanged with one thermal bath at equilibrium. (When the total system is at equilibrium, the net heat amount $\Heat_1+\Heat_2$ received by the system remains finite at any time, but the variance of the heat amount received from one bath grows linearly in time in the long-time limit).

For a system with a finite number of configurations, the MDB entails a symmetry   of the  generating function for the infinite-time limit of the cumulants of $X_t$ per unit time.
This symmetry takes the generic form \eqref{CumRel} in terms of the thermodynamic force $\Fth$. We then show that, in the infinite-time limit,  any odd cumulant per unit time $ \kappa^{[2n+1]}/t$ can be expressed in terms of even-order
cumulants per unit time through the relation
\be
\label{indepAff0}
\lim_{t\to+\infty}\frac{\kappa^{[2n+1]}(\Fth)}{t}=\sum_{k=0}^{+\infty} d_k
\Fth^{2k+1}\lim_{t\to+\infty}\frac{\kappa^{[2(n+k+1)]} (\Fth) }{t}\qquad\text{ for } n=0,1,\cdots,
\ee
where $d_k$ is given in \eqref{expdk}. The corresponding expressions for the ratios  $(1/\Fth)\times \lim_{t\to+\infty} \kappa^{[2n+1]}/t$ may be viewed as generalized Einstein-Green-Kubo relations, where the 
term ``generalized'' refers to the fact that they are valid both far from equilibrium and  for all cumulants.

We also express the response of any odd cumulant per unit time at any order in the thermodynamic force $\Fth$ near equilibrium in terms of non-linear response coefficients for even cumulants per unit time at lower orders in $\Fth$ near equilibrium (see \eqref{indepCoeffBis}). At first order, namely at the  level of linear response, one gets the generalized fluctuation-dissipation relations 
\eqref{GeneralizedFDR} where the 
term ``generalized'' refers to the fact that they are valid for any odd cumulant.

In the more general  situation where there are several independent mean currents between reservoirs, we derive the corresponding  generalized Einstein-Green-Kubo relations \eqref{GeneralizeGBSeveralJ}. From the latter equations one can derive relations between non-linear response coefficients. The latter relations have already been settled by another method by Andrieux and Gaspard \cite{AndrieuxGaspard2007JStatMech}. As noticed by these authors some of them lead to symmetries which are generalizations of  the Onsager reciprocity relation.

\clearpage
\section{Constraints upon transitions rates}

\label{sectionConstraints}

In the present section we review some of the constraints that ergodic
deterministic energy-conserving microscopic dynamics puts on the statistical
mesoscopic description of a finite system $\syst$ which establishes thermal contact
between energy reservoirs ${\cal B}_a$'s with $a=1,\ldots,A$.

Indeed, the following situation occurs commonly : the interactions in the whole
system allow to define one small part $\syst$ in contact with otherwise independent
large parts ${\cal B}_a$'s.  The large parts, which involve a huge number of degrees of
freedom, do not interact directly among each other (this gives a criterion to
identify the distinct large parts), but are in contact with the small part,
which involves only a few degrees of freedom.  Moreover each degree of freedom
in the small part is directly in contact with at most one large part and can vary  only through its interaction 
with the latter large part. This
results in a star-shaped interaction pattern.  It is convenient then to forget
about the microscopic description of the large parts, and turn to a statistical
description of their interactions with the small part. Some general features of
the statistical description can be inferred from microscopic ergodicity.

\subsection{Ergodicity}
\label{Ergodicity}

\subsubsection{Ergodicity in  classical Hamiltonian dynamics}
\label{HamiltonianErgodicity}

In classical mechanics, the time evolution of a system in phase space is
described by a Hamiltonian $H$. If the system is made of several interacting
parts, the Hamiltonian is then referred to as the total Hamiltonian, $H\equiv H_\text{tot}$, and it splits as $H_\text{tot}=H_\text{dec} +H_\text{int} $, where
$H_\text{dec} $ accounts for the dynamics if the different parts were decoupled
and $H_\text{int} $ accounts for interactions. The energy hypersurface $H_\text{tot}=E$,
usually a compact set, is invariant under the time evolution, and in a generic
situation, this will be the only conserved quantity.

The ergodic hypothesis
states that a generic trajectory of the system will asymptotically cover the
energy hypersurface uniformly. To be more precise, phase space is endowed with
the Liouville measure (i.e. in most standard cases the Lebesgue measure for the
product of couples made by every coordinate and its conjugate momentum), which
induces a natural measure on the energy hypersurface, and ergodicity means that,
in the long run, the time spent by the system in each open set of the energy
hypersurface will be proportional to its Liouville measure. Ergodicity can
sometimes be built in the dynamics, or proved, but this usually requires immense
efforts.

Ergodicity depends crucially on the fact that the different parts are coupled :
if $H_\text{int} =0$, each part will have its energy conserved, and motion will
take place on a lower-dimensional surface. If $H_\text{int} $ is very small, the
system will spend a long time very nearby this lower-dimensional surface, but, at
even longer time scales, ergodicity can be restored. By taking limits in a
suitable order (first infinite time and then vanishing coupling among the parts)
one can argue that the consequences of ergodicity can be exploited by reasoning
only on $H_\text{dec}$.

Notice that in this procedure, we have in fact some kind of dichotomy : $H_\text{dec}
$ defines the energy hypersurface, but cannot be used to define the ergodic
motion, which is obtained from $H_\text{tot}=H_\text{dec} +H_\text{int} $ via a limiting
procedure.  So the dynamics conserves $H_\text{dec} $ but is not determined by
$H_\text{dec} $.

\subsubsection{Ergodicity in  deterministic  dynamics for discrete variables}
\label{ErgodicityDiscrete}

Our aim is to translate the above considerations in the context of a large but
finite system described by discrete variables such as classical Ising spins.

In the case of discrete dynamical variables, one can still talk about the energy
$E_\text{tot}$ of a configuration, but there is no phase space and no Hamiltonian dynamics
available.  So there is no obvious canonical time evolution. This is where we
exploit the previously mentioned dichotomy: we do not define the time evolution
in terms of $E_\text{tot}$, but simply impose that the deterministic time evolution
preserves $E_\text{dec}$ and that it respects the  star-shaped interaction pattern between the
small part and the large parts. Besides we also impose that the time evolution is ergodic.

We consider that time is discrete as well, because  ergodicity is most simply expressed in discrete time. Then deterministic dynamics is given by a bijective
map, denoted by $\Udisc$ in what follows, on configuration space, applied at
each time step to get a new configuration from the previous one. 
As the
configuration space is finite, the trajectories are bound to be closed. Then, for a given initial value $E$ of $E_\text{dec}$, a specific dynamics $\Udisc$ conserving $E_\text{dec}$ corresponds to a periodic evolution of the microscopic configuration of the full  system inside the energy level $E_\text{dec}=E$.

Ergodicity entails that the corresponding closed trajectory covers fully the energy level 
$E_\text{dec}=E$, and then it must cover it exactly once during a period because the dynamics is
one to one. As a consequence the period of the ergodic evolution inside a given energy level 
of $E_\text{dec}$ is the same for all choices of ergodic dynamics $\Udisc$ that conserves $E_\text{dec}$. This period, denoted by $N$ in time step units, depends only on the value $E$ of $E_\text{dec}$,
\be
\label{valuePeriod}
N=\Omega_\text{dec}(E),
\ee
where  $\Omega_\text{dec}(E)$ is the total number of microscopic configurations in the level $E_\text{dec}=E$.
This is reminiscent of the microcanonical ensemble. Let us note that
in classical mechanics, there is a time reversal symmetry, related to an
involution of phase space, changing the momenta to their opposites while leaving
the positions fixed (see \eqref{fTimeInvariance}). In the discrete setting, involutions $J$ such that
$J\Udisc J=\Udisc ^{-1}$ always exist, but there is no obvious candidate among
them for representing time reversal and allowing to draw conclusions from it.

In the  context of discrete variables  the star-shaped interaction pattern is implemented as follows. We may naively assume that the energy
conserved by the dynamics $\Udisc$ is simply $E_\text{dec}$, as if there were no energy
for the interactions between the small part and the large ones, but  $\Udisc$ must
reflect the fact that the large parts interact only indirectly: there is an
internal interaction energy $\En(\C)$ for every configuration $\C$ of the small
part and each change in the small part can be associated with an elementary
energy exchange with one of the large parts. If the small part can jump from
configuration $\C$ to configuration $\C'$ in a single time step by exchanging
energy with large part $a$, we use the notation $\C'\in\F_a(\C)$. In this
configuration jump the energy $E_\text{dec}$ of the global system is conserved and the energy
of  large part $a$ is changed from $E_a$ to $E'_a$ according to the
conservation law
\be
 \label{conserE}
E'_a-E_a=
\begin{cases}
  -\left[\En(\C')-\En(\C)\right]   &\textrm{if}\quad  \C'\in\F_a(\C) \\
  0 &\textrm{otherwise},
\end{cases}
\ee 
while the energies of the other large parts are unchanged. Apart from these energy exchange constraints and from ergodicity, the deterministic dynamics $\Udisc$ is  supposed to obey some other natural physical constraints which will be specified later (see subsection \ref{MarkovianApproximationConseq}).

As a final remark, we mention how, in a very simple case,  some kind of deterministic map $\Udisc$ that preserves $E_\text{dec}$ and obeys the star-shaped interaction pattern can be associated with  a deterministic map  $\widetilde{\Udisc}$ that conserves $E_\text{tot}$. We consider the case where the energy exchange between every large part $a$ and the small part is ensured by an interaction energy $E_\text{int}^{(a)}$ between a classical spin 
$\sigma_a^\star$ in part $a$ and a classical spin $\sigma_a$ in the small part and we consider only maps $\widetilde{\Udisc}$ that not only conserve  $E_\text{tot}$ but also satisfy the following rules for all large parts :  (1)  spins $\sigma_a^\star$ and $\sigma_a$ are always flipped at successive time steps (in an order depending on the precise dynamics $\widetilde{\Udisc}$) ; (2) the variations of the interaction energy $E_\text{int}^{(a)}$ associated  with these successive two flips are opposite to each other. Then the map $\Udisc$ that conserves $E_\text{dec}$ is defined from the map $\widetilde{\Udisc}$ by merging  every pair of time steps where $\sigma_a^\star$ and $\sigma_a$ are successively flipped into a single time step where $\sigma_a^\star$ and $\sigma_a$ are  simultaneously flipped. Indeed, in the latter pair of time steps of $\widetilde{\Udisc}$, by virtue of hypothesis (2),  the successive two variations of $E_\text{int}^{(a)}$ cancel each other and the variation of  $E_\text{tot}$ after these two time steps coincides with the variation of $E_\text{dec}$, since by definition the latter variation is $\Delta E_\text{dec}=\Delta E_\text{tot}- \Delta E_\text{int}^{(a)}$. 
Therefore, if map $\widetilde{\Udisc}$ conserves $E_\text{tot}$ at every time step, then the corresponding map $\Udisc$ where the latter  successive two flips occur in a single time step preserves   $E_\text{dec}$ : the conservation rule \eqref{conserE} is indeed satisfied.. The suppression of time steps in the procedure that defines $\Udisc$ from  $\widetilde{\Udisc}$ corresponds to a modification of the accessible configuration space that reflects the fact that  the energy level $E_\text{tot}=E$ and $E_\text{dec}=E$ do not coincide.

\subsubsection{Constraints from ergodicity and interaction pattern upon coarse-grained evolution}

We start from the the familiar observation that keeping
track of what happens in detail in the large parts is out of our abilities, and
often not very interesting anyway. Our ultimate interest is in the evolution of the configuration $\C$ of
the small part in an appropriate limit. As an intermediate step, we keep track
also of the energies $E_a$'s  in the large parts, but not of the detailed configurations
in the large parts.

The coarse graining that  keeps track only of the time evolution of the configuration $\C$ of the small part and the energies $E_a$'s of the large parts is defined as follows.
With each microscopic configuration $\xi$  of the full system we
can associate the corresponding configuration $\C=\C(\xi)$ of the small part and
the corresponding energy $E_a=E_a(\xi)$ carried by part $a$. To simplify the notation, we let $\underline{E}$ denote
the collection of $E_a$'s, so $\underline{E}$ is a vector with as many
coordinates as there are large parts. 

As shown in previous subsubsection,  if  the  initial value of the energy $E_\text{dec}$ is equal  to $E$, then, over a period equal to $N=\Omega_\text{dec}(E)$ in time step units, the trajectory of $\xi$ under any  microscopic ergodic dynamics $\Udisc$ corresponds to a cyclic permutation of all the microscopic configurations  in the  energy level $E_\text{dec}=E$. 
Thus ergodicity entails that at the coarse grained level,  if $N_{(\C,\underline{E})}$ denotes the occurrence number  of
$(\C,\underline{E})$ during the  period of $N$ time steps,
$N_{(\C,\underline{E})}$ is nothing but 
$\Omega_\text{dec}(\C, \underline{E})$, the number
of microscopic configurations of the full system when the small part is in configuration $\C$ and the large parts have energies $E_a$'s in the energy level $E=E_\text{dec}$, namely
\be
\label{Nxvalue}
N(\C, \underline{E})=\Omega_\text{dec}(\C, \underline{E})
\ee
with
\be
\label{EnergyConstraint}
E=E_\text{dec}\equiv\En(\C)+\sum_a E_a.
\ee
In the following we fix the value  of $E$ and  $N$ is called the period of the dynamics, while the  energy constraint  $E=\En(\C)+\sum_a E_a.$ is often implicit in the notations.

Another crucial point is that, since any specific dynamics $\Udisc$ under consideration respects both the conservation of $E_\text{dec}$ and the interaction pattern specified at the end of subsubsection \ref{ErgodicityDiscrete},   the number
of jumps from $(\C,\underline{E})$ to $(\C',\underline{E'})$ over the  period of $N$ time steps, denoted by 
$N_{(\C,\underline{E}),(\C',\underline{E}')}$, is equal to the number of the reversed jumps 
from $(\C',\underline{E'})$ to $(\C,\underline{E})$ during the same time interval, namely
\be
\label{NumbersEqual}
N_{(\C,\underline{E}),(\C',\underline{E}')}=
N_{(\C',\underline{E}'),(\C,\underline{E})}. 
\ee 
In Appendix \ref{CoarseGrainedProperty}
we  give graph-theoretic conditions, not related to ergodicity, that ensure
this property, and show that they are fulfilled in one relevant example, as a consequence of the star-shaped interaction pattern.

\subsection{Markovian approximation for the  mesoscopic dynamics}
\label{MarkovianApproximationConseq}

\subsubsection{Definition of a Markovian approximation for the  mesoscopic dynamics}
\label{DefinitionMarkovApprox}

For our purpose, we first rephrase the coarse-grained evolution as follows. As already noticed, over the period of $N=\Omega_\text{dec}(E)$ time steps, a trajectory under any  microscopic ergodic dynamics $\Udisc$ in the energy level $E_\text{dec}=E$ corresponds to a cyclic permutation of  the N microscopic configurations $\xi$'s in the  energy level. Therefore, if the configuration at some initial time is denoted by  $\xi_1$, then the trajectory  is represented by the sequence
$\omega=\xi_1\xi_2\cdots \xi_N$ where $\xi_{i+1}=\Udisc \xi_i$ with
$\xi_{N+1}= \xi_1$. By the coarse-graining procedure that retains only the mesoscopic variable  $x\equiv(\C,\underline{E})$, the succession of distinct microscopic configurations $\omega$ is  replaced by  $w=x_1x_2\cdots x_N$ where
$x_i=x(\xi_i)=(\C(\xi_i),\underline{E}(\xi_i))$. 
In  $w$ various $x_i$'s take the same value, and a so-called transition corresponds to the case $x_{i+1}\not=x_i$,  namely the case where the configuration $\C$ of the small system is changed in the jump of the microscopic configuration of the full system from $\xi_i$ to  $\xi_{i+1}=\Udisc \xi_i$.

Since the large parts involve many degrees of freedom, the number of times some given value  $x=(\C,\underline{E})$ appears  in the coarse-grained sequence  $w$ is huge, and
even if $\omega$ is given by the deterministic rule $\xi_{i+1}=\Udisc \xi_i$,
there is no such rule to describe the sequence $w$. Moreover the  microscopic configuration 
at the initial time, $\xi_1$, is not known  so that  the coarse-grained sequence that actually appears in the course of time is in in fact a  sequence deduced from $w$ by a translation of all indices.

As explained in Appendix \ref{DiscreteTMarkov}, one may associate to the sequence $w$
a (discrete time) Markov chain such that the mean occurrence frequencies  of the patterns $x$ and $xx'$ in a stationary stochastic sample are equal to the corresponding values, $N_x/N$ and  $N_{xx'}/N$, in the sequence $w$ determined by the dynamics $\Udisc$ (up to a translation of all indices corresponding to a different value of the initial microscopic configuration). Whether this Markovian effective description is accurate depends on several things: the choice of $\Udisc$, the kind of statistical properties of $w$ one wants to check, etc.

We may also argue (see Appendix \ref{ContinousTimeApproximation}) that a continuous time description is enough if we restrict our attention to microscopic dynamical maps $\Udisc$'s such  that transitions, namely the  patterns $xx'$ with $x'\not= x$, are rare and of comparable mean occurrence frequencies over the  period of $N$ time steps. In other words, most of the steps in the dynamics amount to reshuffle the
 configurations of the large parts without changing their energies, leaving the
 configuration of the small part untouched. The latter physical constraint and the hypothesis of  the validity of the Markovian approximation  select a particular class of dynamics $\Udisc$.

With these assumptions,  we associate to the sequence $w$ of coarse grained variables 
$(\C, \underline{E})$ a Markov process  whose stationary measure shares some  of the
 statistical properties  of $w$, namely the values of the mean occurrence frequencies of length $1$ and length $2$ patterns. The  transition rate from  $(\C,\underline{E})$ to 
 $(\C',\underline{E}')$ with $(\C,\underline{E}) \neq (\C',\underline{E}')$ in the approximated Markov process  is given by \eqref{Wxxprim}, where we just have to make the substitutions $N_x=N_{(\C,\underline{E})}$ and 
 $N_{xx'}=N_{(\C,\underline{E}),(\C',\underline{E}')}$, with the result
 \be
\label{TauxMesoW0}
W(\C',\underline{E}'\leftarrow \C,\underline{E})=
\frac{N_{(\C,\underline{E}),(\C',\underline{E}')}}{\tau \,N_{(\C,\underline{E})}}\quad
\text{ for } (\C,\underline{E}) \neq (\C',\underline{E}'),
\ee 
where $\tau$ is a time scale such that  $W(\C',\underline{E}'\leftarrow \C,\underline{E})$ is of order unity. 
The corresponding stationary distribution  by given by \eqref{propequiv1},  
\be
\label{Probmicro0}
\Probst^W(\C,\underline{E})=\frac{N_{(\C,\underline{E})}}{N}.
\ee 
We recall that  $N=\sum_{(\C,\underline{E})}N_{(\C,\underline{E})}$ and 
$N_{(\C,\underline{E})}=\sum_{(\C',\underline{E'})}N_{(\C,\underline{E}),(\C',\underline{E}')}$.

\subsubsection{Microcanonical detailed balance and other properties}
\label{Microcansubsubsection}

By virtue of the ergodicity property 
 \eqref{Nxvalue} at the coarse-grained level,  the transition rate in the approximated Markov process reads
 \be
\label{TauxMesoW}
W(\C',\underline{E}'\leftarrow \C,\underline{E})=
\frac{N_{(\C,\underline{E}),(\C',\underline{E}')}}{\tau \,\Omega_\text{dec}(\C, \underline{E})}\quad
\text{ for } (\C,\underline{E}) \neq (\C',\underline{E}').
\ee
 Meanwhile, by virtue of the ergodicity properties  \eqref{valuePeriod}  and \eqref{Nxvalue}, the corresponding stationary distribution 
 \eqref{Probmicro0} is merely the microcanonical distribution
\be
\label{Probmicro}
\Probst^W(\C,\underline{E})=\frac{\Omega_\text{dec}(\C, \underline{E})}{\Omega_\text{dec}(E)}\equiv\Probmc(\C, \underline{E}).
\ee

Ergodicity also entails that, since all microscopic configurations $\xi$'s in the energy level appear in the sequence $\omega$, all possible values of $x$ also appear in the coarse-grained sequence $w$ : so  any mesoscopic state $(\C,\underline{E})$ can be reached from any other mesoscopic state $(\C',\underline{E}')$ by a succession of elementary transitions, even if they are not involved in an elementary transition (i.e. if $N_{(\C,\underline{E}),(\C',\underline{E}')}=0$);
in other words the graph associated with the transition rates $W(\C',\underline{E}'\leftarrow \C,\underline{E})$ is connected, or,
equivalently, the Markov matrix defined from the transition rates (see
definition below in \eqref{defMarkovM}) is \textit{irreducible}.

The constraint \eqref{NumbersEqual} imposed by the interaction pattern upon the coarse-grained evolution over a period of $N$ time steps entails that the transition rates of the approximated  Markov process defined in \eqref{TauxMesoW} obey two properties. First,
 \be
\label{ergodicRev}
W( \C',\underline{E}' \leftarrow \C,\underline{E})\quad \text{and}\quad
W( \C,\underline{E} \leftarrow \C',\underline{E}') 
\ee 
are
either both $=0$ or both $\neq 0$. This property may be called \textit{microreversibility}.
Second, if the transition rates do not vanish, they obey the equality
\be
\label{ratioWmc}
 \frac{W(\C',\underline{E}'\leftarrow \C,\underline{E})}
  {W(\C,\underline{E}\leftarrow \C',\underline{E}')}=
\frac{\Omega_\text{dec}(\C',\underline{E}')}
{\Omega_\text{dec}(\C,\underline{E})}
=\frac{\Probmc(\C',\underline{E'})}{\Probmc(\C,\underline{E})}.
\ee
Observe that the arbitrary time scale $\tau$ has disappeared in this equation.
The equality between the ratio  of transition rates and the ratio of  probabilities in the corresponding stationary distribution is the so-called detailed balance relation. Here the 
stationary distribution is that of the microcanonical ensemble, and we will refer to relation 
\eqref{ratioWmc} as the \textit{microcanonical detailed balance}.

In appendix \ref{TimeReversal}  we rederive a  microcanonical detailed balance similar to \eqref{ratioWmc}  when the underlying microscopic  dynamics is Hamiltonian and invariant under time reversal. 
The evolution of the probability distribution of the mesoscopic  variables is approximated by a Markov process according to the same scheme as that introduced in subsubsection \ref {DefinitionMarkovApprox}.
Eventually the comparison between the ways in which the microcanonical detailed balance arises in that case and in our previous argument  is the following.

- When the microscopic variables are continuous coordinates in phase space and evolve according to a Hamiltonian dynamics, in the framework of statistical ensemble theory  the stationary measure for the  mesoscopic variables  is the measure that is preserved under the microscopic dynamics ; the fact that it coincides with the microcanonical distribution is enforced by   the invariance of the Liouville measure in phase space under the Hamiltonian evolution ; the microcanonical detailed balance   for mesoscopic variables that are even functions of microscopic momenta mainly arises from the  invariance under time reversal of the  trajectories in  phase space  (see \eqref{ProbJointHBis}).

- When the microscopic dynamical variables are discrete and evolve under an  energy-conserving map,  the stationary measure for  mesoscopic variables is  defined as the average over a period of the microscopic dynamics ; the fact that it is equal to the microcanonical distribution arises from the ergodicity imposed on the microscopic map $\Udisc$ (see \eqref{valuePeriod} and  \eqref{Nxvalue}) ; the microcanonical detailed balance emerges from  the  equality between the  frequencies of a given transition and the reversed one (over the period needed for  the microscopic  map to cover the energy level),  equality which is enforced by the  star-shaped interaction pattern (see \eqref{NumbersEqual}).

\subsubsection{Further consequence of the interaction pattern}

 In the interaction pattern large parts do not interact directly with one another and $\Omega_\text{dec}(\C, \underline{E})=\prod_a \Omega_a(E_a)$, where $\Omega_a(E_a)$ denotes the number of configurations in large part $a$ with energy
$E_a$ when it is isolated. Moreover the energy of a single large part is changed  in a given transition, so  if $(\C',\underline{E}')$ is obtained from
$(\C,\underline{E})$ by an energy exchange with bath $a$ that makes $\C$ jump to 
$\C'$ we have the result, with the notation introduced in \eqref{conserE},
\be
\label{SimplifiedRatio}
\textrm{if $\C'\in\F_a(\C)$} \quad
\frac{\Omega_\text{dec}(\C',\underline{E}')}
{\Omega_\text{dec}(\C,\underline{E})}= \frac{\prod_b
  \Omega_b(E'_b)}{\prod_b \Omega_b(E_b)}=\frac{\Omega_a(E'_a)}{\Omega_a(E_a)}.
\ee 
We have used the energy conservation rule  $E'_b =E_b -\delta_{a,b}\left[\En(\C')-\En(\C)\right]$, so
that for $b \neq a$ the multiplicity factors are unchanged in the transition.

The latter ratio of microstate numbers  can be expressed in terms of
the Boltzmann entropies when each part $a$ is isolated. When Boltzmann constant $\kB$ is set to 1, the dimensionless
Boltzmann entropy 
$\SB_a(E_a)$ for the isolated part $a$ when its energy is equal to $E_a$ is
defined by $\Omega_a(E_a)\equiv\exp \SB_a(E_a)$. With these notations, 
if the transition rate $W( \C',\underline{E}' \overset{\F_a}{\leftarrow}
  \C,\underline{E})$, where $\C'\in\F_a(\C)$, is nonzero, then the transition rate for the reversed jump $W( \C,\underline{E} \overset{\F_a}{\leftarrow} \C',\underline{E}')$, where $\C\in\F_a(\C')$, is also non zero (see \eqref{ergodicRev}) and, by virtue of \eqref{SimplifiedRatio} the relation \eqref{ratioWmc} is reduced to
\be
\label{ergodic}
\frac{W( \C',\underline{E}' \overset{\F_a}{\leftarrow}
  \C,\underline{E})}{W( \C,\underline{E} \overset{\F_a}{\leftarrow}
  \C',\underline{E}')}=\frac{\Omega_a(E'_a)}{\Omega_a(E_a)}
 \equiv e^{\SB_a(E'_a)-\SB_a(E_a)}.
\ee
The latter formula is the first important stage of the argument.

We stress that the present argument does not involve any kind of underlying time reversal. Here the time reversal symmetry arises only at the statistical level of description
represented by the Markov evolution ruled by the transition rates. 

Notice also  that, as only certain ratios are fixed, different ergodic
deterministic microscopic dynamics can lead to very different transition rates,
a remnant of the fact that the coupling between a large part and the small part
can take any value a priori.

Formula \eqref{ergodic} is also a clue to understand a contrario what kind of
physical input is needed for the homogeneous Markov approximation to be valid.
Indeed, why didn't we do the homogeneous Markov approximation directly on the
small part ? We could certainly imagine dynamics making this a valid choice.
However, it is in general incompatible with the pattern of interactions (see the
end of subsubsection \ref{ErgodicityDiscrete}) which is
the basis of our argument. For instance, if in the coarse-graining procedure we had retained only the configurations $\C$'s of the small part, then the corresponding graph introduced in Appendix \ref{CoarseGrainedProperty}  would have been a cycle instead of a tree in the case of a small part made of two spins (see paper II), and the crucial property \eqref{NumbersEqual} would have been  lost : over the period of $N$ time steps of the microscopic dynamics $N_{\C,\C'}\not=N_{\C',\C}$. 
In fact, we may expect, or impose on physical
grounds, that $\Omega_a$ will be exponentially large in the size of  large
part $a$ (i.e.  its number of degrees of freedom $\N_a$), so that even the ratio
$\Omega_a(E'_a)/\Omega_a(E_a)=\Omega_a(E_a-\left[\En(\C') -\En(\C)\right])/
\Omega_a(E_a)$ will vary significantly over the trajectory, meaning that
transition probabilities involving only the small part  cannot be taken to be constant
along the trajectory: the energies of the large parts are relevant variables.

\subsection{Transient regime when large parts are described in the thermodynamic limit}
\label{TransientSec}

\subsubsection{Large parts in the thermodynamic limit}

We now assume that the large parts are large enough that they are accurately
described by a thermodynamic limit, which we take at the most naive level. To
recall what we mean by that, we concentrate on one large part for a while, and
suppress the index used to label it. Suppose this large part has $\N$  degrees of freedom, and suppose that energies are close to an energy $E$ for which the Boltzmann
entropy is $\SB(E)$. That the thermodynamic limit exists means that if one lets
$\N \rightarrow +\infty$ while the ratio $E/\N$ goes to a finite limit
$\epsilon$, there is a differentiable function $s^B(\epsilon)$ such that the
ratio $\SB(E)/\N$ goes to $s^B(\epsilon)$. The quantity 
\be
\label{defbeta}
\frac{ds^B}{d\epsilon}\equiv \beta
\ee
is nothing but the inverse temperature. In that case, as long as $\Delta E \ll E$
 (where $E$ scales as $\N\Escale$ with $\Escale$ some finite energy scale), $\SB(E+\Delta E)-\SB(E) \rightarrow \beta \Delta E$ when $\N \rightarrow
+\infty$. For $\N$ large enough, the relation $\SB(E+\Delta E)-\SB(E) \sim \beta
\Delta E$ is a good approximation.

\subsubsection{Transient regime and modified detailed balance (MDB)}

Notice that when transitions occur, which, by the definition of $\tau$ in \eqref{TauxMesoW0}, happens typically once  
on the macroscopic time scale, the changes in the
energies of the large parts are finite, so that over long windows of time
evolution, involving many changes in the small part,  the relation 
\be
\label{DifferenceSB}
\SB_a(E'_a)-\SB_a(E_a) \sim \beta_a
\left[E'_a-E_a\right]
\ee
is not spoiled, where $E'_a$ and $E_a$ are the energies in large part $a$ at
any moment within the window.

In fact the larger the large parts, the longer
the time window for which \eqref{DifferenceSB} remains valid. The relation between the sizes 
$\N_a$'s  of the large parts and of the length of the
time window depends on the details of $\Udisc$ (which still has to fulfill the
imposed physical conditions). This relation also depends on the values of the energy per degree of freedom in every  large part,
$E_a/\N_a$,  which are essentially constant in such a window.

Because of the ergodicity hypothesis, the largest window (of size
comparable to the period of $\Udisc$ to logarithmic precision) has the property that the
energies $E_a$'s in the large parts will be such that all $\beta_a$'s are close to
each other and  the system will be at equilibrium. Indeed,
inside the largest time window, the
system remains in the region of the energy level where the
energies $E_a$'s   are the most probable, and in the  thermodynamic limit the most probable values for the $E_a$'s in the microcanonical ensemble are the values $E_a^\star$'s  that maximize the product $\prod_{a}\Omega_a(E_a)$ under the constraint $E=\sum_aE_a$ (since the system energies are negligible with respect to those of the large parts). The latter  maximization condition is equivalent to the equalities $ds^B/d\epsilon_a(\epsilon_a^\star)=ds^B/d\epsilon_b(\epsilon_b^\star)$
for all pairs of large parts.

However, if the system
starts in a configuration such that the $\beta_a$'s are distinct, the time window
over which $E_a/\N_a$ and $\beta_a$ are constant (to a good approximation)
will be short with respect to the period of the microscopic dynamics, but long enough that \eqref{DifferenceSB} still holds for a long time  interval.
Then by putting together
the information  on the ratio of transition rates  in terms of  Boltzmann entropies  \eqref{ergodic}, the transient regime approximation \eqref{DifferenceSB} and  the energy conservation \eqref{conserE},  we get
\be
\label{ergodic2}
\frac{W(\C',\underline{E}' \overset{\F_a}{\leftarrow}
  \C,\underline{E})}{W( \C,\underline{E} \overset{\F_a}{\leftarrow}
  \C',\underline{E}')}\sim e^{-\beta_a\left[\En(\C')-\En(\C)\right]}.  
\ee 
Now the right-hand side depends only on the configurations of the small system,
and the parameters $\beta_a$'s are constants. 
Letting the large parts get larger and larger while adjusting the physical
properties adequately, we can ensure that the
time over which \eqref{ergodic2} remains valid gets longer and longer, so, in
the thermodynamic limit for the large parts, the transient regime lasts
forever. This situation is our main interest in what follows. 

The transition rates in the transient regime correspond to a Markov matrix $\M$ defined by 
\be
 \label{defMarkovM}
(\C'\vert \M\vert \C)=
\begin{cases}
(\C'\vert \Trans\vert \C)   &\textrm{if}\quad  \C'\not=\C \\
-\sum_{\C''}(\C''\vert \Trans\vert \C)  &\textrm{if}\quad  \C'=\C.
\end{cases}
\ee
As well as the transition rates
$W(\C',\underline{E}'\overset{\F_a}{\leftarrow}\C,\underline{E})$  they must satisfy the three consequences derived from the properties of the underlying microscopic deterministic dynamics pointed out in subsubsection \ref{ErgodicityDiscrete}, namely ergodicity,  energy conservation and specific interaction pattern.
 First, as shown in subsubsection  \ref{Microcansubsubsection},  the Markov matrix  is irreducible,
or in other words the  graph associated with the transition rates is connected  (see \eqref{Mirreducible0}).
Second, from \eqref{ergodicRev} the transition rates must obey
 the microscopic reversibility condition \eqref{MicroRevCond}
 for any couple of configurations $(\C,\C')$. Third,
from \eqref{ergodic2} one gets 
a constraint  obeyed by
the ratio of transition rates in the transient regime,
\be
\label{ergodicThermo}
\textrm{for $\C'\in\F_a(\C)$}\quad
\frac{(\C'\vert \Trans \vert \C)}
{(\C\vert \Trans \vert \C')}=e^{-\beta_a\left[\En(\C')-\En(\C)\right]}.
\ee
The latter relation is the  so-called modified detailed balance (MDB), which is also referred to in the literature as the ``generalized detailed balance''.

We stress that, by selecting a time window while taking the thermodynamic limit for  the large parts, the microcanonical detailed balance \eqref{ratioWmc} is replaced by the modified detailed balance \eqref{ergodic}, except in the case of the largest time window where all $\beta_a$'s are equal. In the latter case, the microcanonical detailed balance \eqref{ratioWmc} is replaced by the canonical detailed balance and the statistical time reversal symmetry is preserved. Indeed, the equilibrium thermodynamic  regime is reached either if the we start from a situation in
which $\prod_a \Omega_a(E_a)$ is close to its maximum along the trajectory in the energy level $E_\text{dec}=E$, or if we 
wait long enough so that $\prod_a \Omega_a(E_a)$ becomes close to this maximum.
As recalled above, this is true for most of the period of the microscopic dynamics, but reaching this
situation may however take a huge number of time steps if the starting point was
far from the maximum.
By an argument similar to that used in the derivation of \eqref{DifferenceSB}, when $\prod_a \Omega_a(E_a)$ is closed to its maximum and the large parts are considered in the thermodynamic limit, all $\beta_a$'s are equal to the same value $\beta$ and  the relative weight of two configurations in the microcanonical ensemble, $\Probmc(\C',\underline{E}')/\Probmc(\C,\underline{E})$ given by \eqref{Probmicro}, is shown to tend to $\exp\left(-\beta[\En(\C')-\En(\C)]\right)$. Then the equilibrium microcanonical distribution $\Probmc(\C,\underline{E})$  tends to the canonical distribution   
\be
\label{expProbcan}
\Probcan^{\beta}(\C)\equiv\frac{e^{-\beta \En(\C)}}{Z(\beta)},
\ee
where $Z(\beta)$ is the canonical partition function at the inverse temperature $\beta$. 
Meanwhile, 
the detailed balance relation \eqref{ratioWmc} in the
microcanonical equilibrium ensemble for the transition rates
$W(\C',\underline{E}'\leftarrow\C,\underline{E})$ 
becomes a detailed balance relation in the canonical ensemble
at the inverse temperature $\beta$ of the whole system for the transition rates $(\C'\vert
\Trans\vert \C)$, namely
\be
\label{DetailedBalanceCan}
\frac{(\C'\vert \Trans \vert \C)}
{(\C\vert \Trans \vert \C')}=e^{-\beta\left[\En(\C')-\En(\C)\right]}
=\frac{\Probcan^{\beta}(\C')}{\Probcan^{\beta}(\C)}.
\ee
The modified detailed balance \eqref{ergodicThermo}, valid in transient regimes, differs from the  latter detailed balance  in the canonical ensemble  by two features : the various $\beta_a$'s of the distinct large parts appear in place of the
common equilibrium inverse temperature $\beta$, and the stationary distribution for the transition rates is not known a priori.

We conclude this discussion with the following remarks. We have not tried to
exhibit explicit physical descriptions of the large parts, or explicit
formul\ae\ for the dynamical map $\Udisc$. Though it is not too difficult to give
examples for fixed sizes of the large parts, it is harder to get a family of
such descriptions sharing identical physical properties for varying large part
sizes, a feature which is crucial to really make sense of the limits we took
blindly in our derivation. It is certainly doable, but
cumbersome, and we have not tried to pursue this idea.
Let us note also that in
principle, taking large parts of increasing sizes can be used to enhance  the
validity of the approximation of the (discrete-time) Markov chain by a
(continuous-time) Markov process. As the physics of the continuous time limit
does not seem to be related to the physics of convergence towards a heat bath
description we have preferred to keep the discussion separate, taking a
continuous-time description as starting point.

\subsubsection{Expression of MDB in terms of exchange entropy variation}

Observe that though we have given no detailed analysis of the dynamics or the
statistical properties of the large parts, their influence on the effective
Markov dynamics of the small system enters only through the inverse temperatures
$\beta_a$ defined in \eqref{defbeta}.  So we can consistently assume that each
large part becomes a thermal bath with its own temperature. The leading term in
$\SB_a(E'_a)-\SB_a(E_a)$ is the variation $\delta S_a^{\scriptscriptstyle TH}
(\C'\leftarrow \C)$ of the thermodynamic entropy of bath $a$ when it flips the small
system from configuration $\C$ to configuration $\C'$,
\be
\label{defdeltaStherm}
\delta S_a^{\scriptscriptstyle TH}
(\C'\leftarrow \C)=
\begin{cases}
\beta_a \left[E'_a-E_a\right]  &\textrm{if}\quad  \C'\in\F_a(\C) \\
0 &\textrm{otherwise}.
\end{cases}
\ee 
Then we have an idealized description of a thermal contact
between  heat baths.  This is the situation on which we concentrate in this
paper.
 
By definition of a heat source, the variation $\delta S_a^{\scriptscriptstyle TH}(\C'\leftarrow \C)$ of the thermodynamic entropy of bath $a$ when it flips the system from configuration $\C$ to configuration $\C'$  reads
\be
\label{deltaSatherm}
\delta \STH_a
(\C'\leftarrow \C)=- \beta_a \delta\Heatm_a(\C'\leftarrow\, \C),
\ee
where $\delta\Heatm_a(\C'\leftarrow \C)$ is the heat received by the small system from part $a$. According the expression \eqref{defdeltaStherm} and to the energy conservation relation \eqref{conserE},
\be
\label{defdeltaq}
\begin{cases}
\delta\Heatm_a(\C'\leftarrow\, \C)=\En(\C')-\En(\C)  &\textrm{if}\quad  \C'\in\F_a(\C) \\
 \delta\Heatm_a(\C'\leftarrow\, \C)=0 &\textrm{otherwise}.
  \end{cases}
\ee

  Let us introduce $\deltaexch S(\C'\leftarrow\, \C
)$ the exchange entropy variation of the small system (see footnote \ref{exchangename})  that  is associated with the heat exchanges  with the thermostats when the small system goes from configuration $\C$ to configuration $\C'$. Thanks to the definition \eqref{defdeltaStherm}
\be
\deltaexch S(\C'\leftarrow\, \C)
\equiv
-\sum_a\delta \STH_a(\C'\leftarrow\, \C),
\ee
namely, in the case of a pure  energy reservoir (which does not exchange particles)
\be
\label{defdeltaexchSbis}
\deltaexch S(\C'\leftarrow\, \C)\equiv
\sum_a\beta_a\delta\Heatm_a(\C'\leftarrow\, \C).
\ee
Then the modified detailed balance \eqref{ergodicThermo} can be rewritten  in a form which does not involve explicitly the heat bath responsible for the transition from $\C$ to $\C'$,
\be
\label{MDBexch}
\frac{(\C'\vert \Trans \vert C)}{(\C\vert \Trans \vert C')}=e^{-\deltaexch S(\C'\leftarrow\, \C)}.
\ee

\section{Master equation, exchange entropy flow and various entropy variations}

\label{Entropysection}

\subsection{Evolution of the probability distribution without MDB (known results)}
\label{EvolutionSec}

In this subsection we recall previously known results which are important milestones to our original results and which are consequences of the first two properties 
\eqref{Mirreducible0} and \eqref{MicroRevCond}, among the three mesoscopic conditions derived from the ergodicity of the underlying conservative deterministic dynamics.

The starting point is that the evolution of the probability $ \Prob(\C;t)$ that the  system is in configuration $\C$ at time $t$ is ruled by the master equation
\be
\label{MasterEquation0}
\frac{d \Prob(\C;t)}{dt}=
\sum_{\C'\not=\C}(\C\vert \Trans\vert \C') \Prob(\C';t)-\sum_{\C'
\not=\C}(\C'\vert \Trans\vert \C) \Prob(\C;t)=\sum_{\C'}(\C\vert \M\vert \C')\Prob(\C';t),
\ee
where $\M$ is defined in \eqref{defMarkovM}.
Since $\Prob(\C;t)$ is to be interpreted as a probability distribution, 
 it has to satisfy the  positivity and  normalization conditions,
\be
\label{NormProb}
\forall \C\qquad  \Prob(\C;t)\geq0\quad\textrm{and}\quad\sum_{\C}\Prob(\C;t)=1.
\ee

\subsubsection{Generic properties}

According to the theory of systems of ordinary differential equations with constant coefficients, the solution $\Prob(t)$ of the  equation \eqref{MasterEquation0} exists and is unique for a given initial function $\Prob(t=0)$. The Markov matrix $\M$ obeys the property $\sum_{\C,\C'}(\C\vert \M\vert \C')\Prob(\C';t)=0$ for any $\Prob$ and this ensures that the normalization constraint is preserved under the time evolution,
\be
\label{NormConservation}
\sum_{\C}\Prob(C;t=0)=1\qquad \Rightarrow\qquad \forall t>0\quad\sum_{\C}\Prob(C;t)=1.
\ee
The above property of $\M$ also ensures the existence of at least one stationary solution, but there is no argument for every stationary solution to obey the positivity constraint in  \eqref{NormProb} without further assumptions.

\subsubsection{Properties arising from the irreducibility condition}

When the transition rates  satisfy the irreducibility condition
\eqref{Mirreducible0}, and if positivity and normalization \eqref{NormProb} are satisfied at the initial time, the solution of the master equation \eqref{MasterEquation0} not only meets these conditions  for being a probability distribution at any subsequent time, but it  
even has the more stringent property that any configuration has a strictly non-vanishing weight,
\be
\label{WeightNonVanish}
\forall t>0\quad\forall \C\qquad  \Prob(\C;t)>0.
\ee
This result can be derived  in the framework of the theory of ordinary differential equations with constant coefficients \cite{Schnakenberg1976}.

Moreover the irreducibility condition \eqref{Mirreducible0} allows one to build at least formally a stationary solution which fulfills the conditions \eqref{NormProb} for being a probability (with the even more stringent property \eqref{WeightNonVanish}). This solution is obtained in the framework of graph theory 
by using the network representation of the master equation; the  corresponding  
 expression  is called Kirchhoff's theorem in Ref.\cite{Schnakenberg1976}.

The irreducibility condition \eqref{Mirreducible0} also entails that  the  stationary solution of the master equation is unique, and therefore coincides with the expression given by Kirchhoff's theorem.
The uniqueness of the stationary solution can be derived either in the framework of  matrix theory by using the Perron-Frobenius theorem (see for instance \cite{CoxMiller1965}) or in the framework of the theory of  ordinary differential equations with constant coefficients   \cite{Schnakenberg1976}.

The latter derivation uses the stability criterion introduced by Schögl \cite{Schlogl1971}. As noticed by Schlögl a good candidate for the Liapunov function involved in a stability criterion  is  the  relative entropy introduced by Kullback and Leibler \cite{KullbackLeibler1951} in the context of information theory, namely
\be
\label{defSrel}
S_\text{\mdseries rel}[\Prob(t)\vert \Probst]
\equiv \sum_{\C}\Prob(\C;t)\ln \frac{\Prob(\C;t)}{\Probst(\C)}.
\ee
The  relative entropy  is well-defined at any time according to \eqref{WeightNonVanish}. 
(In the generic case $S_\text{\mdseries rel}[\Prob\vert \Prob_0]$ is well defined  if $\Prob_0(\C)=0$ implies $\Prob(\C)=0$.)
The  definition of $S_\text{\mdseries rel}[\Prob\vert \Prob_0]$, with any given $\Prob_0$,  entails that $S_\text{\mdseries rel}[\Prob\vert \Prob_0]$ is positive for any $\Prob$
and  vanishes only when $\Prob$ is equal to $\Prob_0$ (by virtue of the inequality $\ln x< x-1$ if $x>0$ and $x\not=1$). Therefore
\be
\label{SrelPositive}
S_\text{\mdseries rel}[\Prob\vert \Probst]> 0\quad\textrm{if $\Prob\not=\Probst$ and}\quad
S_\text{\mdseries rel}[\Probst\vert \Probst]=0.
\ee
 The definition also ensures that $S_\text{\mdseries rel}[\Prob\vert \Prob_0]$ is convex (i.e. concave upward) for any $\Prob$, so that
\be
\label{SchloglAssumption}
\delta^{(2)}S_\text{\mdseries rel}[\Prob\vert \Probst]>0\quad\textrm{for any $\Prob$},
\ee
with 
\be
\delta^{(2)}f[\Prob]\equiv \frac{1}{2}\sum_{\C,\C'}\left.\frac{\partial^2 f}{\partial \widetilde{\Prob}(\C)\partial \widetilde{\Prob}(\C')}\right\vert_{\widetilde{\Prob}=\Prob}\delta \Prob(\C)\delta \Prob (\C')
\ee
and $\delta \Prob(\C)\equiv\widetilde{\Prob}(\C)-\Prob(\C)$.
Moreover, because of  the structure of the master equation \eqref{MasterEquation0} combined with the properties $\ln x\leq x-1$ for any $x>0$ and $\sum_{\C'}(\C\vert \M\vert \C')\Probst(\C')=0$, the  time derivative of $S_\text{\mdseries rel}[\Prob(t)\vert \Probst]$ is negative at any time \cite{Schnakenberg1976},
\be
\label{dSreldt}
\frac{dS_\text{\mdseries rel}[\Prob(t)\vert \Probst]}{dt}\leq 0 .
\ee
The properties \eqref{SrelPositive} and \eqref{dSreldt} define a Liapunov function  and 
ensure that for  any initial distribution $\Prob(t_0)$ \textit{in the vicinity} of $\Probst$
$\lim_{t\to+\infty}\Prob(t)=\Probst$ (because the property \eqref{SrelPositive} ensures that 
$\delta^{(2)}S_\text{\mdseries rel}[\Prob\vert \Probst]>0$ for any $\Prob$  in the vicinity of $\Probst$). With the extra property  \eqref{SchloglAssumption} one can apply the stability theorem by Schlögl which states that
 $\lim_{t\to+\infty}\Prob(t)=\Probst$  \textit{for any} initial distribution $\Prob(t=0)$.
The interpretation given by  Schögl  of the stability condition $d S_\text{\mdseries rel}[\Prob(t)\vert \Probst]/dt\leq 0$ can be rephrased as follows. If the observer does not know more than that the system  was initially  in some unknown state $\Prob_0$ of the stability region, then an unbiased estimate for the  state  $\Prob(t)$ at time $t$ would be $\Probst$. But if the observer knows $\Prob_0$ by measurements, then his excess knowledge at initial time is equal to $S_\text{\mdseries rel}[\Prob_0\vert \Probst]$, and the property $d S_\text{\mdseries rel}[\Prob(t)\vert \Probst]/dt\leq 0$ reflects the fact that the spontaneous development of the states after the last observation can only go such that this knowledge does not increase.

\subsection{Microscopic currents}
\label{MicroCurrentSec}

\subsubsection{Definitions}

Our results in the framework of Markovian stochastic dynamics described by a master equation can be written in compact forms if we introduce the generic current $\jinst(\C)$ for a microscopic variation when the system goes out of a given configuration $\C$. Such a current is 
associated either with the variation $\Oobs(\C')-\Oobs(\C)$ of a configuration observable $\Oobs(\C)$ and then
\be
\label{defjinst0}
\jinst_{\Oobs}(\C)\equiv \sum_{\C'}(\C'\vert \Trans \vert
\C)\left[\Oobs(\C')-\Oobs(\C)\right], \ee 
or more generally with some exchange quantity $\delta K(\C'\leftarrow\, \C)$ such
as a heat amount or the variation of the reservoir entropies when the system
jumps from configuration $\C$ ot $\C'$, and then 
\be
\label{defjinstdeltaK}
\jinst_{\delta K}(\C)\equiv \sum_{\C'}(\C'\vert \Trans \vert \C)\delta
K(\C'\leftarrow\, \C).  \ee 
We include in the definition of an exchange quantity that it obeys the
antisymmetry property $\delta K(\C'\leftarrow\, \C)=-\delta K(\C\leftarrow\,
\C')$. We notice that the above definitions are valid even if the observable
$\Oobs$ (or the variation $\delta K(\C'\leftarrow\, \C)$) depends explicitly on
time.

The average of a current $\jinst(\C)$ at time $t$  is given by the generic formula   for the  mean value $\Esp{\Oobs}_t$ of a configuration observable $\Oobs(\C)$ at time $t$,  namely  \be
\label{defEspOb}
\Esp{\Oobs}_t\equiv\sum_{\C}\Oobs(\C)\Prob(\C;t).
\ee
According to the master equation \eqref{MasterEquation0} the time derivative of the mean value $\Esp{\Oobs}_t$ of a configuration observable $\Oobs(\C)$ which does not depend explicitly on time is equal to the mean  value $\Esp{\jinst_\Oobs}_t$ of the associated current $\jinst_\Oobs(\C)$,
\be
\label{MeanAEvolution}
\frac{d \Esp{\Oobs}_t}{dt}=\Esp{\jinst_\Oobs}_t,
\ee
where $\jinst_\Oobs(\C)$ is the current \eqref{defjinst0} of observable $\Oobs$ associated with the stochastic jumps going out of configuration $\C$.

\subsubsection{Probabilistic interpretation of microscopic observable currents}

With every observable i.e. with every real-valued function $\Oobs(\C)$ on the
configuration space, one can associate the random process $\Oobs(\C _t)$ where $\C
_t$ denotes the configuration of the system at time $t$. Note that
$\Prob(\C;t)$ is nothing but $\Prob(\C _t = \C)$, so that $\Esp{\Oobs}_t$ as
defined above could also be written $\Esp{\Oobs(\C _t)}$.

The process associated
with the current observable $\jinst_\Oobs(\C)$ at time $t$ has a clear probabilistic
meaning. Without explaining the details, let us just say that the decomposition
\be
\Oobs(\C _t)=\left(\Oobs(\C _t)-\int_0^t ds \jinst_\Oobs(\C _s)\right) +
\left(\int_0^t ds \jinst_\Oobs(\C _s)\right)\equiv M_t+N_t,
\ee
called the Doob-Meyer decomposition of the process $\Oobs(\C _t)$, expresses
$\Oobs(\C _t)$ as the sum of a martingale $M_t$ and a predictible process $N_t$
vanishing at $t=0$. Such a decomposition is unique. Informally a martingale is a
process whose expectation in the future knowning all the past up to now is equal
to its present value. In particular, the expectation $\Esp{M_t}$ is equal to
$0$.  Taking this as a fact, one gets immediately \eqref{MeanAEvolution}.
Informally again, $N_t$ is predictible because its value at $t+dt$, $N_{t+dt}$,
is not sensitive to the randomness (i.e. to a possible jump occuring) between
$t$ and $t+dt$.

Even if $\Oobs(\C _t)$ depends only on the configuration at
time $t$, this is not true anymore of $M_t$ and $N_t$ which in general depend on
the history up to time $t$. Moreover, the decomposition does not behave
trivially under nonlinear maps, so that $N_t^2$ is not the predictible process that
appears in the Doob-Meyer decomposition of $[\Oobs(\C _t)]^2$.

\subsubsection{Mean energy time derivative}

As an example of \eqref{MeanAEvolution}, the time derivative of the mean energy $\Esp{\En}_t$ is equal to the mean value of the energy current $\jinst_{\En}(\C)$. In the present model, energy variations are only due to heat exchanges with the two thermostats according to the conservation rule
\be
\label{ConservEnS}
\En(\C')-\En(\C)=\delta\Heatm_1(\C'\leftarrow\, \C)+\delta\Heatm_2(\C'\leftarrow\, \C),
\ee
where $\delta\Heatm_a(\C'\leftarrow \C)$ is the heat received from thermal bath $a$ for a jump of configuration as defined in \eqref{defdeltaq}.
Therefore the energy current $\jinst_{\En}(\C)$ can be split into two heat currents $j_{\delta q_1}(\C)$ and $j_{\delta q_2}(\C)$ received from the thermal  baths $1$ and $2$ respectively.
Then the evolution equation \eqref{MeanAEvolution} applied to the  energy observable $\En$ can be rewritten as
\be
\label{derivEav}
\frac{d \Esp{\En}_t}{dt}=\Esp{\jinst_1}_t+\Esp{\jinst_2}_t,
\ee
where $\jinst_a$ is a short notation for the instantaneous heat current  received from  thermal bath $a$ by the system when it leaves configuration $\C$ :
\be
\label{defjinstHeat}
\jinst_a(\C)\equiv \jinst_{\delta q_a}(\C)\equiv
\sum_{\C'}(\C'\vert \Trans\vert \C) \,\delta\Heatm_a(\C'\leftarrow \C).
\ee

Note that, unless $\beta_1=\beta_2$,
 there is no observable ${\cal O}_a$ for which $\jinst_{\delta q_a}(\C)$ would be equal to $\jinst_{\Oobs_a}(\C)$ with the definition of $\jinst_{\Oobs_a}(\C)$ given in \eqref{defjinst0}. Moreover, under the microscopic reversibility condition \eqref{MicroRevCond},  if $\beta_1=\beta_2=\beta$  the modified detailed balance \eqref{ergodicThermo} becomes the canonical detailed balance \eqref{DetailedBalanceCan}.
Then the current associated with any observable $\Oobs_a$ (more generally any
exchange quantity) has a zero mean in the stationary equilibrium state with
distribution $\Probcan^{\beta}$ : $\Esp{\jinst_{\Oobs_a}}_{\Probcan^{\beta}}=0$.

\subsubsection{Exchange entropy flow}

The heat currents are associated with an exchange entropy variation of the system (see footnote \ref{exchangename} for the meaning) according to the relation \eqref{defdeltaexchSbis}. Similarly to the definition \eqref{defjinstHeat} of the microscopic heat current $\jinst_a(\C)$ in terms of the heat amount $\delta\Heatm_a(\C'\leftarrow \C)$, the microscopic exchange entropy current $\jinst_{\deltaexch S}(\C)$ when  the system  goes out of  configuration $\C$ is defined from the exchange entropy variation $\deltaexch S(\C'\leftarrow\,\C)$  as
\be
\label{defjinstexch}
\jinst_{\deltaexch S}(\C)\equiv\sum_{\C'} (\C'\vert \Trans\vert \C) \deltaexch S(\C'\leftarrow\,\C).
\ee
The  exchange entropy flow   is defined as
\be
\label{exchangeentropyflow0}
\frac{\dexch S}{dt}\equiv\sum_{\C,\C'}(\C'\vert\Trans\vert\C)\deltaexch S(\C'\leftarrow\, \C)\Prob(\C;t).
\ee
It  can be expressed as the average of the current $\jinst_{\deltaexch S}(\C)$,
\be
\label{exchangeentropyflow}
\frac{\dexch S}{dt}=\Esp{\jinst_{\deltaexch S}}_t.
\ee
The current definitions \eqref{defjinstHeat} and \eqref{defjinstexch} together with the relation \eqref{defdeltaexchSbis} between the exchange entropy variation and heat transfers imply that
\be
\label{SexGDB}
\frac{\dexch S}{dt}=\beta_1\Esp{\jinst_1}_t+\beta_2\Esp{\jinst_2}_t.
\ee

In the stationary state, the time derivative of the mean energy vanishes, namely ${d <\En>_\text{st}/dt=0}$, where  $\Espst{\cdots}$ denotes an average with the stationary distribution $\Probst$. Henceforth, according to the evolution equation \eqref{derivEav} of $<\En>_\text{st}$,
\be
\label{balancejins}
\Espst{\jinst_1}+\Espst{\jinst_2}=0.
\ee
By inserting the current balance \eqref{balancejins} into the expression \eqref{SexGDB} of the  exchange entropy flow, we  get
\be
\label{dexchSst}
\left.\frac{\dexch S}{dt}\right\vert_\text{st}=-(\beta_1-\beta_2)\Espst{\jinst_2}.
\ee

\subsection{Evolution of the Shannon-Gibbs entropy}
\label{EvolSSGSec}

\subsubsection{Definition of the entropy production rate}

The dimensionless  Shannon-Gibbs entropy (where the Boltzmann constant is set equal to 1) is defined from the configuration  probability distribution $\Prob(\C;t)$ as
\be
\label{defSG}
\SG[\Prob(t)]\equiv- \sum_{\C}\Prob(\C;t)\ln\Prob(\C;t)=-\Esp{\ln \Prob(t)}_t.
\ee
When the evolution is ruled by the master equation \eqref{MasterEquation0}, the time derivative of $\SG(t)$ takes the form
\be
\label{dSGdt0}
\frac{d \SG}{dt}=-\sum_{\C,\C'}(\C'\vert\Trans\vert\C) \Prob(\C;t)\ln \frac{\Prob(\C';t)}{\Prob(\C;t)}=
-\Esp{\jinst_{\ln\Prob(t)}}_t,
\ee
where we have used the definition \eqref{defjinst0} of an observable current $\jinst_{\Oobs}(\C;t)$, also  valid in the case of an observable which depends explicitly on time.

As it is done for the phenomenological entropy introduced in the thermodynamics of irreversible processes \cite{DeGroot1952,Pottier2007} the time derivative of $\SG(t)$  can be split into two contributions, an exchange (or external) part arising from exchanges with the external reservoirs, $\dexch S/dt$,
 and an internal  (or  irreversible) part   due to the internal irreversible processes in the system, $\dint \SG/dt$,
\be
\label{defdirrSG0}
\frac{d \SG}{dt} \equiv \frac{\dexch S}{dt}+\frac{\dint \SG}{dt}.
\ee
By virtue of its definition  \eqref{exchangeentropyflow0}, $\dexch S/dt$ is expressed in terms of the exchange entropy variation $\deltaexch S(\C'\leftarrow\, \C)$, associated with a jump  of the  system from a microscopic configuration  $\C$ to another one $\C'$.
By virtue of the definition \eqref{defdirrSG0} $\dint \SG/dt$ is a functional of $\Prob(\C;t)$ determined from $d \SG/dt$ given  in \eqref {dSGdt0} and $\dexch S/dt$ given in
\eqref{exchangeentropyflow0}.  

\subsubsection{Comparison with the thermodynamics of irreversible processes}

An implicit postulate in the modern literature is that, when the system is out of equilibrium and evolves on time scales far smaller than the reservoirs, one can still define some universe entropy with the following properties. It coincides with its equilibrium statistical  expression when the system and the reservoirs are at equilibrium; when equilibrium conditions are not fulfilled, the equilibrium Gibbs entropy of the system is replaced by its instantaneous Shannon-Gibbs entropy while the reservoir entropies  can be approximated by  their thermodynamic entropies. In other words, the variation of the out-of-equilibrium universe entropy   is the sum of the variation of  the Shannon-Gibbs entropy of the system  and the variation of the total thermodynamic entropy  of the reservoirs, namely 
\be
\label{PostulateSG}
\frac{d S_\text{Univ}}{dt}=\frac{d \SG}{dt}+\frac{d S^{\scriptscriptstyle TH}_\text{res}}{dt}.
\ee
On the other hand, 
 by definition, the exchange entropy flow $\dexch S/dt$
received by the system from the reservoirs  
is  the opposite of the time-derivative of the   thermodynamic entropy of the reservoirs
$\dexch S/dt\equiv - d S^{\scriptscriptstyle TH}_\text{res}/dt$.
 As a consequence, by virtue of definition \eqref{defdirrSG0},
\be
\label{PostulatedirrSG0}
 \frac{d S_\text{Univ}}{dt}=
\frac{\dint \SG}{dt}.
\ee
The interpretation of the latter equality is that, since the reservoirs are at thermodynamic equilibrium,  the  variation rate of the universe entropy, $dS_\text{Univ}/dt$, is equal to the production rate of the Shannon-Gibbs entropy  in the system, namely the internal part $\dint \SG/dt$ in the time-derivative $d \SG/dt$  of the system Shannon-Gibbs entropy. 
As shown below, if the modified detailed balance is obeyed, the universe entropy increases when the system is out-of-equilibrium, as   the universe entropy increases between two equilibrium states, according to the second principle of thermodynamics.

In the stationary state the time derivative of the Shannon-Gibbs entropy \eqref{defSG}  vanishes,
$d\SG[\Probst]/dt=0$, and the decomposition \eqref{defdirrSG0} leads to
\be
\label{reldintdexchst}
\left.\frac{\dint \SG}{dt}\right\vert_\text{st}=-\left.\frac{\dexch S}{dt}\right\vert_\text{st}.
\ee
Then according to \eqref{dexchSst}, $\dint \SG/dt\vert_\text{st}$ can be written as the entropy production rate introduced in the framework of irreversible thermodynamics when there is only one independent mean instantaneous current $\Jcumav$ (see for instance 
Ref.\cite{Jaynes1980}), namely in the form
\be
\label{defAffBis}
\left.\frac{\dint \SG}{dt}\right\vert_\text{st}=\Fth\Jcumav,
\ee
where $\Fth$ is the so-called thermodynamic force  associated with the heat current $\Jcumav$. In the present case $\Espst{\jinst_2}=-\Espst{\jinst_1}$ and in \eqref{defAffBis} we can make the identification
\be
\label{defAffBisDetail}
\Jcumav=\Espst{\jinst_2}
\quad\textrm{and}\quad \Fth=\beta_1-\beta_2.
\ee
(Another identification might have been $\Jcumav=\Espst{\jinst_1}$ and $\Fth=\beta_2-\beta_1$.)
If $T_2>T_1$, $\Fth$ is  positive and the positivity of  
$\dint \SG_\text{st}/dt$ (settled in next subsubsection  when the modified detailed balance is satisfied) ensures that the mean current $\Espst{\jinst_2}$ received from  heat bath $2$ is also positive.

\subsubsection{Entropy production rate under MDB}
\label{EntropyProductionMDB}

When the transition rates obey the modified detailed balance 
\eqref{MDBexch}, the microscopic exchange entropy current defined in 
\eqref{defjinstexch} can be rewritten as
\be
\label{expjinstdeltaexchS}
\jinst_{\deltaexch S}(\C)\underset{MDB}{=}-\sum_{\C'}
 (\C'\vert \Trans \vert \C)\ln\frac{(\C'\vert \Trans\vert \C)}{(\C\vert \Trans\vert \C')}.
\ee
Then the exchange entropy flow defined in \eqref{exchangeentropyflow0}
 becomes equal to
\be
\label{sigexch0}
\frac{\dexch S}{dt}\underset{MDB}{=}-\sum_{\C,\C'}(\C'\vert\Trans\vert\C)\Prob(\C;t)\ln\frac{(\C'\vert\Trans\vert\C) }{(\C\vert\Trans\vert\C')}\equiv \sigexch[\Prob(t)],
\ee
while the entropy production rate determined by 
\eqref{dSGdt0} and \eqref{defdirrSG0} becomes equal to
\be
\label{sigirr0}
\frac{\dint \SG}{dt}\underset{MDB}{=}\sum_{\C,\C'}(\C'\vert\Trans\vert\C) \Prob(\C;t)
 \ln\frac{(\C'\vert\Trans\vert\C) \Prob(\C;t)}{(\C\vert\Trans\vert\C') \Prob(\C';t)}\equiv\sigint[\Prob(t)]
\ee
The latter expression takes the form of a relative entropy so that $\dint \SG/dt$ is positive under  MDB.

We notice that, in the generic case where the modified detailed balance does not necessarily holds,  Lebowitz and Spohn \cite{LebowitzSpohn1999}  have introduced the splitting of $d\SG/dt$ into $\sigexch+\sigint$, where $\sigexch[\Prob(t)]$ is defined in \eqref{sigexch0} and 
$\sigint[\Prob(t)]$ is defined as the symmetrized expression of the definition in \eqref{sigirr0}, 
\be
\label{defsintLS}
\sigint[\Prob(t)]=\frac{1}{2}\sum_{\C,\C'}
\left[(\C'\vert \Trans \vert \C)\Prob(\C;t)-(\C\vert \Trans \vert \C')\Prob(\C';t)\right]
\ln\frac{(\C'\vert \Trans\vert \C)\Prob(\C;t)}{(\C\vert \Trans\vert \C')\Prob(\C';t)},
\ee
which is obviously positive.
Another rewritting for the definition of $\sigint[\Prob(t)]$  given in \eqref{sigirr0}  is
\be
\label{defdirrSGdt}
\sigint[\Prob(t)]=\Esp{\jinst_{\deltaint S}(t)}_t \quad\textrm{with}\quad
\deltaint S(\C'\leftarrow\, \C)\equiv \ln\frac{(\C'\vert \Trans\vert
  \C)\Prob(\C;t)}{(\C\vert \Trans\vert \C')\Prob(\C';t)}.  \ee Then the
positivity of $\sigint[\Prob(t)]$ may be viewed as arising from the property :
for $x>0,$ $-\ln x\geq 1-x$.  We notice that $\jinst_{\deltaint S}(\C;t)$ is not
the current introduced in \cite{LebowitzSpohn1999}. However, the current
$\jinst_{\deltaexch S}(\C)$ is introduced through its expression
\eqref{expjinstdeltaexchS} in the latter reference and it has been used in the literature, possibly under another
denomination (see for instance \cite{LecomteETAL2005}).

\subsubsection{Decomposition of the entropy production rate into two positive contributions}
 
The microscopic variation $\deltaint S(\C'\leftarrow\, \C)$ defined in \eqref{defdirrSGdt} can be decomposed into two contributions as
\be
\label{decompdeltaint}
\deltaint S(\C'\leftarrow\, \C)=
-\left[\ln \frac{\Prob(\C';t)}{\Probst(\C')}-\ln \frac{\Prob(\C;t)}{\Probst(\C)}\right]
+\delta \Aff_{[\Probst]}(\C'\leftarrow\, \C),
\ee
where
\be
\label{deltaAffst}
\delta \Aff_{[\Probst]}(\C'\leftarrow\, \C)\equiv \ln\frac{(\C'\vert \Trans\vert \C)\Probst(\C)}{(\C\vert \Trans\vert \C')\Probst(\C')}.
\ee
By analogy with  chemical reaction kinetics (see for instance Ref.\cite{Schnakenberg1976} and subsubsection \ref{ComparisonGraph} below), $\delta \Aff_{[\Probst]}(\C'\leftarrow\, \C)$ may be viewed as the affinity of the 
elementary reversible reaction (or phase change) $\C\rightleftharpoons \C'$ in the stationary state resulting from all possible reversible pair reactions between all configurations. 
We notice that $\deltaint S(\C'\leftarrow\, \C)$ depends explicitly on time, contrarily to
$\delta \Aff_{[\Probst]}(\C'\leftarrow\, \C)$.  According to the definitions \eqref{defjinst0} and \eqref{defjinstdeltaK}
 of the currents associated respectively with an observable $\Oobs$ or with some exchange quantity $\delta K(\C'\leftarrow\, \C)$,   the decomposition \eqref{decompdeltaint} allows to rewrite the expression \eqref {defdirrSGdt} for $\sigint[\Prob(t)]$ as
\be
\label{jinstdeltaintSdecomp}
\sigint[\Prob(t)]=-
\Esp{\jinst_{\ln[\Prob(t)/\Probst]}}_t+\Esp{\jinst_{\delta \Aff[\Probst]}}_t.
\ee
We stress that, by virtue of the already used inequality   $-\ln x\geq 1-x$ for $x>0$ and according to the stationary condition $\sum_{\C'\not=\C}(\C\vert \Trans\vert \C') \Probst(\C')-\sum_{\C'\not=\C}(\C'\vert \Trans\vert \C) \Probst(\C)=0$ (see \eqref{MasterEquation0}), the microscopic current  $\jinst_{\delta \Aff[\Probst]}(\C)$ that we have introduced is positive,
\be
\label{jinspositiv}
\jinst_{\delta \Aff[\Probst]}(\C)\geq 0 \quad\textrm{for any $\C$}.
\ee
The latter positivity may be viewed as a consequence of the fact that
$\jinst_{\delta \Aff[\Probst]}(\C)$ is the relative entropy (for a given $\C$) of the rate $(\C'\vert \Trans \vert \C)$ with respect to the rate 
$(\C'\vert \Trans^{\dag} \vert \C)\equiv 
\left[\Probst(\C)\right]^{-1}(\C\vert \Trans \vert \C')\Probst(\C')$. 

According to the master equation \eqref{MasterEquation0} and the definition \eqref{defSrel},  the first contribution in the r.h.s. of \eqref{jinstdeltaintSdecomp} is  the opposite of the time derivative of 
 the relative entropy of the probability distribution $\Prob(t)$ with respect to the stationary solution $\Probst$. Therefore
the expression  \eqref{defdirrSGdt} of the entropy production rate $\sigint[\Prob(t)]$ can be split  into two contributions,  both of which are positive,
\be
\label{sigintSplit}
\sigint[\Prob(t)]=- \frac{dS_\text{\mdseries rel}[\Prob(t)\vert \Probst]}{dt}+\Esp{\jinst_{\delta \Aff[\Probst]}}_t.
\ee
The positivity of the first term has played a role in the discussion of the uniqueness of the stationary state (see Eq.\eqref{dSreldt}) and the positivity of the second term arises from the positivity of the current  pointed out in  \eqref{jinspositiv}.

For the sake of completeness, we notice that,  when the transition rates depend on time  the master equation \eqref{MasterEquation0} remains unchanged, so that the decomposition of the time-derivative of the Shannon-Gibbs entropy into $\sigexch+\sigint$  as well as the 
 decomposition \eqref{jinstdeltaintSdecomp} of $\sigint$ remain unchanged, except that $\Probst$ is to be replaced by a function of time $\Prob^{(0)}(t)$, namely the probability distribution that is the zero right-eigenvector of the Markov matrix  at a given time $t$. 
Then the splitting  \eqref{jinstdeltaintSdecomp} of $\sigint$ corresponds to  that introduced in  Ref.\cite{EspositoVanDenBroeck2010PREa} for an evolution where both the bath temperature and the energy levels may change with time. In the splitting of the latter reference $\Probst$  is to be replaced by $\Prob^{(0)}(t)$, and  then the so-called non-adiabatic term corresponds to minus the average of the current associated with the observable $\ln[\Prob(t)/\Prob^{(0)}(t)]$ while the so-called adiabatic term corresponds to the average of the  affinity current $\Esp{\jinst_{\delta \Aff[\Prob^{(0)}(t)]}}_t$.

\subsubsection{Comparison with NESS characterization from  graph theory}
\label{ComparisonGraph}

The r.h.s. of the master equation \eqref{MasterEquation0} can be rewritten as a sum of probability currents between configurations, 
\be
\label{MasterCurrent}
\frac{d \Prob(\C;t)}{dt}=\sum_{\C'}\Jcumav_{[\Prob]}(\C,\C';t),
\ee
where 
\be
\label{defJedge}
\Jcumav_{[\Prob]}(\C',\C;t)\equiv (\C'\vert\Trans\vert\C) \Prob(\C;t)-(\C\vert\Trans\vert\C') \Prob(\C';t).
\ee
The probability current  $\Jcumav_{[\Prob]}(\C',\C;t)$ has the form of the  chemical reaction rate associated with the reversible reaction (or phase change) $\C\rightleftharpoons \C'$ with the species concentrations replaced by the configuration probabilities $\Prob(\C;t)$ and $\Prob(\C';t)$ and the reaction rate constant for $\C\rightharpoonup \C'$ replaced by
the transition rate $(\C'\vert\Trans\vert \C)$.
The entropy production rate $\sigint[\Prob(t)]$, introduced in Ref.\cite{LebowitzSpohn1999} in the generic case (namely when the MDB is not necessarily  satisfied) by the definition \eqref{defsintLS}, can be rewritten (with  notations similar to those of Ref.\cite{Schnakenberg1976}) as
\be
\label{sigirrCurr0}
\sigint[\Prob(t)]=\frac{1}{2}\sum_{\C,\C'}\delta\Aff_{[\Prob]}(\C', \C;t)\Jcumav_{[\Prob]}(\C',\C;t), 
\ee
where the dimensionless affinity $\delta\Aff_{[\Prob]}(\C', \C;t)$ of an oriented pair (already introduced in the case of the stationary distribution $\Probst$  in  \eqref{deltaAffst})  is defined  as
\be
\label{deltaAff}
\delta\Aff_{[\Prob]}(\C', \C;t)\equiv \ln\frac{(\C'\vert \Trans\vert \C)\Prob(\C;t)}{(\C\vert \Trans\vert \C')\Prob(\C';t)}.
\ee
Indeed, in the irreversible thermodynamics of chemical reactions at a temperature $T$ fixed by a thermostat (see for instance \cite{Prigogine1968}),  the  affinity of the  reaction $\C \rightleftharpoons \C'$ is equal to $\kB T$ times an expression similar to $\delta\Aff_{[\Prob]}(\C', \C;t)$, with the concentration of species $\C$ ($\C'$) in place of $\Prob(\C;t)$ ($\Prob(\C';t)$)  and the reaction rate constant for $\C\rightharpoonup \C'$ in place of
the transition rate $(\C'\vert\Trans\vert \C)$.

 The previous  rewritings are convenient to handle  the master equation  in the framework of network theory where the master equation is represented by a graph $G$ as follows. Each vertex of the graph corresponds to a given configuration $\C$ and there exists an edge between two vertices if  at least one of the transition rates $(\C'\vert \Trans \vert \C)$ or $(\C\vert \Trans \vert \C')$ does not vanish.  
Moreover an arbitrary orientation is chosen for every edge in the graph $G$ so that the transition rate  $(\C'\vert \Trans \vert \C)$ (or $(\C\vert \Trans \vert \C')$) can be shortly referred to as the transition rate along the edge in either the positive sense or the negative sense.

 From a connected graph $G$ one can define   several possible fundamental sets of $N_c$ circuits (or closed paths)  on the graph. 
 The number $N_c$ of circuits is only determined by the edge number $N_e$ and the vertex number $N_v$  through the relation $N_c=N_e-N_v+1$. 
 (The final results are independent  of the specific fundamental set used in intermediate algebraic calculations.)
 A  fundamental set is built from one among the 
 various possible maximal (or spanning) trees which are defined by removing $N_c$ edges of $G$. 
  For a given maximal tree $T(G)$ the corresponding removed edges are called chords and indexed by $\alpha=1,\ldots,N_c$.
For each $\alpha$  the  circuit $C_\alpha$ is obtained  by first considering the  graph made by adding the chord $\alpha$ to $T(G)$   and then   removing all edges which are not part of the circuit closed by the insertion of the chord $\alpha$.
A cycle $\vec{C}_\alpha$ is  associated with every circuit $C_\alpha$ by
choosing an arbitrary orientation to go around the circuit, so ``cycle'' is a
synonymous for ``oriented circuit''. 

The affinity $\Aff(\vec{C}_\alpha)$ associated with a cycle $\vec{C}_\alpha$  is defined as an algebraic sum of  all edge affinities which is calculated as follows:  each edge affinity, whose expression is given by   \eqref{deltaAff} for a positive orientation from $\C$ to $\C'$ in the graph $G$, is multiplied by the sign of the relative orientation of the edge in the cycle $\vec{C}_\alpha$ and in the graph $G$. Because of the cyclic structure of $\vec{C}_\alpha$ the affinity $\Aff(\vec{C}_\alpha)$ does not depend explicitly on  the configuration probability distribution $\Prob$ and  it is  determined only from   the transition rates $(\C'\vert\Trans\vert \C)$. Indeed, if $N({\vec{C}_\alpha})$ is the number of configurations involved in the cycle 
$\vec{C}_\alpha$, and if the configurations are labeled with indices increasing with one unit when one goes from one configuration to the next one in the sense chosen for the orientation of the cycle $\vec{C}_\alpha$, then
\be
\label{valueAffalpha}
\Aff(\vec{C}_\alpha)=
\ln \prod_{i=1}^{N(\vec{C}_\alpha)}
\frac{(\C_{i+1}\vert \Trans \vert \C_{i}) }{(\C_{i}\vert \Trans\vert \C_{i+1}) },
\ee
with the notational convention $\C_{N(\vec{C}_\alpha)+1}\equiv \C_1$.
The probability current $\Jcumav_{[\Prob]}(\vec{C}_\alpha;t)$ associated with the cycle  is defined as the probability current in the chord  $\alpha$ in the sense of the cycle orientation : it is given by  definition \eqref{defJedge} where $\C$ and $\C'$ are respectively the initial and final configurations
in the sense of the cycle orientation.

As shown in Ref.\cite{Schnakenberg1976}, the stationary state  has the following properties.  First,
 the vanishing of all cycle affinities $A(\vec{C}_\alpha)$ is equivalent to 
 the vanishing of all cycle probability currents $\Jcumav_{[\Probst]}(\vec{C}_\alpha)$ in the stationary state,
\be
\label{VanishAJcum}
\forall \alpha \quad A(\vec{C}_\alpha) =0
 \qquad \Leftrightarrow \qquad 
\forall \alpha \quad\Jcumav_{[\Probst]}(\vec{C}_\alpha)=0.
\ee
Second, the vanishing of all stationary cycle probability currents $\Jcumav_{[\Probst]}(\vec{C}_\alpha)$ is equivalent to the fact that the stationary state obeys the detailed balance condition, namely
\be
\label{detailedbalanceG}
\forall \alpha \quad\Jcumav_{[\Probst]}(\vec{C}_\alpha)=0 \qquad \Leftrightarrow 
\qquad \forall (\C,\C')\quad
\frac{(\C'\vert\Trans\vert\C) \Probst(\C)}{(\C\vert\Trans\vert\C') \Probst(\C')}=1.
\ee
Schnakenberg specifies that it is in fact  a ``complete detailed balance'' condition  in the sense that if there exist several kinds of independent transitions between two configurations $\C$ and $\C'$, the detailed balance must be satisfied by every kind of transition.
Since equilibrium is characterized at the mesoscopic level by the detailed balance (see the derivation of  \eqref{DetailedBalanceCan}), 
the  properties  \eqref{VanishAJcum} and \eqref{detailedbalanceG} entail that the equilibrium is characterized by either the vanishing of all stationary cycle currents or the vanishing of all cycle affinities.
Moreover, the entropy production rate \eqref{sigirrCurr0}  in the stationary state reads
\be
\label{sigintcycles}
\sigint[\Probst]=\sum_{\alpha=1}^{N_c}A(\vec{C}_\alpha) )\Jcumav_{[\Probst]}(\vec{C}_\alpha),
\ee
where $A(\vec{C}_\alpha)$ is given directly in terms of the transition rates by the formula \eqref{valueAffalpha}. By virtue of the definitions \eqref{sigexch0} and \eqref{sigirr0}, 
$\sigexch[\Prob(t)]+\sigint[\Prob(t)]=d \SG/dt$ and in the stationary state 
$\sigint[\Probst]=-\sigexch[\Probst]$.

 When the MDB is satisfied,  according to \eqref{sigexch0}, $\sigexch[\Probst]\underset{MDB}{=}\dexch S/dt\vert_\text{st}$  so that 
 $\sigint[\Probst]\underset{MDB}{=}-\dexch S/dt\vert_\text{st}$ and \eqref{sigintcycles} becomes
 \be
 \label{sigexchcycles}
\left.\frac{\dexch S}{dt}\right\vert_\text{st}\underset{MDB}{=}-\sum_{\alpha=1}^{N_c}A(\vec{C}_\alpha) )\Jcumav_{[\Probst]}(\vec{C}_\alpha).
 \ee
Meanwhile, by virtue of \eqref{MDBexch0} the affinity of a cycle $A(\vec{C}_\alpha)$ given by  
 \eqref{valueAffalpha} becomes equal to
\be
\label{valueAffalphaMDB}
\Aff(\vec{C}_\alpha)\underset{MDB}{=} -\sum_{i=1}^{N({\vec{C}_\alpha})}
 \deltaexch S(\C_{i+1}\leftarrow\, \C_{i}),
\ee
where the latter expression depends on the thermodynamic parameters of the reservoirs and on the quanta of microscopic quantities that the system exchanges with the reservoirs.
In the generic case there may be several cycles corresponding to the same current between two reservoirs and there is no straightforward correspondence between the expression of
 $-\dexch S/dt\vert_\text{st}$ given by \eqref{sigexchcycles}  in terms of the $A(\vec{C}_\alpha)$'s and $\Jcumav_{[\Probst]}(\vec{C}_\alpha)$'s  and the expression of
 $-\dexch S/dt\vert_\text{st}$ given by  the entropy production rate for the phenomenological  entropy $S$ in thermodynamics of irreversible processes
$-\dexch S/dt\vert_\text{st}=\dint S/dt\vert_\text{st}=\sum_\gamma \Fth_\gamma \Jcumav^\star_\gamma$,
where the $ \Jcumav^\star_\gamma$'s are independent macroscopic currents.

 However in the case of the simple solvable model considered in paper II, where the thermal contact between two thermostats is settled by a set of independent  two-spin systems,  where each
  spin $\sigma_a$ is flipped by  thermostat $a$ ($a=1,2$), only one cycle is involved for every spin pair; then one can make the correspondence between, on the one hand, the affinity 
 $\Aff(\vec{C})$ of the cycle and the stationary probability current $\Jcumav_{[\Probst]}(\vec{C})$ that goes through it,  and,
on the other hand, the thermodynamic force $\Fth$ and the mean heat current $\Jcumav$.
Indeed, in this model, the graph associated with the master equation is itself a cycle, which can be orientated to read
 \be 
 \begin{array}{ccc} (+,+) & \rightarrow & (-,+) \\ \uparrow  & &
  \downarrow \\ (+,-) & \leftarrow & (-,-) \end{array}.
  \ee
The cycle may be rewritten as $\C_1\to \C_2\to\C_3\to\C_4\to \C_1$
 where  the configurations are labeled in the positive sense of the cycle orientation. Then, since the model obeys the MDB,  the affinity of the cycle is given by \eqref{valueAffalphaMDB}, and by virtue of the definitions \eqref{defdeltaq} and  \eqref{defdeltaexchSbis} it reads
$\Aff(\vec{C})=-\beta_1 \Heatm_1(\vec{C})-\beta_2 \Heatm_2(\vec{C})$, where $\Heatm_a(\vec{C})$ is the heat received from the thermostat $a$ when the system configuration performs the cycle once in the positive sense. Moreover, according to the energy conservation law, 
$ \Heatm_1(\vec{C})+\Heatm_2(\vec{C})$ is equal to the energy difference between the final and initial states when the cycle is performed once : this difference vanishes for a cycle so that
\be
\Aff(\vec{C})=(\beta_1 -\beta_2) \Heatm_2(\vec{C}).
\ee
On the other hand, since the graph is exactly a cycle, the current along an edge defined in \eqref{defJedge} has the same value for all edges in the stationary state (because, by virtue of \eqref{MasterCurrent}, the stationary condition $d \Prob(\C;t)/dt\vert_\text{st}=0$
is equivalent to Kirchoff's current law at every vertex of the graph). As a consequence, the stationary current associated with  the cycle reads
\be
\Jcumav_{[\Probst]}(\vec{C})=(\C_{i+1}\vert\Trans\vert\C_i) \Probst(\C_i)-(\C_i\vert\Trans\vert\C_{i+1}) \Probst(\C_{i+1}),
\ee
where $i$ is any label in $\{1,2,3,4\}$. The  exchange entropy flow given by \eqref{sigexchcycles} reads 
$\dexch S/dt\vert_\text{st}=-(\beta_1 -\beta_2) \Heatm_2(\vec{C})\Jcumav_{[\Probst]}(\vec{C})$. Comparison with the expression $\dexch S/dt=-\Fth \Jcumav$  in  irreversible processes thermodynamics (see \eqref{reldintdexchst}-\eqref{defAffBis}-\eqref{defAffBisDetail}) leads to the following identification of the macroscopic heat current $\Jcumav=\Espst{\jinst_2}$ received from  heat bath $2$ (and given to  heat bath $1$),
\be
\Jcumav= \Heatm_2(\vec{C})\Jcumav_{[\Probst]}(\vec{C}),
\ee
while the thermodynamic force $\Fth=\beta_1-\beta_2$ is to be identified with 
\be
\Fth=\frac{\Aff(\vec{C})}{\Heatm_2(\vec{C})}.
\ee
The probabilistic  interpretation of the affinity $\Aff(\vec{C})$ is given  in paper II.

\section{Exchange entropy variation and symmetries at finite time under MDB}

\label{FiniteTimeSymmetry}

\subsection{Exchange entropy variation for a history}

For a history $\Hist$ where the system starts in configuration $\C_0$ at time $t_0=0$ and  ends in configuration $\C_f$ at time $t$ after going through successive configurations $\C_0$, $\C_1$,\ldots, $\C_N=\C_f$, the exchange entropy variation $\Deltaexch S[\Hist]$  corresponding to the history is defined from the heat amounts $\Heat_1[\Hist]$ and $\Heat_2[\Hist]$ received from two thermal baths as
\be
\label{DefSexch}
\Deltaexch S[\Hist]\equiv\beta_1 \Heat_1[\Hist]+\beta_2 \Heat_2[\Hist]
\quad\textrm{with}\quad  \Heat_a[\Hist]\equiv\sum_{i=0}^{N -1}\delta \Heatm_a(\C_{i+1}\leftarrow\C_i).
\ee
The  expectation value of $\Deltaexch S[\Hist]$ with respect to the measure over all possible histories starting from configurations distributed according to  some initial probability distribution (see appendix \ref{DefHistories}) is equal to the time integral of the mean exchange entropy current calculated with the instantaneous configuration probability distribution, 
\be
\Esp{\Deltaexch S}=\int_0^t dt'
\Esp{\jinst_{\deltaexch S}}_{t'}=\int_0^t dt' \frac{\dexch S}{dt'}.
\ee
The second equality arises from  \eqref{exchangeentropyflow}.

\subsection{MDB and symmetry between time-reversed histories}
\label{SymmetryHistories}

Let  $\TimeRev$ be the time reversal operator for histories.
 If $\Hist$ is a history that starts at time $t_0=0$ in $\C_0$ and ends  at time $t$ in $\C_f$ after $N$ jumps from $\C_{i-1}$ to  $\C_i$ at time $T_i$,
$\TimeRev\Hist$ is a history that starts at time $t_0=0$ in $\C_f$ and ends at $\C_0$ at time $t$ after $N$  jumps from $\C'_{i-1}$ to  $\C'_i$ at time $T'_i$  with $\C'_i=\C_{N-i}$ and  $T'_i= t-T_{N-i+1}$ , namely
\bea
\Hist:\qquad\C_0\,\textrm{at}\, t_0=0 \quad 
&&\C_0 \overset{T_1}{\longrightarrow}\C_1\cdots  \C_{N-1}\overset{T_{N}}{\longrightarrow}\C_f
\\
\nonumber
\TimeRev\Hist:\qquad
\C_f\,\textrm{at}\, t_0=0 \quad
&&\C_f \overset{T'_1}{\longrightarrow}\C_{N-1}\cdots  \C_{1}\overset{T'_{N}}{\longrightarrow}\C_0.
\eea
From the definition of the measure $d\Prob_{\C_0, \C_f} $ over histories starting in configuration $\C_0$ and ending in configuration $\C_f$ (see Appendix \ref{DefHistories}),
\be
\label{defratio}
\frac{d\Prob_{\C_f, \C_0}\left[\Hist \right] }{d\Prob_{\C_0, \C_f} \left[\TimeRev\Hist \right]}=
 \prod_{i=0}^{N-1}\frac{(\C_{i+1}\vert \Trans \vert \C_i)}{(\C_{i}\vert \Trans \vert \C_{i+1})}.
\ee

When the transition rates obey the modified detailed balance \eqref{MDBexch} written in terms of $\deltaexch S(\C'\leftarrow\,\C)$, the exchange entropy variation for the history, defined in \eqref{DefSexch}, can be rewritten as 
\be
\label{SexchActionF}
\Deltaexch S[\Hist]
\underset{MDB}{=}-\ln \prod_{i=0}^{N-1}\frac{(\C_{i+1}\vert \Trans \vert \C_i)}{(\C_{i}\vert \Trans \vert \C_{i+1})},
\ee
and equation \eqref{defratio} can be rewritten as
\be
\label{historypropertySexch}
\frac{d\Prob_{\C_f, \C_0}\left[\Hist \right] }{d\Prob_{\C_0, \C_f} \left[\TimeRev\Hist \right]}
\underset{MDB}{=}e^{-\Deltaexch S[\Hist]}.
\ee
We stress that, according to \eqref{SexchActionF},   when the MDB is satisfied  the expression   of the exchange entropy variation for a history defined in \eqref{DefSexch}  coincides with the opposite of
 the ``action functional'' introduced by Lebowitz and Spohn in Ref.\cite{LebowitzSpohn1999} in the generic case where the MDB does not necessarily hold.

\subsection{Symmetry between time-reversed evolutions with fixed heat amounts}
\label{SymmetryConfigMesoSec}

The probability $\Prob\left(\C_f \vert \Heat_1,\Heat_2, t\vert  \C_0\right)$ that the system has evolved from  configuration $\C_0$ at $t_0=0$ to  configuration $\C_f$ at $t$ while receiving the heat amounts $\Heat_1$ and $\Heat_2$  from the thermostats 1 and 2 reads
\be
\Prob\left(\C_f \vert \Heat_1,\Heat_2; t\vert  \C_0\right)\equiv
\int d\Prob_{\C_f, \C_0}\left[\Hist \right]\delta\left( \Heat_1[\Hist]-\Heat_1\right)\delta\left( \Heat_2[\Hist]-\Heat_2\right),
\ee
where $\int d\Prob_{\C_f, \C_0}$ denotes the ``summation'' over the histories from $\C_0$ to $\C_f$.
The time-reversal symmetry property \eqref{historypropertySexch} for the history measure $d\Prob_{\C_f, \C_0}\left[\Hist \right]$ implies the  following relation between probabilities of forward and backward evolutions where initial and final configurations are exchanged (and  heat amounts are changed into their opposite values),
\be
\label{TimeRevPQ1PQ2CfC0}
\frac{\Prob\left(\C_f \vert \Heat_1,\Heat_2; t\vert  \C_0\right)}{\Prob\left(\C_0\vert -\Heat_1,-\Heat_2; t\vert \C_f\right)}
=e^{-\Deltaexch S(\Heat_1,\Heat_2)},
\ee
with the definition
\be
\Deltaexch S(\Heat_1,\Heat_2)\equiv \beta_1\Heat_1+\beta_2\Heat_2.
\ee
An analogous relation for $\Deltaexch S$ in place of $(\Heat_1,\Heat_2)$ is derived in \cite{Jarzynski2000} in the case where the microscopic dynamics of the heat baths is assumed to be Hamiltonian.

\subsection{Symmetries in protocols starting from an equilibrium state}
\label{SymProtocoleEqSec}

We consider a protocol where the system is prepared in an equilibrium state  at the inverse temperature $\beta_0$ and suddenly put at time $t_0=0$ in thermal contact  with the two thermostats at the inverse temperatures $\beta_1$ and $\beta_2$ respectively. Then the system evolution is a relaxation from an equilibrium state to a stationary non-equilibrium state.

The initial  equilibrium distribution  at the inverse temperature $\beta_0$ is the canonical  distribution \eqref{expProbcan}. $Z(\beta_0)$ cancels in the ratio $\Probcan^{\beta_0}(\C_0)/\Probcan^{\beta_0}(\C_f)$ and
\be
 \label{lnProb}
\ln \frac{\Probcan^{\beta_0}(\C_0)}{\Probcan^{\beta_0}(\C_f)}=\beta_0\left[\En(\C_f)-\En(\C_0)\right]=\beta_0 (\Heat_1+\Heat_2),
\ee
where the last equality is enforced by energy conservation. 
Then the time-reversal symmetry \eqref{TimeRevPQ1PQ2CfC0} and the specific form
\eqref{lnProb} for $\ln[\Probcan^{\beta_0}(\C_0)/\Probcan^{\beta_0}(\C_f)]$   imply that
\be
\label{TimeRevPQ1PQ2CfC0Eq}
\frac{\Prob\left(\C_f \vert \Heat_1,\Heat_2; t\vert  \C_0\right)\Probcan^{\beta_0}(\C_0)}{\Prob\left(\C_0\vert -\Heat_1,-\Heat_2; t\vert \C_f\right)\Probcan^{\beta_0}(\C_f)}
=e^{-\Deltaexchexcess S(\Heat_1,\Heat_2)},
\ee
where the excess exchange entropy variation $\Deltaexchexcess S(\Heat_1,\Heat_2)$  is defined as the difference between the  exchange entropy variation in an evolution under the non-equilibrium constraint  $\beta_1\neq\beta_2$ where the system receives heat amounts $\Heat_1$ and $\Heat_2$ and that in an evolution under the equilibrium condition $\beta_1=\beta_2=\beta_0$ where the system would received the same heat amounts. It reads
\be
\label{expDeltaexchexcess}
\Deltaexchexcess S(\Heat_1,\Heat_2)=\Deltaexch S(\Heat_1,\Heat_2)-\beta_0(\Heat_1+\Heat_2)=
(\beta_1-\beta_0)\Heat_1+(\beta_2-\beta_0)\Heat_2.
\ee
A crucial point is that $\Deltaexchexcess S(\Heat_1,\Heat_2)$ does not depend explicitly on the  initial and final configurations and is only a function of the heat amounts received from the thermal baths,

As a consequence, the measurable joint distribution $\Prob_{\Probcan^{\beta_0}}\left(\Heat_1,\Heat_2; t\right)$ for the heat amounts $\Heat_1$ and $\Heat_2$ received between $t_0=0$ and $t$ when the initial configuration of the system is distributed according to $\Probcan^{\beta_0}$, namely
$\Prob_{\Probcan^{\beta_0}}\left(\Heat_1,\Heat_2; t\right)=\sum_{\C_0,\C_f}\Prob\left(\C_f \vert \Heat_1,\Heat_2, t\vert  \C_0\right)\Prob_{\Probcan^{\beta_0}}(\C_0)$, satisfies the identity
\be
\label{ratioProbQ1Q2}
\frac{\Prob_{\Probcan^{\beta_0}}\left(\Heat_1,\Heat_2; t\right)}{\Prob_{\Probcan^{\beta_0}}
\left(-\Heat_1,-\Heat_2; t\right)}=e^{-\Deltaexchexcess S(\Heat_1,\Heat_2)}.
\ee
Subsequently the  measurable quantity $\Deltaexchexcess S(\Heat_1,\Heat_2)$, with the distribution probability
\\
$\Prob_{\Probcan^{\beta_0}}\left(\Deltaexchexcess S\right)=\sum_{\Heat_1,\Heat_2}
\delta\left(\Deltaexchexcess S-(\beta_1-\beta_0)\Heat_1-(\beta_2-\beta_0)\Heat_2\right)
\Prob_{\Probcan^{\beta_0}}\left(\Heat_1,\Heat_2; t\right)$ 
obeys  the symmetry relation at any finite time, which may be referred to as a detailed fluctuation relation,
\be
\label{DFRDeltaexchexcess}
\frac{\Prob_{\Probcan^{\beta_0}}\left(\Deltaexchexcess S\right)}{\Prob_{\Probcan^{\beta_0}}\left(-\Deltaexchexcess S\right)}=e^{-\Deltaexchexcess S}.
\ee
The latter relation itself entails the identity, which may be referred to as an integral fluctuation relation,
\be
\label{IFRDeltaexchexcess}
\Esp{e^{\Deltaexchexcess S}}_{\Probcan^{\beta_0}}=1.
\ee
To our knowledge these two relations have not appeared explicitly in the literature. 
We notice that, from a purely technical point of view, the derivation has similarities with the argument first exhibited by Crooks \cite{Crooks1999} and then Seifert \cite{Seifert2005, Seifert2008} for the entropy production along a stochastic trajectory when the system is in thermal contact with only one heat bath and is driven out of equilibrium by a time-dependent external parameter. (In Crooks' argument the initial  configurations for the forward and backward evolutions are distributed with different equilibrium probabilities, $\Probcan^{\beta_0}$ and $\Probcan^{\beta_f}$, whereas forward and backward evolutions with the same initial distribution had already been considered in 
\cite{BochkovKuzovlev1981a,BochkovKuzovlev1981b}).

\subsection{Symmetries in protocols starting from a stationary state with a canonical distribution}
\label{SymProtocolNESSSec}

For some systems, such as the two-spin model studied in paper II, the stationary distribution 
when the thermostats are at the inverse temperatures $\beta_1$ and $\beta_2$ proves to be a canonical distribution at the effective inverse temperature $\betaeff(\beta_1,\beta_2)$. 

When the system is  prepared in a stationary state  between two heat baths at the inverse temperatures $\beta_1^0$ and $\beta_2^0$ and then put in thermal contact with two thermostats at the inverse temperatures $\beta_1$ and $\beta_2$ at time $t_0=0$,
 the protocol describes the relaxation from a given stationary state corresponding to $(\beta^0_1,\beta_2^0)$ to another stationary state
corresponding to $(\beta_1,\beta_2)$.
When the initial stationary state has  the canonical  distribution at the effective inverse temperature $\beta_{\star}^0=\betaeff(\beta_1^0,\beta_2^0)$, the argument of the previous subsection can be repeated and the equalities \eqref{DFRDeltaexchexcess} and \eqref{IFRDeltaexchexcess}  
 still hold with  $\beta_0$ replaced by $\beta_\star^0$ and $\Deltaexchexcess S$ replaced by
\be
\Delta_\text{\mdseries exch}^{\text{\mdseries excs},\beta_\star^0} S(\Heat_1,\Heat_2)=(\beta_1-\beta_{\star}^0)\Heat_1+(\beta_2-\beta_{\star}^0)\Heat_2.
\ee

When the system is  already  in the stationary state  corresponding to  the inverse temperatures $\beta_1$ and $\beta_2$ at time $t_0=0$,  the equalities \eqref{DFRDeltaexchexcess} and \eqref{IFRDeltaexchexcess} for $\Deltaexchexcess S$ still hold with $\beta_\star(\beta_1,\beta_2)$ in place of $\beta_0$ :
\be
\frac{\Probst\left(\Delta_\text{\mdseries exch}^{\text{\mdseries excs},\beta_\star} S\right)}{\Probst\left(-\Delta_\text{\mdseries exch}^{\text{\mdseries excs},\beta_\star} S \right)}=e^{-\Delta_\text{\mdseries exch}^{\text{\mdseries excs},\beta_\star} S},
\ee
where the subscript ``st'' in the notation for the probability is  a reminder of the fact that  the initial configurations are distributed according to the stationary measure, which is equal to
$\Probcan^{\betaeff}$ in the present case. Another detailed fluctuation relation involving the forward histories for the original dynamics and the backward histories for the dual reversed dynamics is 
derived in \cite{EspositoVanDenBroeck2010PRL} for the case where the external parameters also vary during the time interval $]t_0,t]$;   these considerations are out of the scope of the present paper.

\section{Long-time symmetries : non-equilibrium stationary state with MDB}

\subsection{Fluctuation relations}

\label{FluctuationRelations}

\subsubsection{Fluctuation relation for the cumulative exchange  entropy variation}
\label{FluctuationRelationSexch}

As recalled with some details in section \ref{Entropysection}, if the Markov
matrix is irreducible and the system has a finite number of configurations, then
the Perron-Frobenius theorem entails that there exists a single stationary state
$\Probst$ and every configuration $\C$ has a non-zero probability $\Probst(\C)$. Let us call $\Probst^\text{min}$ and $\Probst^\text{max}$ the minimum and maximum values taken by $\Probst$. Since
$\Probst\left(\Heat_1,\Heat_2; t\right)=\sum_{\C_0,\C_f}\Prob\left(\C_f \vert \Heat_1,\Heat_2; t\vert  \C_0\right)\Probst(\C_0)$, and 
$\frac{\Probst^\text{min}}{\Probst^\text{max}}\leq \frac{\Probst(\C_f)}{\Probst(\C_0)}\leq\frac{\Probst^\text{max}}{\Probst^\text{min}}$,
the time-reversal symmetry \eqref{TimeRevPQ1PQ2CfC0} for
$\Prob\left(\C_f \vert \Heat_1,\Heat_2; t\vert  \C_0\right)$ entails that  
\be
\label{InequalityHeats}
\frac{\Probst^\text{min}}{\Probst^\text{max}}\leq
\frac{\Probst\left(\Heat_1,\Heat_2; t\right)}{\Probst\left(-\Heat_1,-\Heat_2; t\right)e^{-\Deltaexch S(\Heat_1,\Heat_2)}}
\leq\frac{\Probst^\text{max}}{\Probst^\text{min}}.
\ee
The distribution probability  for the exchange entropy variation $\Deltaexch S$ can be determined from  measurements of heat amounts through the relation
\be
\Probst\left(\Deltaexch S;t\right)=\sum_{\Heat_1,\Heat_2}\delta\left(\Deltaexch S-\beta_1\Heat_1-\beta_2\Heat_2\right)\Probst\left(\Heat_1,\Heat_2;t\right)
.\ee
The inequalities \eqref{InequalityHeats} imply that 
\be
\label{InequalitySexch}
\frac{\Probst^\text{min}}{\Probst^\text{max}}\leq
\frac{\Probst\left(\Deltaexch S;t\right)}{\Probst\left(-\Deltaexch S;t\right)e^{-\Deltaexch S}}
\leq\frac{\Probst^\text{max}}{\Probst^\text{min}}.
\ee

In the long-time limit the system reaches its non-equilibrium stationary state exponentially fast, 
so the existence and  value of a large deviation function for the cumulative current $\Jcum\equiv \Deltaexch S/t$ are not expected to depend on the initial distribution. Appendix \ref{LDRemarks} contains several definitions of large deviation functions  for a cumulative random variable $X_t$ with the sign convention used in mathematical literature (see subsections 
\ref{MathDefSec} and  \ref{MathDefAltSec}). 
In the present paper we use the opposite sign convention and we consider the large deviation function  $f_X(\Jcum)\equiv -R_X(\Jcum)$  where    $R_X$ denotes  the rate function introduced by the proper mathematical definition \eqref{defRXa}-\eqref{defRXb} for the values $\Jcum$   taken  by $X_t/t$.  When $f_X(\Jcum)$ is  continuous over some interval $I$,  the  property \eqref{CharacterizationRContinuousBis} reads
\be
\label{deffX}
\lim_{t\to+\infty} \frac{1}{t}\ln \Prob\left(\frac{X_t}{t}\in I\right)=\sup_{\Jcum \in I} f_X(\Jcum).
\ee
The presence of the upper bound reflects the fact that, if $I$ is split into small intervals $I_k$, only the interval where $\Prob\left(\frac{X_t}{t}\in I_k\right)$ is maximum contributes to 
the limit in the left-hand side of \eqref{deffX}, and    this limit can indeed be rewritten as the upper bound over $I$ of a function $f_X(\Jcum)$ that depends only on a single variable (see the heuristic argument in subsection \ref{MathDefSec}). The supremum in \eqref{deffX} is also crucial in the derivation of the Gärtner-Ellis theorem which, in its simplified version, allows to compute $f_X(\Jcum)$ from the generating function for infinite-time  cumulants of $X_t$ per unit time, namely $\lim_{t\to+\infty} \frac{1}{t}\ln \Esp{e^{\lambda X_t}}$, when the latter generating function exists and is differentiable for all $\lambda$ in $\mathbb{R}$ (see subsection \ref{GET} below).  A consequence of \eqref{deffX} is that
\be
\label{ExpfXcontinuous}
\lim_{\epsilon \to 0} \lim_{t\to+\infty} \frac{1}{t}\ln \Prob\left(\frac{X_t}{t}\in \left[\Jcum-\epsilon, \Jcum +\epsilon\right]\right)= f_X(\Jcum).
\ee
Eventually we recall an expression often encountered in the literature. In the case where $f_X(\Jcum)$ is strictly convex downward, with a maximum value at $\Jcumav=\lim_{t\to+\infty} \frac{X_t}{t}$ where 
$f_X(\Jcum)$ vanishes, then $f_X(\Jcum)$ can be expressed in terms of the cumulative distribution function  and the complementary cumulative distribution function  as
\be
\label{LDfuntionSupInf}
f_X(\Jcum)\equiv
\begin{cases}
\lim_{t\to\infty}\frac{1}{t}\ln \Prob(\frac{X_t}{t}>  \Jcum  ;t)  &\textrm{for $\Jcum>\Jcumav$}   \\
\lim_{t\to\infty}\frac{1}{t}\ln \Prob(\frac{X_t}{t}<  \Jcum ;t)  &\textrm{for $\Jcum<\Jcumav$} 
\end{cases}.
\ee 
We notice that the latter expression is not to be generalized to the case of several independent currents, contrarily to the definition \eqref{deffX}.

Subsection \ref{IneqLargeDeviation} contains  a proof that, whatever the definition among the three ones discussed in Appendix \ref{LDRemarks}, a relation like \eqref{InequalitySexch} implies that, if 
$\Deltaexch S$ has a large deviation function $f_{\Deltaexch S}(\Jcum)$ -- under $\Probst$ but then also under any initial probability distribution -- then
\be
\label{FTSexch}
f_{\Deltaexch S}(\Jcum)-f_{\Deltaexch S}(-\Jcum)=-\Jcum.
\ee
Our derivation of the fluctuation relation is close to the argument given in
Ref.\cite{BodineauDerrida2007} for the large deviation of the cumulative heat
current (see \eqref{FRHeatCurrent}). It relies on the MDB obeyed by the
transition rates. Since the opposite of the exchange entropy variation is the
specific form that the Lebowitz-Spohn action functional
\cite{LebowitzSpohn1999} takes in the presence of MDB, the fluctuation relation
\eqref{FTSexch} is in fact a special case of the fluctuation relation satisfied
by the action functional for a system with a finite number of configurations
under the assumption \eqref{MicroRevCond} of reversibility for configuration
jumps and the assumption \eqref{Mirreducible0} that the Markov matrix is
irreducible, without the extra assumption of MDB \eqref{MDBexch0}.

\subsubsection{Constraints from the bound upon the system energy}
\label{ConstraintEnergy}

In a system with a finite number of configurations, $\Heat_1+\Heat_2=\En(\C_f)-\En(\C_0)$ is bounded and this entails several properties   upon the  large deviation functions of the cumulative currents $\Jcum_1=\Heat_1/t$ and $\Jcum_2=\Heat_2/t$.

When $\Heat_1+\Heat_2$ is bounded,  the consequences for the cumulative heat
currents are conveniently investigated if one considers the couple of variables
$\left(\Heatd_1, \Heat_2\right)$ where $\Heatd_1=-\Heat_1$ is the heat amount
dissipated towards  thermal bath 1.  The fact that the difference $\Heatd_1-\Heat_2$ is
bounded means that there exists some (time-independent) cosntant $M>0$ such that
\be
\label{boundedness}
\vert \Heatd_1-\Heat_2 \vert <M.
\ee
The first straightforward consequence upon the cumulative heat currents $\Jcumd_1=\Heatd_1/t$ and $\Jcum_2=\Heat_2/t$ is that
$\lim_{t\to+\infty}\Esp{\Jcumd_1}=\lim_{t\to+\infty}\Esp{\Jcum_2}\equiv \Jcumav$, namely
\be
\label{defJcumaHeat2}
\lim_{t\to+\infty} \frac{-\Esp{\Heat_1}}{t}=\lim_{t\to+\infty} \frac{\Esp{\Heat_2}}{t}=\Jcumav.
\ee

The second consequence of the fact that $\Heatd_1-\Heat_2$ is sub-extensive is
that, according to \eqref{alternative}, $\Heatd_1$ and $\Heat_2$ have the same
large deviation function, $f_{\Heatd_1}(\Jcum)=f_{\Heat_2}(\Jcum)$, namely \be
f_{\Heat_1}(\Jcum)=f_{\Heat_2}(-\Jcum).  \ee Similarly the difference between
$\Deltaexch S$ and $-(\beta_1-\beta_2)\Heat_2$, which is equal to
$\beta_1[\En(\C_f)-\En(\C_0)]$, is bounded. 
Therefore the difference  is sub-extensive, so that according to
\eqref{alternative} ${f_{\Deltaexch
  S}(\Jcum)=f_{-(\beta_1-\beta_2)\Heat_2}(\Jcum)}$, namely \be f_{\Deltaexch
  S}(\Jcum)=f_{\Heat_2}\left(-\frac{\Jcum}{\beta_1-\beta_2}\right).  \ee As a
consequence, the fact that the exchange entropy variation $\Deltaexch S$ obeys
the fluctuation relation \eqref{FTSexch} is equivalent to the fact that the 
cumulative heat current received from  heat bath $2$ obeys the fluctuation
relation \be
\label{FRHeatCurrent}
f_{\Heat_2}(\Jcum)-f_{\Heat_2}(-\Jcum)= \left(\beta_1-\beta_2\right)\Jcum.
\ee
The latter equation is the long-time limit of the relation  first exhibited by
Jarzynski and Wojcik \cite{JarzynskiWojcik2004} for the thermal contact between
two bodies initially prepared at different inverse temperatures
$\beta_1$ and $\beta_2$ and whose microscopic Hamiltonian dynamics involves an interaction turned on at time $t_0=0$  and turned off at time $t$ and which is assumed to be
 negligible with respect to the  heat quantity that goes from one body to the other during time $t$.
 The fluctuation relation \eqref{FRHeatCurrent} is also derived in the
framework of the master equation approach in
Refs.\cite{Derrida2007,BodineauDerrida2007}.

We notice that, since the function $\Deltaexchexcess S(\Heat_1,\Heat_2)$ defined
in \eqref{expDeltaexchexcess} is equal to $\Deltaexch S$ plus a term
$-\beta_0(\Heat_1+\Heat_2)$ which is bounded by virtue of energy conservation,
the large deviation functions for $\Deltaexchexcess S$ and $\Deltaexch S$
coincide (see \eqref{alternative}). As a consequence, \eqref{FTSexch} entails
that $\Deltaexchexcess(\Heat_1,\Heat_2)$ obeys the fluctuation relation
\be
f_{\Deltaexchexcess}(\Jcum)-f_{\Deltaexchexcess}(-\Jcum)=-\Jcum.
\ee

The relation \eqref{FRHeatCurrent} can be written in a more generic form by the following argument. The fact that  the difference between $\Deltaexch S$ and $-(\beta_1-\beta_2)\Heat_2$ is bounded yields the following relation between the infinite-time mean values of the corresponding cumulative currents,
\be
\label{meanDeltaexchS}
 \lim_{t\to +\infty}\frac{\Esp{\Deltaexch S(t)}}{t}=-(\beta_1-\beta_2)\lim_{t\to+\infty}\frac{\Esp{\Heat_2(t)}}{t}.
\ee
Since the infinite-time limit of the mean value of a cumulative current $\Jcum_t$ measured during the interval $[0,t]$ is equal to the mean value of the corresponding instantaneous current in the stationary state, namely
\be
\label{defJcumav}
\lim_{t\to\infty} \Esp{\Jcum_t}=\Espst{\jinst}\equiv\Jcumav, \ee we get
$\lim_{t\to +\infty}\Esp{\Deltaexch S}/t= \Espst{\jinst_{\deltaexch S}}=\dexch
S/dt\vert_\text{st}$ as well as $\lim_{t\to
  +\infty}\Esp{\Heat_2}/t=\Espst{\jinst_2}$ (with the instantaneous current
definitions \eqref{defjinstexch} and \eqref{defjinstHeat} respectively). With
these identifications the comparison of \eqref{meanDeltaexchS} with the property
\eqref{defFth} (which exhibits the analogy with the thermodynamics of
irreversible processes) shows that $\beta_1-\beta_2$ in \eqref{meanDeltaexchS}
is to be interpreted as the thermodynamic force $\Fth$. Therefore the fact that
the fluctuation relation \eqref{FRHeatCurrent} arises from the boundedness of
the difference between $\Deltaexch S$ and $-(\beta_1-\beta_2)\Heat_2$, as the
relation between the mean values \eqref{meanDeltaexchS} does, implies that the
fluctuation relation \eqref{FRHeatCurrent} for $f_{\Heat_2}$ is a special case
of the more generic fluctuation relation 
\be
\label{FRJcumAff}
f(\Jcum;\Fth)-f(-\Jcum;\Fth)=\Fth\Jcum, 
\ee 
where $\Fth$ is the thermodynamic force that appears in the stationary exchange
entropy flow \break $\dexch S/dt\vert_\text{st}=-\Fth\Jcumav$. 
(We recall that in the case where 
$f(\Jcum;\Fth)=-\infty$, the relation \eqref{FRJcumAff} makes sense when written as
$f(\Jcum;\Fth)=f(-\Jcum;\Fth)-\Fth\Jcum$.)

\subsubsection{Gärtner-Ellis theorem and some of its consequences}
\label{GET}

As exemplified in the previous subsubsection, large deviation functions may
depend on auxiliary parameters. For certain questions, they are simply
spectators. This is the case in the following discussion, so we do not mention
possible auxiliary parameters explicitly. However, we nevertheless write all
derivatives as partial derivatives.

In this short subsubsection, we introduce an important tool to study the
existence and properties of a large deviation function $f(\Jcum)$: the
infinite-time limit of the generating function for the cumulants of
$X_t=t\Jcum_t $ per unit time, namely $\alpha(\lambda)\equiv \lim_{t\to+\infty}
(1/t) \ln\Esp{e^{\lambda X_t}}$. 
$\alpha(\lambda)$ is known as the scaled cumulant generating function in the literature about large deviations.
This function, if it exists, is automatically
convex (downward).

According to a simplified version of the Gärtner-Ellis theorem (see e.g. the
review for physicists \cite{Touchette2009} or the mathematical point of view
\cite{DemboZeitouni1998}), if $\alpha(\lambda)$ exists and is differentiable for
all $\lambda$ in $\mathbb{R}$, then the large deviation function $f$ of the
current $\Jcum$ exists and it can be calculated as the Legendre-Fenchel
transform of $\alpha(\lambda)$, namely, with the signs chosen in the definitions
used in the present paper, 
\be f(\Jcum)=\min_{\lambda\in
  \mathbb{R}}\{\alpha(\lambda)-\lambda \Jcum\}.  \ee

If $\alpha(\lambda)$ is strictly convex and continuously differentiable, then for
each $\Jcum$ the minimum is achieved for a single value of $\lambda$, which is
the unique solution $\lambda_c$ of $\frac{\partial \alpha}{\partial
  \lambda}=\Jcum$ i.e. the Legendre-Fenchel transform reduces to the
usual Legendre transform, and the duality relation,
$\lambda_c(\Jcum)=-\frac{\partial f}{\partial \Jcum}$, holds, i.e.
\be \left. \frac{\partial \alpha}{\partial
    \lambda}\right\vert_{\lambda=-\frac{\partial f}{\partial \Jcum}}=\Jcum . \ee

If moreover $\alpha(\lambda)$ is differentiable twice, taking the derivative of
this relation with respect to  $\Jcum$ one obtains
\be 
\frac{\partial^2 f}{(\partial \Jcum)^2} \left. \frac{\partial^2
  \alpha}{(\partial \lambda)^2}\right\vert_{\lambda=\lambda_c(\Jcum)}=-1. \ee

From the fundamental properties of any large deviation function, $f(\Jcum)$ is
maximum at $\Jcum=\Jcumav$ defined in \eqref{defJcumav}, so by construction
$\lambda_c(\Jcumav)=0$. But derivatives of $\alpha(\lambda)$ at
$\lambda=0$ are related to cumulants of $X_t$ at large $t$. For instance 
\be
\left.\frac{\partial^2 \alpha}{(\partial \lambda)^2}\right\vert_{\lambda=0} =\lim_{t\to+\infty}
\frac{\Espst{X_t^2}-\Espst{X_t}^2}{t},
\ee  
and one gets: 
\be \label{eq:hesscumul}
\left.\frac{\partial^2 f}{(\partial
    \Jcum)^2}\right\vert_{\Jcum=\Jcumav}=- \left[\lim_{t\to+\infty}
  \frac{\Espst{X_t^2}-\Espst{X_t}^2}{t}\right]^{-1}. 
\ee

We now apply these considerations to linear response.

\subsubsection{Linear response far from equilibrium}
\label{LRLD}

In the case at hand, the large deviation function $f$ depends on other
parameters. The one relevant for the discussion of linear response is the
thermodynamic force $\Fth$, and we write $f(\Jcum;\Fth)$. In general $f$ depends
on other variables which we do not mention explicitly, and which are supposed to
be kept constant whenever a partial derivative is taken in the sequel. 

We start from the property recalled in the previous subsubsection:
\be
\label{JcumderivLDfJmean}
\left.\frac{\partial f(\Jcum;\Fth)}{\partial \Jcum}\right\vert_{\Jcum=\Jcumav}=0.
\ee 
Taking the derivative with respect to $\Fth$ leads to
\be
\label{AffJcumderivLDfJmean}
\left.\frac{\partial \Jcumav}{\partial \Fth}\frac{\partial^2 f(\Jcum;\Fth)}{(\partial \Jcum)^2}\right\vert_{\Jcum=\Jcumav}
=-\left.\frac{\partial^2 f(\Jcum;\Fth)}{\partial \Fth\partial \Jcum}\right\vert_{\Jcum=\Jcumav}.
\ee

According to \eqref{eq:hesscumul}, $\left.\frac{\partial^2
    f(\Jcum;\Fth)}{(\partial \Jcum)^2}\right\vert_{\Jcum=\Jcumav}$ can be
expressed in terms of the second cumulant of $X_t$ in the infinite-time limit, and we can
rewrite the equality \eqref{AffJcumderivLDfJmean} as:
\be
\label{genericlinearresponse}
\frac{\partial \Jcumav}{\partial \Fth}=
\left.\frac{\partial^2 f(\Jcum;\Fth)}{\partial \Fth\partial \Jcum}\right\vert_{\Jcum=\Jcumav}
\times \lim_{t\to+\infty} \frac{\Espst{X_t^2}-\Espst{X_t}^2}{t}.
\ee
This is the  generic expression of the  linear response in the non-equilibrium state far from equilibrium. In the case of thermal contact $X_t$ is the cumulative  heat $\Heat_2(t)$ received from  heat bath 2.

\clearpage
\subsubsection{Einstein-Green-Kubo relation as a consequence of the fluctuation relation for the large deviation function}
\label{GreenKuboFR}

When the large deviation function obeys the fluctuation relation \eqref{FRJcumAff} arising from the MDB,
successive partial derivatives of the fluctuation relation entail that
\be
\label{AffJcumderivFRJcumAff}
\frac{\partial^2 f(\Jcum;\Fth)}{\partial \Fth\partial \Jcum}+
\left.\frac{\partial^2 f(\Jcum';\Fth)}{\partial \Fth\partial \Jcum'}\right\vert_{\Jcum'=-\Jcum}=1.
\ee
Then the relation  \eqref{AffJcumderivFRJcumAff}  for $\Jcum=0$ yields
$\left.\frac{\partial^2 f(\Jcum;\Fth)}{\partial \Fth\partial \Jcum}\right\vert_{\Jcum=0}=\frac{1}{2}$ and, since the stationary state with $\Fth=0$ is in fact the equilibrium state, the linear response relation \eqref{genericlinearresponse} becomes
\be
\label{GreenKuboGeneric}
  \left.\frac{\partial \Jcumav}{\partial \Fth}\right\vert_{\Fth=0}=
\frac{1}{2}
\times \lim_{t\to+\infty} \frac{\Espeq{X_t^2}-\Espeq{X_t}^2}{t}.
\ee
In the following the latter fluctuation-dissipation relation will be referred to as the Einstein-Green-Kubo relation. The precise terminology has been pointed out in  Ref.\cite{AndrieuxGaspard2007JStatMech}. When the equilibrium fluctuations are written in terms of a second-order cumulant, namely the  mean value  of the squared Helfand moment $X_t-\Esp{X_t}$, as it is the case for Einstein relation, the fluctuation-dissipation relations are called  Einstein-Helfand formulae \cite{Einstein1956, Helfand1960}.  When the equilibrium fluctuations are written in terms of the time-correlation function of the instantaneous current, they are are known as the Green-Kubo relations 
\cite{Green1952,Green1954, Kubo1957, KuboETAL1957} or the Yamamoto-Zwanzig formulae in the context of chemical relations \cite{Yamamoto1960, Zwanzig1965}.

The Einstein-Green-Kubo relation \eqref{GreenKuboGeneric} can be rephrased in terms of the Onsager coefficient $L$ defined   in the framework of thermodynamics of  irreversible processes  near equilibrium as
\be
\label{defOnsagerCoeff}
L\equiv \left.\frac{\partial \Jcumav}{\partial \Fth}\right\vert_{\Fth=0}
 =\lim_{\Fth\to 0}\frac{\Jcumav}{\Fth},
\ee
where the last equality is valid when there is only one nonzero mean current in the stationary state, in which case this single current $\Jcumav$ vanishes as $\Fth$ goes to zero.

In the case of thermal contact, the limit $\Fth\to 0$ in \eqref{defOnsagerCoeff} can be stated more precisely according to the following argument. If $\beta_1=\beta_2$ the stationary state is the equilibrium state and $\Jcumav(\beta_1,\beta_2)$ becomes $\Jcumav(\beta,\beta)=\Espeq{\jinst_2}=0$  (see the comment after \eqref{defjinstHeat}). 
As a consequence ${\Jcumav(\beta+d\beta,\beta+d\beta)-\Jcumav(\beta,\beta)=0}$ for any infinitesimal $d \beta$, namely
\be
\left.\frac{\partial \Jcumav}{\partial \beta_1}\right\vert_{\beta_1=\beta_2=\beta}
=-\left.\frac{\partial \Jcumav}{\partial \beta_2}\right\vert_{\beta_1=\beta_2=\beta},
\ee
so that
\be
\Jcumav(\beta_1,\beta_2) \underset{\beta_1,\beta_2\to \beta}{\sim} (\beta_1-\beta_2)
\left.\frac{\partial \Jcumav}{\partial \beta_1}\right\vert_{\beta_1=\beta_2=\beta}
\ee
independently of the way $\beta_1$ and $\beta_2$ go to $\beta$. Subsequently 
the linear response  coefficient $L$ defined in \eqref{defOnsagerCoeff} also reads
\be
L=\left.\frac{\partial \Jcumav}{\partial \beta_1}\right\vert_{\beta_1=\beta_2=\beta},
\ee
while the Einstein-Green-Kubo relation \eqref{GreenKuboGeneric} can also be  stated as
\be
\label{GreenKuboHeat}
\lim_{\beta_1,\beta_2\to \beta}\frac{\Jcumav(\beta_1,\beta_2)}{\beta_1-\beta_2}
=\frac{1}{2}
\lim_{t\to+\infty} \frac{\Espeq{\Heat^2_2(t)}^\beta-\left(\Espeq{\Heat_2(t)}^\beta\right)^2}{t}.
\ee
where $\Jcumav(\beta_1,\beta_2)=\Espst{\jinst_2}^{(\beta_1,\beta_2)}$  and the limit is independent of the way $\beta_1-\beta_2$ vanishes. Eventually in the case of thermal contact the Einstein-Green-Kubo relation relates the proportionality coefficient between the stationary heat current and  the difference between the bath inverse temperatures when the system is weakly out of equilibrium to   the fluctuations of the  heat amount received from one thermal bath  in the equilibrium situation where both thermostats are at the same temperature.

\clearpage
\subsection{Symmetry of the generating function for cumulants per unit time under MDB}
\label{LongTimeCumulants}

Contrarily to the case of the Einstein-Green-Kubo relation \eqref{GreenKuboGeneric}
between the mean current and the equilibrium infinite-time second cumulant per
unit time, relations between higher cumulants per unit time cannot be readily obtained from the fluctuation relation \eqref{FRJcumAff}.
In order to obtain such relations one has to resort to the cumulant generating function.

\subsubsection{Infinite-time cumulants per unit time for $\Heat_1$ and $\Heat_2$}
\label{relcumgeneric}

The $k$th cumulant $\kappa^{[p]}_{\Heat_a}$ for the heat amount $\Heat_a$ received from  bath $a$ or the joint cumulant $\kappa^{[p,q]}_{\Heat_a,\Heat_b}$ for the  heat amounts $\Heat_a$ and $\Heat_b$ can be computed from the ``characteristic function''
\be
\label{defCharFunc}
\Esp{e^{\lambda_1\Heat_1(t)+\lambda_2\Heat_2(t)}}=\sum_{\Heat_1}\sum_{\Heat_2}
e^{\lambda_1\Heat_1+\lambda_2\Heat_2}\sum_{\C_f}\sum_{\C_0}\Prob\left(\C_f \vert \Heat_1,\Heat_2; t\vert  \C_0\right)\Prob(\C_0,t_0=0)
\ee
through the following derivatives
\be
\label{defCumulant}
\kappa^{[p]}_{\Heat_a}(t)=\left.\frac{\partial^p \ln\Esp{e^{\lambda_1\Heat_1(t)+\lambda_2\Heat_2(t)}}}{\partial \lambda_a^p}\right\vert_{\lambda_1=\lambda_2=0}
\quad\textrm{for $a=\{1,2\}$}
\ee
\be
\label{defCumulantDeux}
\kappa^{[p,q]}_{\Heat_1,\Heat_2}(t)=\left.
\frac{\partial^{p+q} \ln\Esp{e^{\lambda_1\Heat_1(t)+\lambda_2\Heat_2(t)}}}
{\partial \lambda_1^p\partial\lambda_2^q}\right\vert_{\lambda_1=\lambda_2=0}.
\ee

For a Markov  process, the leading long-time behaviors of these cumulants are proportional to the time $t$ elapsed from the beginning of the measurements. The asymptotic behavior of the cumulants per unit time are given by the derivatives of 
 \be
 \label{alpha12def}
 \alpha_{1,2}(\lambda_1,\lambda_2)\equiv\lim_{t\to+\infty}\frac{1}{t}
 \ln\Esp{e^{\lambda_1\Heat_1(t)+\lambda_2\Heat_2(t)}}
 \ee
with respect to $\lambda_1$ and $\lambda_2$ at $\lambda_1=\lambda_2=0$.
In other words, 
\be
\label{defCumulantDoubpertime} 
\lim_{t\to+\infty}\frac{\kappa^{[p,q]}_{\Heat_1,\Heat_2}}{t}
=\left.\frac{\partial^{p+q}  \alpha_{1,2}}{\partial \lambda_1^p\partial\lambda_2^q}\right\vert_{\lambda_1=\lambda_2=0}.
\ee
Similarly
\be
\label{defCumulantpertime} 
\lim_{t\to+\infty}\frac{\kappa^{[p]}_{\Heat_a}}{t}
=\left.\frac{\partial^p  \alpha_{a}}{\partial \lambda_a^p}\right\vert_{\lambda_a=0}
\quad\textrm{for $a=\{1,2\}$}
\ee
where
\be
\label{defalphaa}
\alpha_a(\lambda)\equiv\lim_{t\to+\infty}\frac{1}{t}\ln \Esp{e^{\lambda\Heat_a(t)}}.
\ee
The generating function for infinite-time cumulants per unit time, $\alpha_a(\lambda)$,  may also be referred to as the  scaled cumulant generating function.

We notice that, since for any cumulant  $\lim_{t\to\infty}\kappa^{[p]}_{\Heat_2}/t$ is finite, under some technical conditions, the probability distribution of the variable $\left[\Heat_2(t)-\Esp{\Heat_2(t)}\right]/\sqrt{t}$ becomes Gaussian in the long-time limit. Indeed  the logarithm of the characteristic function for the variable $Y_2(t)=\left[\Heat_2(t)-\Esp{\Heat_2(t)}\right]/\sqrt{t}$ reads $\ln\Esp{e^{\lambda Y_2(t)}}=\sum_{p=2}^{+\infty} (1/p!)\left(\lambda/\sqrt{t}\right)^p \kappa^{[p]}_{\Heat_2}(t)$.  If the sum and the $t\to+\infty$ limit can be interchanged, $\ln\Esp{e^{\lambda Y_2(t)}}$ becomes proportional to $\lambda^2$ in the limit where $t$ goes to infinity :
only the second cumulant of $Y_2(t)$ survives in the long-time limit.

\subsubsection{Constraints from the bound upon the system energy}

In the case of a system with a finite number of configurations,
$\Heat_1+\Heat_2=\En(\C_f)-\En(\C_0)$ is restricted to some interval $[-|\Delta\En|_\text{max}, +|\Delta\En|_\text{max}]$, and the definition \eqref{defCharFunc} entails the inequalities
\be
e^{-\lambda_1|\Delta\En|_\text{max}}\leq 
\frac{\Esp{e^{\lambda_1\Heat_1(t)+\lambda_2\Heat_2(t)}}}
{\Esp{e^{(\lambda_2-\lambda_1)\Heat_2(t)}}}\leq e^{\lambda_1|\Delta\En|_\text{max}}.
\ee
As a consequence,
\be
\label{alpha12finite}
 \alpha_{1,2}(\lambda_1,\lambda_2)= \alpha_a(\lambda_a-\lambda_b)
 \quad\textrm{with $\{a,b\}=\{1,2\}$}.
\ee
The relation \eqref{alpha12finite} can be rewritten as 
\be
\label{alpha12finiteBis}
\alpha_{1,2}(\lambda_1,\lambda_2)= \alpha_2(\lambda_2-\lambda_1)
\quad\textrm{and}\quad  
\alpha_1(\lambda)=\alpha_2(-\lambda).
\ee
The specific dependence of $\alpha_{12}(\lambda_1,\lambda_2)$ upon $\lambda_2-\lambda_1$ together with the generic formul\ae\  \eqref{defCumulantDoubpertime} -\eqref{defCumulantpertime}  imply the following relations between the infinite-time cumulants per unit time,
\be
\label{longtimekappa}
\lim_{t\to\infty}\frac{\kappa^{[p]}_{\Heat_1}}{t}=(-1)^p\lim_{t\to\infty}\frac{\kappa^{[p]}_{\Heat_2}}{t}
\ee
and
\be
\label{relCumCumJoint}
\lim_{t\to\infty}\frac{\kappa^{[p,q]}_{\Heat_1,\Heat_2}}{t}=(-1)^p\lim_{t\to\infty}\frac{\kappa^{[p+q]}_{\Heat_2}}{t}.
\ee

\subsubsection{MDB and symmetry of the generating function for infinite-time cumulants per unit time}
\label{symgenfunccumul}

The modified detailed balance entails the  time-reversal symmetry \eqref{TimeRevPQ1PQ2CfC0} for $\Prob\left(\C_f \vert \Heat_1,\Heat_2; t\vert  \C_0\right)$ at finite time. Henceforth, according to its definition \eqref{defCharFunc},  the characteristic function can be rewritten as
\be
\label{CharFuncMDB}
\Esp{e^{\lambda_1\Heat_1(t)+\lambda_2\Heat_2(t)}}
=\sum_{\Heat_1}\sum_{\Heat_2}
e^{(\beta_1-\lambda_1)\Heat_1+(\beta_2-\lambda_2)\Heat_2}
\sum_{\C_f}\sum_{\C_0}\Prob\left(\C_0 \vert \Heat_1,\Heat_2; t\vert  \C_f\right)\Prob(\C_0,t_0=0).
\ee
When the Markov matrix is irreducible and the system has a finite number of
configurations, the Perron-Frobenius theorem entails that there exists a single
stationary state $\Probst$ and that every configuration $\C$ has a non-zero probability $\Probst(\C)$. Henceforth, in the case where the initial distribution is the stationary one, the relation \eqref{CharFuncMDB} leads to the inequalities
\be
\label{IneqChar}
\frac{\Probst^\text{min}}{\Probst^\text{max}}
\leq \frac{\Espst{e^{(\beta_1-\lambda_1)\Heat_1+(\beta_2-\lambda_2)\Heat_2}}}
{\Espst{e^{\lambda_1\Heat_1+\lambda_2\Heat_2}}}
\leq \frac{\Probst^\text{max}}{\Probst^\text{min}}.
\ee
(We recall that $\Espst{\cdots}$ denotes an average when the initial configurations are distributed according to the stationary measure $\Probst(\C)$, the maximum and minimum values of which are  $\Probst^\text{max}$ and $\Probst^\text{min}$ respectively (see \eqref{InequalityHeats}).
These inequalities entail that the generating function for the infinite-time limits of the joint cumulants per unit time defined in 
\eqref{alpha12def} obeys the symmetry 
\be
\label{symalpha12}
 \alpha_{1,2}(\lambda_1,\lambda_2)= \alpha_{1,2}(\beta_1-\lambda_1,\beta_2-\lambda_2).
\ee

We notice that the symmetry \eqref{symalpha12} can also be derived by considering the evolution of the Laplace transform of $\Prob(\Heat_1,\Heat_2;t)$.  With the notations of paper II,
$\Espst{e^{\lambda_1\Heat_1(t)+\lambda_2\Heat_2(t)}}=\sum_{\C_f}\sum_{\C_0}
(\C_f\vert\UevG(e^{\lambda_1},e^{\lambda_2};t)\vert\C_0)\Probst(\C_0)$ and  $\alpha_{12}(\lambda_1,\lambda_2)$ coincides with the largest eigenvalue of the operator
$\widetilde{\mathbb A}(\lambda_1,\lambda_2)$ which rules the evolution of
$\UevG(e^{\lambda_1},e^{\lambda_2};t)$ according to 
$d \UevG/dt=\widetilde{\mathbb A}\UevG$. The MDB implies that
the operator $\widetilde{\mathbb A}(\lambda_1,\lambda_2)$ obeys the symmetry
$\widetilde{\mathbb A}(\lambda_1,\lambda_2)=\widetilde{\mathbb A}^T(\beta_1-\lambda_1,\beta_2-\lambda_2)$, where $\widetilde{\mathbb A}^T$ denotes the transposed matrix of $\widetilde{\mathbb A}$ in the configuration basis. Then, by using the Perron-Frobenius theorem, one can prove that the largest eigenvalue of $\widetilde{\mathbb A}(\lambda_1,\lambda_2)$ satisfies the symmetry relation \eqref{symalpha12} (see the argument  in \cite{LebowitzSpohn1999} where an analogous symmetry is exhibited and then used for the derivation of the long-time fluctuation relation obeyed by the action functional defined in the comment after\eqref{SexchActionF}-\eqref{historypropertySexch}).

Moreover, since the system has a finite number of configurations, $\alpha_{12}(\lambda_1,\lambda_2)=\alpha_a(\lambda_a-\lambda_b)$  by virtue of \eqref{alpha12finite}, and the  symmetry property \eqref{symalpha12} of  $ \alpha_{1,2}(\lambda_1,\lambda_2)$ becomes 
\be
\label{symAlphaQ}
\alpha_a(\lambda)=\alpha_a(\beta_a-\beta_b-\lambda)
\quad\textrm{for $\{a,b\}=\{1,2\}$}.
\ee

We can apply the Gärtner-Ellis theorem (see subsubsection \ref{GET} and
references there):  if $\alpha_2(\lambda)$  exists and is differentiable for all $\lambda$ in $\mathbb{R}$, then the large deviation function of the current $\Jcum=\Heat_2/t$
exists and it can be calculated as the  Legendre-Fenchel  transform of $\alpha_2(\lambda)$, namely, with the signs chosen in the definitions used in the present paper,
\be
f_{\Heat_2}(\Jcum)=\min_{\lambda\in \mathbb{R}}\{\alpha_2(\lambda)-\lambda \Jcum\}.
\ee
Then, since $\alpha_2(\lambda)$ obeys the symmetry \eqref{symAlphaQ}, 
$f_{\Heat_2}(\Jcum)$ obeys the fluctuation relation \eqref{FRHeatCurrent}.
We notice that we can similarly retrieve the fluctuation relation obeyed by $f_{\Deltaexch S}$. Indeed,
$\alpha_{\Deltaexch S}(\lambda)\equiv\lim_{t\to+\infty}(1/t)\ln \Esp{e^{\lambda\Deltaexch S(t)}}$ is equal to $\alpha_{12}(\lambda \beta_1,\lambda \beta_2)$ with $\alpha_{12}(\lambda_1,\lambda_2)$ defined in \eqref{alpha12def}, and, according to \eqref{alpha12finite},
$
\alpha_{\Deltaexch S}(\lambda)=\alpha_{a}\left(\lambda(\beta_a-\beta_b)\right)
$.
Therefore the symmetry \eqref{symAlphaQ} implies that
\be
\label{symAlphaSexch}
\alpha_{\Deltaexch S}(1-\lambda)=\alpha_{\Deltaexch S}(\lambda).
\ee
Under the above assumptions, the large deviation function of the exchange entropy cumulative
current can be computed as the Legendre-Fenchel transform of $\alpha_{\Deltaexch
  S}$, namely $f_{\Deltaexch S}(\Jcum)=\min_{\lambda\in
    \mathbb{R}}\{\alpha_{\Deltaexch S}(\lambda)-\lambda \Jcum\}$. Henceforth the 
symmetry \eqref{symAlphaSexch} allows to retrieve the fluctuation relation \eqref{FTSexch}.

Another consequence of the symmetry \eqref{symAlphaQ} is that the Einstein-Green-Kubo
relation near equilibrium, which can be derived from the fluctuation relation
\eqref{FRHeatCurrent} for $f_{\Heat_2}(\Jcum)$ as shown in section
\ref{GreenKuboFR}, can equivalently be directly derived from the symmetry
\eqref{symAlphaQ} of its Legendre-Fenchel transform $\alpha_2(\lambda)$. This can be
checked as follows. Relation \eqref{symAlphaQ} can be rewritten for $a=2$ as 
\be
\label{symAlphaQAff}
\alpha_2(\lambda;\beta_1,\beta_2)=\alpha_2(-\Fth-\lambda;\beta_1,\beta_2),
\ee
where $\Fth=\beta_1-\beta_2$ is the thermodynamic force. The Einstein-Green-Kubo
relation involves the variation of the mean current
$\Jcumav(\beta_1,\beta_2)\equiv \left.\frac{\partial \alpha_2}{\partial
    \lambda}\right\vert_{(\lambda=0;\beta_1,\beta_2)}$ with respect to $\Fth$, so
we need to compute mixed derivatives of $\alpha_2$ with respect to $\lambda$ and
$\Fth$. Notice that in the identity \eqref{symAlphaQAff}, the second and the third
arguments are untouched. So if we make any invertible ($\lambda$-independent)
change of variables $(\beta_1,\beta_2)\leftrightarrow (\Fth,\rho)$ and set
$\alpha_2(\lambda;\beta_1,\beta_2)\equiv \alpha(\lambda;\Fth,\rho)$, we have the
following identity 
\be
\label{symAlphaQAff2}
\alpha(\lambda;\Fth,\rho)=\alpha(-\Fth-\lambda;\Fth,\rho), 
\ee 
where the third variable $\rho$ is purely a spectator. Keeping it fixed, we get
from the symmetry \eqref{symAlphaQAff2}: 
\be \left. \frac{\partial^2
    \alpha}{\partial\Fth\partial \lambda}\right\vert_{(\lambda;\Fth)} =
-\left.\frac{\partial^2 \alpha}{\partial\Fth \partial
    \lambda}\right\vert_{(-\Fth-\lambda;\Fth)}+\left.\frac{\partial^2
    \alpha}{\partial \lambda^2}\right\vert_{(-\Fth-\lambda;\Fth)}.  
\ee 
Taking this relation at $\lambda=\Fth=0$, we obtain 
\be \label{eq:GKprelim} 
2 \left. \frac{\partial^2 \alpha}{\partial\Fth\partial
    \lambda}\right\vert_{(\lambda=0;\Fth=0)} = \left. \frac{\partial^2
    \alpha}{\partial \lambda^2}\right\vert_{(\lambda=0;\Fth=0)}.
\ee
From now on, the discussion parallels the one in section \ref{GreenKuboFR}. We
repeat the argument in a slightly different form to stress again the slight
subtlety involved in variations with respect to $\Fth$ in the context of thermal
contact.  The condition $\Fth=0$ means $\beta_1=\beta_2\equiv \beta$. So the
right-hand side of \eqref{eq:GKprelim} is simply $\left. \frac{\partial^2
    \alpha}{\partial \lambda^2}\right\vert_{(\lambda=0;\Fth=0)}=
\left.\frac{\partial^2 \alpha_2}{\partial
    \lambda^2}\right\vert_{(\lambda=0;\beta_1=\beta_2=\beta)}$.  The left-hand
side of \eqref{eq:GKprelim} can be dealt with as follows. The change of
variables $(\beta_1,\beta_2)\leftrightarrow (\Fth,\rho)$ is
$\lambda$-independent, so we can set $\lambda=0$ once the derivative with
respect to $\lambda$ is taken 
\be 
\left. \frac{\partial \alpha}{\partial
    \lambda}\right\vert_{(\lambda=0;\Fth,\rho)}= \left. \frac{\partial \alpha_2}{\partial
    \lambda}\right\vert_{(\lambda=0;\beta_1,\beta_2)}=\Jcumav(\beta_1,\beta_2).  
\ee 
Of course, taking the partial derivative of $\Jcumav(\beta_1,\beta_2)$ with
respect to $\Fth$ is ambiguous, as it depends on the choice of the variable
$\rho$.  However, $\left.\frac{\partial
    \Jcumav}{\partial\Fth}\right\vert_{\Fth=0}$ has an intrinsic meaning,
independent of the choice of the variable $\rho$. Indeed, by construction
$\Jcumav(\beta,\beta)=0$ (equilibrium) so $\left. \frac{\partial
    \Jcumav}{\partial\beta_1}\right\vert_{\beta_1=\beta_2=\beta}+\left.
  \frac{\partial
    \Jcumav}{\partial\beta_2}\right\vert_{\beta_1=\beta_2=\beta}=0$. Moreover,
from $\Fth=\beta_1-\beta_2$ we get $1=\frac{\partial
  \beta_1}{\partial\Fth}-\frac{\partial \beta_2}{\partial\Fth}$. Now
$\frac{\partial \Jcumav}{\partial\Fth}=\frac{\partial
  \Jcumav}{\partial\beta_1}\frac{\partial \beta_1}{\partial\Fth}+\frac{\partial
  \Jcumav}{\partial\beta_2}\frac{\partial \beta_2}{\partial\Fth}$.  This implies
that $\left.\frac{\partial
    \Jcumav}{\partial\Fth}\right\vert_{\Fth=0}=\left.\frac{\partial
    \Jcumav}{\partial\beta_1}\right\vert_{\beta_1=\beta_2=\beta}$ is intrinsic
and we can write \eqref{eq:GKprelim} as 
\be 
\left.\frac{\partial \Jcumav}{\partial\Fth}\right\vert_{\Fth=0}= \left.
  \frac{1}{2} \frac{\partial^2 \alpha_2}{\partial
    \lambda^2}\right\vert_{(\lambda=0;\beta_1=\beta_2=\beta)}.  
\ee
By using the relation \eqref{defCumulantpertime} for $p=2$ between
$\alpha_2(\lambda)$ and the infinite-time cumulant per unit time together with the
fact that, when $\beta_1=\beta_2$, the stationary state is the equilibrium
state, we get the Einstein-Green-Kubo relation \eqref{GreenKuboGeneric}.  We notice that
the present derivation of the Einstein-Green-Kubo relation is very similar to the
argument developed by Lebowitz and Spohn in Ref.\cite{LebowitzSpohn1999} in the case
where there are several independent currents corresponding to several parameters
which drive the system out of equilibrium.

\subsection{Far from equilibrium  relations for infinite-time heat cumulants per unit time}
\label{RelCumulants}

In the present subsection we settle generic results ; thus we replace the cumulative heat $\Heat_2(t)$ by a generic cumulative quantity $X_t$.
According to the Einstein-Green-Kubo relation \eqref{GreenKuboGeneric} where, by virtue of \eqref{defOnsagerCoeff}, $\partial \Jcumav/\partial \Fth\vert_{\Fth=0}$  can be replaced by $\lim_{\Fth\to 0} \Jcumav/\Fth$,  in the vicinity of equilibrium the ratio $\Jcumav/\Fth$ is equal to the equilibrium value of
the second cumulant of the cumulative quantity $X_t$ per unit time. On the contrary,
when the system is far from equilibrium,  $\partial \Jcumav/\partial \Fth$ and $\Jcumav/\Fth$ are expected not to be equal to $1/2$ times the second cumulant of the cumulative heat $X_t$ per unit time in the stationary state. The property
\be
\frac{\partial \Jcumav}{\partial \Fth}\not=\frac{1}{2}
\lim_{t\to+\infty} \frac{\Espst{X_t^2}-\Espst{X_t}^2}{t}
\ee
arises from \eqref{genericlinearresponse}, while the discrepancy
\be
\label{GreenKubogen}
\frac{ \Jcumav}{\Fth}
\not=\frac{1}{2}
\lim_{t\to+\infty} \frac{\Espst{X_t^2}-\Espst{X_t}^2}{t}.
\ee
is already mentioned in Ref.\cite{LebowitzSpohn1999}. This discrepancy implies
that the generating function $\alpha(\lambda)$ for the infinite-time cumulants of
$X_t$ per unit time is not a quadratic function of its argument $\lambda$,
namely the large deviation function of the current $\Jcum_t=X_t/t$ is not a
quadratic in the generic case, i.e. the probability distribution of $X_t$ is not
asymptotically Gaussian. Indeed, if the generating function $\alpha(\lambda)$ were quadratic in $\lambda$, i.e.  $\alpha(\lambda)=\alpha^{(1)}\lambda + \frac{1}{2}\alpha^{(2)} \lambda^2$, then the symmetry \eqref{symAlphaQAff2} would lead to $\alpha^{(1)}= \frac{1}{2}\alpha^{(2)} \Fth$.
In the present section we derive an equation hierarchy  for the infinite-time cumulants per unit time far from equilibrium, which in particular gives how $\Jcumav/\Fth$ is related to all even cumulants per unit time in the infinite-time limit.

\clearpage
\subsubsection{Generalized Einstein-Green-Kubo relations for cumulants per unit time far from equilibrium}
\label{HierarchySingleCurrent}

The starting point is \eqref{symAlphaQAff} which we rewrite for convenience in the
generic form
\be
\label{CumRel}
\alpha(\lambda;\Fth)=\alpha(-\Fth-\lambda;\Fth),
\ee
where $\Fth$ is the thermodynamical force (an inverse temperature difference
in \eqref{symAlphaQ}). The viewpoint is that the coefficients  in the expansion of
$\alpha(\lambda;\Fth)$ in powers of $\lambda$ at $\lambda=0$ are related to the heat cumulants per unit time in the infinite-time limit,
\be
\Avkappa^{[p]}\equiv \lim_{t\to+\infty} \frac{1}{t}\kappa^{[p]}(t),
\ee
and these cumulants per unit time are  experimentally measurable. This is true at least for the
first few ones.

There are a number of ways to rewrite the symmetry \eqref{CumRel}. Formally it is equivalent
to
\be
\label{Formal}
e^{-\Fth\frac{\partial}{\partial \lambda}}
\alpha(\lambda;\Fth)=\alpha(-\lambda;\Fth).
\ee
Expanding both sides in powers of $\lambda$ and comparing we get that the cumulants per unit time, given by  $\alpha(\lambda;\Fth)\equiv
\sum_{p=0}^{+\infty}\frac{1}{p!}\lambda^p\Avkappa^{[p]}(\Fth)$ obey the equation hierarchy
\be
\label{redontAff}
\Avkappa^{[p]}(\Fth)=\sum_{q=0}^{+\infty}
(-1)^{p+q}\frac{\Fth^q}{q!}\Avkappa^{[p+q]}(\Fth) \qquad\text{ for } p=0,1,\cdots.
\ee
In \eqref{redontAff} $\Avkappa^{[0]}(\Fth)=0$, because $\alpha(\lambda;\Fth)$ is a cumulant generating function. The relation \eqref{redontAff}  can be explicitly checked in the case of the solvable model studied in paper II in a  kinetic regime where the probability distribution of the heat received from the slow thermostat is that of an asymmetric random walk.

A further expansion of the equations \eqref{redontAff} in powers of $\Fth$, with
$\Avkappa^{[p]}(\Fth)\equiv \sum_{q=0}^{+\infty}\frac{1}{q!}\Fth^q\Avkappa^{[p;q]}$, yields another hierarchy of relations, each of which involves only a finite number of derivatives of  cumulants per unit time with respect to $\Fth$,
\be
\label{redontCoeff}
\Avkappa^{[p;q]}=\sum_{r=0}^{q}(-1)^{p+r} \binom{q}{r}\Avkappa^{[p+r;q-r]} \qquad \text{ for
} p,q =0,1,\cdots,
\ee
where $\binom{q}{r}\equiv q!/[r!(q-r)!]$, and $\Avkappa^{[p;0]}=\Avkappa^{[p]}(\Fth=0)$ while
for $q\geq1$
\be
\Avkappa^{[p;q]}\equiv\left.\frac{\partial^q \Avkappa^{[p]}}{\partial \Fth^q}\right\vert_{\Fth=0}.
\ee
From the cumulant generating function interpretation, in \eqref{redontCoeff}  it is to be understood that
$\Avkappa^{[0;q]}=0$ for $q=0,1,\cdots$, since $\Avkappa^{[0]}(\Fth)=0$.

The relation hierarchies \eqref{redontAff} and \eqref{redontCoeff} are pure consequences of the symmetry
\eqref{CumRel} without using any further properties of $\alpha$. Anyway, they express
nothing but a parity relation around $\lambda=\Fth /2$, which fixes half of the
coefficients in terms of the other ones, so the equations are not expected to be all independent. Indeed,
this is particularly transparent at small order in $\Fth$ :
for $q=0$ \eqref{redontCoeff} yields nothing if $p$ is even, but if $p$
is odd it leads to $\Avkappa^{[2n+1;0]}=0$. As for the hierarchy \eqref{redontAff} obtained before expansions in powers of $\Fth$, the dependence between its equations can be exhibited as follows. By specifying the equations to $p=2n$ or $p=2n+1$, the infinite system of equations \eqref{redontAff} can be rewritten as the equivalent hierarchy,
\be
\left.
\begin{array}{ll}
& \displaystyle
\Avkappa^{[2n+1]}(\Fth)=\sum_{q=1}^{+\infty}(-1)^{q+1}\frac{\Fth^q}{(q+1)!}\Avkappa^{[2n+q+1]}(\Fth) 
\\
& \displaystyle
\Avkappa^{[2n+1]}(\Fth)=\frac{1}{2}\sum_{q=1}^{+\infty}(-1)^{q+1}\frac{\Fth^q}{q!}\Avkappa^{[2n+q+1]}(\Fth) 
\end{array}
\right \}\qquad \textrm{for $n=0,1,\cdots$},
\ee
where the right-hand sides of both relations involves both odd and even
cumulants per unit time. The difference between the two expressions for $
\Avkappa^{[2n+1]}(\Fth)$ gives the infinite sum rule \\
$0=\sum_{q=2}^{+\infty}(-1)^{q+1}\left[\frac{1}{(q+1)!}-\frac{1}{2}\frac{1}{q!}\right]\Fth^q\Avkappa^{[2n+q+1]}(\Fth) $.

It is not difficult to obtain independent relations which express odd cumulants  in terms of even cumulants. We rewrite \eqref{Formal}
as
\be
\left(1+e^{-\Fth\frac{\partial}{\partial \lambda}}\right)
\alpha(\lambda;\Fth)=\alpha(\lambda;\Fth)+\alpha(-\lambda;\Fth),
\ee
i.e.
\be
\label{impair}
\alpha(\lambda;\Fth) = \frac{1}{1+e^{-\Fth\frac{\partial}{\partial \lambda}}}
\left[\alpha(\lambda;\Fth)+\alpha(-\lambda;\Fth)\right].
\ee
Note that $\left[1+e^{-x}\right]^{-1}+\left[1+e^{x}\right]^{-1}=1$ so that
$\left[1+e^{-x}\right]^{-1}-\frac{1}{2}$ is an odd function of $x$. Hence the
Taylor expansion in $x$ can be written
\be
\frac{1}{1+e^{-x}}=\frac{1}{2}\left(1+\sum_{k=0}^{+\infty} d_k x^{2k+1}\right)
=\frac{1}{2}\left(1+\frac{x}{2}-\frac{x^3}{24}+\frac{x^5}{240}-\frac{17x^7}{40320}+\cdots\right),
\ee
where the coefficients $d_k$ are related to the classical Bernoulli numbers by
\be
\label{expdk}
d_k=2\frac{4^{k+1}-1}{(2k+2)! }B_{2k+2}.
\ee
From the identity
\be
1=\frac{1}{2}(1+e^{-x})\left(1+\sum_{k=0}^{+\infty} d_k x^{2k+1}\right)
\ee
one infers the recursion relation
\be
d_k=\frac{1}{2}\left(\frac{1}{(2k+1)!}-\sum_{0 \leq l <k }\frac{d_l}{(2(k-l))!}
\right).
\ee
Finally, in the expansion of \eqref{impair} the even terms cancel out and one is
left with
\be
\label{indepAff}
\Avkappa^{[2n+1]}(\Fth)=\sum_{k=0}^{+\infty} d_k
\Fth^{2k+1}\Avkappa^{[2(n+k+1)]} (\Fth) \qquad\text{ for } n=0,1,\cdots,
\ee
which expresses systematically odd cumulants per unit time in terms of even
cumulants per unit time. For instance,  in the case $n=0$, the ratio of the  out-of-equilibrium current  $\Jcumav(\Fth)=\Avkappa^{[1]}$ and the thermodynamic force $\Fth$  is determined by  all even cumulants as
\be
\frac{\Jcumav(\Fth)}{\Fth}=\sum_{q=0}^{+\infty} d_q \Fth^{2q}
\lim_{t\to+\infty} \frac{\kappa_\text{\mdseries }^{[2(q+1)]}(\Fth)}{t} .
\ee
The latter relation may be viewed as a a  far-from-equilibrium generalization of
the Einstein-Green-Kubo relation \eqref{GreenKuboGeneric}-\eqref{defOnsagerCoeff}. 

\clearpage
\subsubsection{Relations between non-linear responses of cumulants per unit time near equilibrium}

It is to be noted that the two equivalent hierarchies \eqref{redontAff} and \eqref{indepAff} for relations between cumulants make sense only if
the cumulants satisfy some growth conditions, the second one being most stringent
(note that $d_q \simeq  4 (-1)^q \pi^{-2q-2}$). Such problem do not arise in the
double expansions in both $\lambda$ and $\Fth$. Expanding \eqref{indepAff} in
powers of $\Fth$ yields for $n=0,1,\cdots$
\bea
\label{indepCoeff}
\Avkappa^{[2n+1;0]}&=&0
\\\nonumber
\Avkappa^{[2n+1;q]}&=&
\sum_{k=0}^{\lfloor(q-1)/2\rfloor} \frac{d_k q!}{[q-(2k+1)]!}\Avkappa^{[2(n+k+1);q-(2k+1)]}
\qquad \textrm{$q=1,2,\cdots$}
\eea
where $\lfloor y\rfloor$ denotes the lower integer part of $y$.
It is easy to check, at least for small values of $q$, that the contents of \eqref{redontCoeff} and
\eqref{indepCoeff}  are the same.
The physical consequences of the latter relations  are more readily inferred by explicitly rewriting the  relations in the case where $q$ is odd or even and in terms of either equilibrium cumulants per unit time or the partial derivatives of out-of-equilibrium cumulants per unit time with respect to the thermodynamic force $\Fth$. Then the formulae
\eqref{indepCoeff} read
\begin{subequations}
\label{indepCoeffBis}
\be
\label{CumRel1}
\Avkappaeq^{[2n+1]}=0
\ee
\be
\label{CumRel2}
\left.\frac{\partial\Avkappa^{[2n+1]}}{\partial \Fth}\right\vert_{\Fth=0}=
\frac{1}{2}\Avkappaeq^{[2(n+1)]}
\ee
\be
\label{CumRel3}
\left.\frac{\partial^{2m}\Avkappa^{[2n+1]}}{\partial \Fth^{2m}}\right\vert_{\Fth=0}
\underset{m\geq1}{=}
\sum_{r=1}^{m}d_{m-r}\frac{(2m)!}{(2r-1)!}
\left.\frac{\partial^{2r-1}\Avkappa^{[2(n+m+1-r)]}}{\partial \Fth^{2r-1}}\right\vert_{\Fth=0}
\ee
\be
\label{CumRel4}
\left.\frac{\partial^{2m+1}\Avkappa^{[2n+1]}}{\partial \Fth^{2m+1}}\right\vert_{\Fth=0}
\underset{m\geq1}{=}
d_m (2m+1)! \,\Avkappaeq^{[2(n+m+1)]}
+\sum_{r=1}^{m}d_{m-r}\frac{(2m+1)!}{(2r)!}
\left.\frac{\partial^{2r}\Avkappa^{[2(n+m+1-r)]}}{\partial \Fth^{2r}}\right\vert_{\Fth=0}
\ee
\end{subequations}
where the values of the $d_k$'s are given in \eqref{expdk}.

The equilibrium statements \eqref{CumRel1} are exemplified in the case of thermal contact as follows. The first cumulant per unit time $\Avkappa^{[1]}$ for the cumulative heat $\Heat_2$ coincides with the average of the  instantaneous current of $\jinst_2$, $\Espst{\jinst_2}^{(\beta_1,\beta_2)}\equiv\Jcumav$, and  from the first equation \eqref{CumRel1} for $n=0$  we retrieve that at equilibrium
 the mean instantaneous current vanishes, $\Jcumav_\text{\mdseries eq}=0$.
More generally,  \eqref{CumRel1} states that  at equilibrium all odd cumulants per unit time of the heat amounts received from one of the two heat baths vanish.

The relations \eqref{CumRel2} are generalized fluctuation-dissipation relations in the vicinity of equilibrium, which are called ``generalized'' in the sense that they express  the linear response of any cumulant per unit time. Indeed, in the case of thermal contact, from \eqref{CumRel2} for $n=0$ one retrieves
the fluctuation-dissipation relation  (also called Einstein-Green-Kubo relation) in the form 
\eqref{GreenKuboGeneric} where $X_t$ is the heat amount received from one of the two heat baths.
More generally, from \eqref{CumRel2} for any value of $n$ one gets the  generalized fluctuation-dissipation relations 
\be
\label{GeneralizedFDR}
\lim_{(\beta_1,\beta_2)\to(\beta,\beta)}\frac{1}{\beta_1-\beta_2}\left(\lim_{t\to+\infty} \frac{\kappa^{[2n+1]}}{t} \right)=\frac{1}{2}
\lim_{t\to+\infty} \frac{\kappa_\text{\mdseries eq}^{[2n+2]}}{t}.
\ee
Analogous  relations have  been derived  in the case of several independent out-of-equilibrium steady currents by 
Andrieux and Gaspard \cite{AndrieuxGaspard2007JStatMech} (see also subsection \ref{DemoSeveralCurrents})
We notice that for $n=1$, \eqref{GeneralizedFDR} means that, in  the limit of vanishing  $\beta_1-\beta_2$, the ratio between 
the out-of-equilibrium  third centered moment (non-normalized skewness) $\Esp{(\Heat_2-\Esp{\Heat_2})^3}$ per unit time and  the thermodynamic force $\beta_1-\beta_2$ tends to half  the equilibrium fourth cumulant (kurtosis multiplied by the square of the variance) 
$\Espeq{(\Heat_2-\Espeq{\Heat_2})^4}-3\Espeq{(\Heat_2-\Espeq{\Heat_2})^2}$ per unit time. These  cumulants are expected to be experimentally measurable.

The relations \eqref{CumRel3} and  \eqref{CumRel4} deal with higher-order
response coefficients. 
The relation \eqref{CumRel3} in the case $m=1$ reads 
$\partial^2 \Avkappa^{[2n+1]}/\partial \Fth^2\vert_{\Fth=0}=
\partial \Avkappa^{[2n+2]}/\partial \Fth\vert_{\Fth=0}$. 
For  instance, for $n=0$ and $n=1$ it leads respectively to
\be
\left.\frac{\partial^2 \Jcumav}{\partial \Fth^2}\right\vert_{\Fth=0}
=\left.\frac{\partial \Avkappa^{[2]}}{\partial \Fth}\right\vert_{\Fth=0},
\ee
and 
\be
\left.\frac{\partial^2 \Avkappa^{[3]}}{\partial \Fth^2}\right\vert_{\Fth=0}
=\left.\frac{\partial \Avkappa^{[4]}}{\partial \Fth}\right\vert_{\Fth=0}.
\ee
The relation \eqref{CumRel4} in the  case $m=1$ gives
$\partial^3\Avkappa^{[2n+1]}\partial \Fth^3\vert_{\Fth=0}
=\frac{3}{2}\partial^2\Avkappa^{[2n+2]}/\partial \Fth^2\vert_{\Fth=0}
-\frac{1}{4}\Avkappa_\text{\mdseries eq}^{[2n+4]}$.
For instance, in the case $n=0$ it reads
\be
\left.\frac{\partial^3 \Jcumav}{\partial \Fth^3}\right\vert_{\Fth=0}
=\frac{3}{2}\left.\frac{\partial^2 \Avkappa^{[2]}}{\partial \Fth^2}\right\vert_{\Fth=0}
-\frac{1}{4}\Avkappa_\text{\mdseries eq}^{[4]}.
\ee

\section{Extension of previous results to a larger class of models}
\label{GeneraliztionSeveralCurrents}

\subsection{Generic expression of exchange entropy variation}
\label{GenericExchS}

In this section we consider the generic case where   the finite-size system $\syst$ is made of $\nu_s$ species of 
mobile elementary constituents   and  can occupy a domain  whose  boundaries may be mobile interfaces. The degrees of freedom of every elementary constituent involve the site where it sits in discretized space and  some possible internal degrees of freedom. In the following we call degrees of freedom of a configuration $\C$ of the system $\syst$  the degrees of freedom of the  elementary constituents in this configuration. (The number of constituents of every species may vary from one configuration to another.) When  some boundaries are mobile interfaces, the description of any configuration $\C$ of the system not only involves the values of its degrees of freedom  
but it is also specified by the positions of the  interfaces that surround the domain, called ${\cal D}(\C)$, that the system $\syst$  can  occupy. For each configuration $\C$ one can  define the following global quantities : the energy $\En(\C)$, the volume $v(\C)$ of ${\cal D}(\C)$ and the total number of elementary constituents $n(\C)=\sum_{s=1}^{\nu_s}n_s(\C)$, where $n_s(\C)$ is the  number of elementary constituents of species $s$ that sit in ${\cal D}(\C)$. All these quantities are assumed to take a finite number of values. The system $\syst$ is in contact with several macroscopic bodies ${\cal B}_a$'s.

A crucial assumption is that in the course of the ergodic deterministic microscopic  dynamics of the whole system ($\syst$ and the large parts ${\cal B}_a$), for a given configuration $\C$, the domain ${\cal D}(\C)$ can be divided in several disjoint subdomains ${\cal D}_a(\C)$'s such that some  boundary portion of ${\cal D}_a(\C)$ can move only thanks to a corresponding volume variation of large part ${\cal B}_a$  and  the values of the degrees of freedom that sit inside ${\cal D}_a(\C)$ can vary only by  exchanging microscopically conserved  quantities (energy  and/or matter)  with  the corresponding large part ${\cal B}_a$. Then  a   jump of  system $\syst$  from  configuration $\C$  to another one $\C'$  is allowed only if ${\cal D}(\C)$ and ${\cal D}(\C')$ differ by a displacement of some  boundary portion of only one ${\cal D}_a(\C)$ and by different values of the degrees of freedom inside ${\cal D}_a(\C)$ and ${\cal D}_a(\C')$ ; then we use the notation $\C'\in \F_{a}(\C)$. 
Moreover  the corresponding jump of the microscopic configuration of ${\cal B}_a$ is such that   conservation rules hold for the sum  of the energies of $\syst$ and ${\cal B}_a$, $\En(\C')+E'_a=\En(\C)+E_a$, for the sum of the volumes that they occupy, 
$v(\C')+V'_a=v(\C)+V_a$, and for the sum of the numbers of elementary constituents of species $s$ that they contain, $n_s(\C')+N'_{a,s}=n_s(\C)+N_{a,s}$, where $E_a$, $V_a$ and the $N_{a,s}$ are the values of the extensive parameters that characterize body ${\cal B}_a$ (and the prime denotes their values after the configuration jump). For instance if   system  $\syst$ is a mobile  diathermal thin solid wall separating a vessel in two parts filled with gases kept at different temperatures and pressures, then $\syst$ can be viewed as made of two layers of constituents, each of which interacts respectively  with body
${\cal B}_1$ and  ${\cal B}_2$. Then an infinitesimal displacement of $\syst$ such that the volume of ${\cal B}_1$ increases while that of ${\cal B}_2$ decreases by the same absolute amount can be  decomposed into two microscopic configuration jumps of the global system :  in the first (second) jump only the layer in contact with ${\cal B}_2$ (${\cal B}_1$) moves and the volume of $\syst$ increases (decreases), and after two  jumps the volume of $\syst$  has retrieved its initial value. 

Then the mesoscopic Markovian dynamics defined according to the prescription of subsection \ref{MarkovianApproximationConseq} is such that, when energy, volume and matter are exchanged, the transition rates obey the microcanonical  detailed balance
 \be
\label{ratioWmcgen}
 \frac{W(\C',E'_a,V'_a, \{N'_{a,s}\}\leftarrow \C,E_a,V_a, \{N_{a,s}\})}
  {W(\C,E_a,V_a, \{N_{a,s}\}\leftarrow \C',E'_a,V'_a, \{N'_{a,s}\})}=
\frac{\Omega_a(E'_a,V'_a, \{N'_{a,s}\})}
{\Omega_a(E_a,V_a, \{N_{a,s}\})}
\ee
with
\be
\label{ratioOmegaPattern}
\frac{\Omega_a(E'_a,V'_a, \{N'_{a,s}\})}
{\Omega_a(E_a,V_a, \{N_{a,s}\})}
\equiv e^{\SB_a(E'_a,V'_a, \{N'_{a,s}\})-\SB_a(E_a,V_a, \{N_{a,s}\})},
 \ee
where $\Omega_a(E_a,V_a, \{N_{a,s}\})$ is the number of configurations (or microstates) of large part ${\cal B}_a$ when it is isolated. 
In the following, in the spirit of the notations used by Callen \cite{Callen1960} we denote by $X_i^{(a)}$ the extensive macroscopic parameters of ${\cal B}_a$, namely in the generic case
\be
\label{examplesX}
X^{(a)}_0\equiv E_a, \quad  X^{(a)}_1\equiv V_a, \quad X^{(a)}_{1+s}=N_{a,s}.
\ee
The total number of elementary constituents in ${\cal B}_a$ is $N_a=\sum_sN_{a,s}$. The microscopic conservation rules that are associated with \eqref{ratioWmcgen} read
\be
\label{ConservationRuleGen}
X'^{(a)}_i-X^{(a)}_i=-\left[x_i(\C')-x_i(\C)\right] \quad   \text{if}\quad \C'\in\F_a(\C),
\ee
 with the same notations for  system $\syst$ as those introduced in \eqref{examplesX} for the macroscopic bodies.

If the  bodies ${\cal B}_a$ are so large that they can be  described by a thermodynamic limit, then, in a transient regime  where the macroscopic extensive parameters $X^{(a)}_i$'s have negligible relative variations, 
  ${\cal B}_a$ remains at thermodynamic equilibrium. 
The thermodynamic entropy  per elementary constituent, $\STH_a/N_a$, coincides with the thermodynamic limit of the Boltzmann entropy per  elementary constituent, $\SB_a/N_a$. The  intensive thermodynamic parameter $F^{(a)}_i$ conjugate to the extensive quantity $X_i^{(a)}$  by $F^{(a)}_i\equiv \partial \STH_a /\partial X^{(a)}_i$ (with $X^{(a)}_i$ defined in \eqref{examplesX}) is given by
\be
\label{examplesF}
F^{(a)}_0\equiv \beta_a, \quad  F^{(a)}_1\equiv \beta_a P_a, \quad F^{(a)}_{1+s}=-\beta_a \mu_{a,s},
\ee
where $\beta_a$ is the inverse thermodynamic temperature, $P_a$ the thermodynamic pressure and $\mu_{a,s}$ the chemical potential of species $s$ in ${\cal B}_a$. Then from the relations \eqref{ratioWmcgen} and \eqref{ratioOmegaPattern}, 
one can show, as in subsection \ref{TransientSec},  that in the  transient regime   the transition rates obey the modified  detailed balance \eqref{MDBexch} where $\deltaexch S(\C'\leftarrow\, \C)$ is opposite to
 the infinitesimal variation at fixed intensive parameters of the thermodynamic entropy of the reservoir ${\cal B}_a$ that causes the jump   from $\C$ to $\C'$ under the conservation rules \eqref{ConservationRuleGen}. Therefore, if $\C'\in\F_a(\C)$
\be  
\label{ExplicitDeltaSexchGen}
 \deltaexch S(\C'\leftarrow\, \C)=\beta_a [\En(\C')-\En(\C)]
+\beta_a  P_a [v(\C')-v(\C)]- \beta_a\sum_s \mu_{a,s} [n_s(\C')-n_s(\C)].
\ee
In the case of pure thermal contact, the volume of  system $\syst$ does not vary, $v(\C)=v(\C')$, and there is no matter exchange ; then
   $\deltaexch S(\C'\leftarrow\, \C)=\beta_a [\En(\C')-\En(\C)]$ coincides with  $\beta_a$ times the opposite of the variation of the internal energy  of ${\cal B}_a$, which is equal in that case to the heat given by ${\cal B}_a$ at constant volume.
If system $\syst$ and macroscopic body ${\cal B}_a$ are compressible then, during energy exchanges such that the thermodynamic pressure $P_a$ of ${\cal B}_a$ remains fixed,   the variation of the internal energy of the macroscopic body ${\cal B}_a$ involves both heat and   pressure work ; then
$
\deltaexch S(\C'\leftarrow\, \C)=\beta_a [\En(\C')-\En(\C)]
+\beta_a  P_a [v(\C')-v(\C)]
$
 coincides with $\beta_a$ times the opposite of the variation of the enthalpy  of ${\cal B}_a$, which is equal in that case to the heat given by ${\cal B}_a$ at constant pressure.
 Moreover  system $\syst$ may  receive work from some conservative external forces $ f^\text{ext}_b$ (such as gravitational or electrical fields), each of which causes the variation of some global  coordinate $Z_b(\C)$ of the system (such as its mass center or its electrical  barycenter) but does not act upon the macroscopic bodies. In that case the mechanical energy $\En(\C)$ of system $\syst$ in configuration $\C$ is the sum of its internal energy  $\En_\text{int}(\C)$ and an external potential energy  $\En_\text{ext}(\C)$.   In all these situations the distances between the possible  energy levels $\En$ are time-independent, and  during the evolution  only the occupation of the energy levels  is modified, contrarily to the case where a time-dependent force acts on the system by changing the spacing between energy levels.

In order to handle compact notations for the successive variations  $\deltaexch S$ in the course of a history of system $\syst$, let us  introduce
the notation
$\delta  \chi_i^{(a)}(\C'\leftarrow\, \C)$ for the quantity with index $i$
received by the system from  reservoir ${\cal B}_a$ when the system jumps from $\C$ to
$\C'$, with the same convention as in definition \eqref{defdeltaq}, namely 
\be
\label{defdeltax}
\begin{cases}
\delta \chi_i^{(a)}(\C'\leftarrow\, \C)=x_i(\C')-x_i(\C)  &\textrm{if}\quad  \C'\in\F_a(\C) \\
\delta \chi_i^{(a)}(\C'\leftarrow\, \C)=0 &\textrm{otherwise}
  \end{cases}.
\ee
(In the case where $ \C'\in\F_a(\C)$ but where in fact reservoir ${\cal B}_a$ does not exchange quantity with index $i$ in the jump from $\C$ to $\C'$, $x_i(\C')-x_i(\C)=0$). Then the exchange entropy variation in a jump takes the form
\be
\label{ExplicitGenericdeltaexchS}
\deltaexch S(\C'\leftarrow\, \C)=\sum_{i}\sum_{a\in R(i)} F_i^{(a)}\delta \chi_i^{(a)}(\C'\leftarrow\, \C),
\ee 
where the first sum runs over the indices $i$ of the extensive quantities  defined in \eqref{examplesX}, the second sum runs  over the indices of the reservoirs that indeed can exchange quantity $i$ with the system and $R(i)$ denotes the set of the latter reservoirs.

\subsection{Consequences of MDB  at finite time}

In the form \eqref{MDBexch} where it involves the microscopic exchange entropy variation $\deltaexch S(\C'\leftarrow\, \C)$ associated with a  jump from configuration $\C$ to  configuration $\C'$, the modified  detailed balance entails the symmetry \eqref{historypropertySexch} between the probabilities for time-reversed histories. At a more mesoscopic level,
  let us compare the probability of all evolutions from  configuration $\C_0$ to  configuration $\C_f$, in the course of which the system receives given cumulative quantities  with index $i$ ${\cal X}^{(a)}_i=\sum \delta \chi^{(a)}_i$ from each reservoir ${\cal B}_a$,  and the probability of the reversed evolutions, namely evolutions from $\C_f$ to $\C_0$ where the cumulative quantities are $-{\cal X}^{(a)}_i$'s.  The symmetry \eqref{TimeRevPQ1PQ2CfC0} written in the case of thermal contact takes the following form in the generic case
\be
\label{TimeRevPQ1PQ2CfC0Gen}
\frac{\Prob\left(\C_f \vert\{{\cal X}^{(a)}_i\}; t\vert  \C_0\right)}{\Prob\left(\C_0\vert \{-{\cal X}^{(a)}_i\}; t\vert \C_f\right)}
=e^{-\Deltaexch S(\{{\cal X}^{(a)}_i\})},
\ee
with, according to  \eqref{ExplicitGenericdeltaexchS},
\be 
\label{defgenDeltaexchS}
\Deltaexch S(\{{\cal X}^{(a)}_i\})=\sum_i\sum_{a\in R(i)} F^{(a)}_i{\cal X}^{(a)}_i.
\ee
In \eqref{TimeRevPQ1PQ2CfC0Gen} and in the formul\ae\, derived from it in the following, 
${\cal X}^{(a)}_i$ occurs only if reservoir ${\cal B}_a$ indeed exchanges quantity with index $i$.
The ratio of probabilities in \eqref{TimeRevPQ1PQ2CfC0Gen} does not  explicitly  depend on the initial and final configurations $\C_0$ and $\C_f$. There is only an implicit dependence on these configurations through the conservation rules that the  cumulative exchange quantities ${\cal X}^{(a)}_i$ for a given history must satisfy, namely $\sum_{a\in R(i)} {\cal X}^{(a)}_i=x_i(\C_f)-x_i(\C_0)$.

At the  macroscopic level, namely when only the exchanges of extensive quantities with the reservoirs are measured, there appears a symmetry for transient regimes where the system  is initially prepared in some equilibrium state with a fixed intensive parameter $F_i^0$ for each configuration observable $x_i$ and then is suddenly put into contact with reservoirs with thermodynamic parameters $F^{(a)}_i$'s that drive the system into a non-equilibrium state. 
The symmetry involves the excess exchange entropy variation $\Deltaexchexcessgen S$, defined as the difference between the exchange entropy variation under the non-equilibrium external constraints $F^{(a)}_i$'s, defined in \eqref{defgenDeltaexchS}, and the corresponding variation under the equilibrium conditions where for  all reservoirs  that exchange  quantity with index $i$  the  thermodynamic parameter has the same value
$F_i^0$, 
 \be
\Deltaexchexcessgen(\{{\cal X}^{(a)}_i\})\equiv\Deltaexch S (\{{\cal X}^{(a)}_i\})-\sum_i F_i^0\sum_{a\in R(i)} {\cal X}^{(a)}_i.
\ee 
When the system is prepared in the equilibrium state with probability distribution $\Probeq(\{F_i^0\})$ and put into contact with reservoirs with thermodynamic parameters $F^{(a)}_i$'s at the initial time of the measurements of exchanged quantities, the joint probability for the cumulative exchange quantities ${\cal X}^{(a)}_i$ obeys the symmetry relation at any finite time,
\be
\label{DFRJointProbChi}
\frac{\Prob_{\Probeq(\{F_i^0\})}\left(\{{\cal X}^{(a)}_i\}\right)}
{\Prob_{\Probeq(\{F_i^0\})}\left(\{-{\cal X}^{(a)}_i\}\right)}=e^{-\Deltaexchexcessgen S(\{{\cal X}^{(a)}_i\})}.
\ee
As a consequence, the excess exchange entropy variation $\Deltaexchexcessgen S$   obeys  the symmetry relation at any finite time, or ``detailed fluctuation relation'', 
\be
\label{DFRDeltaexchexcess0gen}
\frac{\Prob_{\Probeq(\{F_i^0\})}\left(\Deltaexchexcessgen S\right)}
{\Prob_{\Probeq(\{F_i^0\})}\left(-\Deltaexchexcessgen S\right)}=e^{-\Deltaexchexcessgen S}.
\ee
The latter relation itself entails the identity, or ``integral fluctuation relation'', 
\be
\Esp{e^{\Deltaexchexcessgen S}}_{\Probeq(\{F_i^0\})}=1.
\ee

\subsection{Consequences of MDB in the infinite-time limit}

A generalization of the argument in  subsubsection \ref{FluctuationRelationSexch} shows that  the symmetry relation \eqref{TimeRevPQ1PQ2CfC0Gen} enforced by the MDB at finite time leads to the existence of  lower and upper bounds for the ratio between  the finite-time probability  to measure cumulative quantities with values $\{{\cal X}^{(a)}_i\}$ and the corresponding probability to measure the opposite values $\{-{\cal X}^{(a)}_i\}$, when the system is in its stationary state. 
The finite-time  inequalities \eqref{InequalityHeats} become in the generic case
\be
\label{InequalityHeatsGen}
\frac{\Probst^\text{min}}{\Probst^\text{max}}\leq
\frac{\Probst\left(\{{\cal X}^{(a)}_i\}; t\right)}{\Probst\left(\{-{\cal X}^{(a)}_i\}; t\right)e^{-\Deltaexch S(\{{\cal X}^{(a)}_i\})}}
\leq\frac{\Probst^\text{max}}{\Probst^\text{min}}.
\ee
As a consequence, the  dimensionless  exchange entropy variation $\Delta_\text{exch}S$ given by \eqref{defgenDeltaexchS} obeys the fluctuation relation \eqref{FTSexch0}.
 
As for the ``characteristic function'' of the extensive exchanged quantities ${\cal X}^{(a)}_i$'s in the stationary state, namely
$\Espst{e^{\sum_a\lambda_i^{(a)} {\cal X}^{(a)}_i(t)}}$, the symmetry \eqref{TimeRevPQ1PQ2CfC0Gen}-\eqref{defgenDeltaexchS} arising from the  MDB entails that the characteristic function obeys an inequality similar to \eqref{IneqChar}.
As a consequence, the generating function of the infinite-time limit of the joint cumulants per unit time obeys a symmetry which is a generalization of \eqref{symalpha12}
\be
\alpha_{\{{\cal X}^{(a)}_i\}}(\{\lambda_i^{(a)}\})=\alpha_{\{{\cal X}^{(a)}_i\}}(\{-F^{(a)}_i-\lambda_i^{(a)}\}).
\ee

However, because of the conservation laws for the microscopic quantities exchanged with the reservoirs, the joint cumulants of the ${\cal X}^{(a)}_i$'s per unit time are not independent in the infinite-time limit. For instance, if $R(i)$ denotes the set of reservoirs which exchange the  quantity with index $i$, because of the conservation law for every species of exchanged quantity, $\sum_{a\in R(i)} {\cal X}^{(a)}_i$  is equal to the difference between the values of some observable of the system in the final and initial configurations of the evolution.
When the system has a finite number of configurations, $\sum_{a\in R(i)} {\cal X}^{(a)}_i$  is bounded and
 the  mean currents $\Jcumav_i^{(a)}$'s defined as $\Jcumav_i^{(a)}=\lim_{t\to+\infty}\Esp{{\cal X}^{(a)}_i(t)}/t$ are related by
\be
\sum_{a\in R(i)} \Jcumav_i^{(a)}=0.
\ee
Other relations may arise from other microscopic conservation rules determined by the specific forms of the transitions rates.  
We consider the generic (but not universal) case where there exists a set of cumulative quantities $Y_\gamma$'s, which are linear combinations of the ${\cal X}^{(a)}_i$'s, but less numerous than the ${\cal X}^{(a)}_i$'s, and a set of parameters $\Fth_\gamma$'s such that $\sum_i \sum_{a\in R(i)} F^{(a)}_i {\cal X}^{(a)}_i+\sum_\gamma \Fth_\gamma Y_\gamma$ is  bounded by a constant independent of time. 
With these assumptions the expression for the exchange entropy flow in the stationary state  takes the generic form
\be
\label{defAffalphagenIndep}
\left.\frac{\dexch S}{dt}\right\vert_\text{st}=
\sum_i \sum_{a\in R(i)} F^{(a)}_i \Jcumav_i^{(a)}=-\sum_\gamma \Fth_\gamma \Jcumav^\star_\gamma,
\ee
where the mean currents $ \Jcumav^\star_\gamma$'s, defined as $\Jcumav^\star_\gamma\equiv\lim_{t\to+\infty}\Esp{Y_\gamma(t)}/t$, are less numerous than the $\Jcumav_i^{(a)}$'s.
Moreover, these assumptions entail that the symmetry \eqref{TimeRevPQ1PQ2CfC0Gen}-\eqref{defgenDeltaexchS} leads to the inequality
\be
\label{TimeRevPQ1PQ2CfC0GenBis}
m\leq \frac{\Prob\left(\C_f \vert\{Y_\gamma\}; t\vert  \C_0\right)}{\Prob\left(\C_0\vert \{-Y_\gamma\}; t\vert \C_f\right) e^{\sum_\gamma \Fth_\gamma Y_\gamma}},
\leq M,
\ee
where $m$ and $M$ are constants.
As a consequence, the characteristic function for the $Y_\gamma$'s obeys an inequality similar to \eqref{IneqChar} and subsequently the generating function of the infinite-time limits of their joint cumulants per unit time obeys a symmetry which is a generalization of \eqref{symAlphaQAff},
\be
\label{CumRelGen}
\alpha_{\{Y_\gamma\}}(\{\lambda_\gamma\};\{\Fth_\gamma\})=\alpha_{\{Y_\gamma\}}(\{-\Fth_\gamma-\lambda_\gamma\};\{\Fth_\gamma\}).
\ee
In the framework of graph theory (see subsubsection \ref{ComparisonGraph})  Andrieux and Gaspard \cite{AndrieuxGaspard2007JStatPhys} have shown an analogous symmetry  for dimensionless cumulative quantities $Y_\gamma$'s, where each $Y_\gamma$ is defined as the sum of the cumulative probability currents through the chords of all cycles in the graph that have the same affinity 
$\Aff_\gamma$ and where the $\Fth_\gamma$'s are replaced by the cycle affinities $\Aff_\gamma$'s. A similar result is also obtained in Ref.\cite{FaggionatoDiPietro2011}. We also notice that an example of the symmetry relation \eqref{CumRelGen} is derived in another context in Ref.\cite{CleurenVanDenBroeckKawai2006}.

\subsection{Generalized Einstein-Green-Kubo relations for several independent currents}
\label{DemoSeveralCurrents}

In standard fluctuation-dissipation Einstein-Green-Kubo formulae, valid near equilibrium, the relevant out-of-equilibrium quantity that is related to equilibrium fluctuations is the Onsager coefficient $L_{\alpha\gamma}$ introduced in the phenomenological thermodynamics of irreversible phenomena :  ``thermodynamic fluxes'' $\Jcumav^\star_\gamma$'s and ''thermodynamic forces''  $\Fth_\gamma$'s are defined by pairs from the expression of the entropy production rate $\dint S/dt\vert_\text{st}=-\dexch S/dt\vert_\text{st}$ given by \eqref{defAffalphagenIndep}, where the $\Jcumav^\star_\gamma$'s are independent currents  between reservoirs,  and the Onsager coefficient is defined as
 \be
 \label{defLalphagen}
 L_{\alpha\gamma}\equiv \left.\frac{\partial \Jcumav^\star_\alpha}{\partial \Fth_\gamma}\right\vert_{\{\Fth_{\gamma'}=0\}}.
 \ee
The generic statement of Einstein-Green-Kubo relations reads 
\be
L_{\alpha\gamma}= \frac{1}{2}\lim_{t\to+\infty}
\frac{\Espeq{Y_\alpha(t) Y_\gamma(t)}-\Espeq{Y_\alpha(t)}\Espeq{Y_\gamma(t)}}{t}.
\ee 
Therefore in the case where the non equilibrium stationary state  involves several independent  stationary currents, the symmetry 
$\Espeq{Y_\alpha (t)Y_\gamma(t)}=\Espeq{Y_\gamma(t) Y_\alpha(t)}$
allows to retrieve the phenomenological Onsager symmetry for the off-diagonal Onsager coefficients, namely  $L_{\alpha\gamma}=L_{\gamma\alpha}$.

Far from equilibrium generalized Einstein-Green-Kubo relations can be derived from the symmetry 
\eqref{CumRelGen} of the generating function for the infinite-time cumulants per unit time.
The generalization of the combinatorics considerations of subsection \ref{RelCumulants}
is straightforward. 
Suppose the
independent currents are indexed by $\gamma \in \Gamma$ so that instead of a
single parameter $\lambda$ one deals with a collection
$\underline{\lambda}\equiv (\lambda_{\gamma})_{\gamma \in \Gamma}$. In the same way,
one has a collection $\underline{\Fth}\equiv (\Fth_{\gamma})_{\gamma \in
 \Gamma}$. However, one considers a single function $\alpha$ and we rewrite 
 \eqref{CumRelGen} as
 \be
\label{CumRelMult}
\alpha(\underline{\lambda};\underline{\Fth})=\alpha(-\underline{\Fth}-
\underline{\lambda};\underline{\Fth}).
\ee
One could redo all the  derivations performed in subsection \ref{RelCumulants}. We content to express all odd
cumulants in terms of the even ones.
Write the expansion in powers of the $ \lambda_{\gamma}$'s in a compact way as
\be
\alpha(\underline{\lambda};\underline{\Fth})\equiv
\sum_{\underline{p}\geq
\underline{0}}\frac{1}{\underline{p}!}\underline{\lambda}^{\underline{p}}\Avkappa^{[\underline{p}]}(\underline{\Fth}),
\ee
where the summation is over $\Gamma$-tuples $\underline{p}\equiv
(p_{\gamma})_{\gamma \in \Gamma}$ of non-negative integers,
$\underline{p}! \equiv \prod_{\gamma \in \Gamma} p_{\gamma}!$ and
$\underline{\lambda}^{\underline{p}} \equiv \prod_{\gamma \in \Gamma}
\lambda_{\gamma}^{p_{\gamma}}$. Note that
\be
\Avkappa^{[\underline{p}]}(\underline{\Fth}) \equiv
\left.\frac{\partial^{|\underline{p}|}\alpha
 (\underline{\lambda};\underline{\Fth})}
{\partial\underline{\lambda}^{\underline{p}}}\right\vert
_{\underline{\lambda}=\underline{0}},
\ee
where $|\underline{p}|\equiv \sum_{\gamma \in \Gamma}p_{\gamma}$ and
$\partial\underline{\lambda}^{\underline{p}}\equiv \prod_{\gamma \in \Gamma}
\partial\lambda_{\gamma}^{p_{\gamma}}$.
We expand the symmetry relation written in the form
\be
\label{impairMult}
\alpha(\underline{\lambda};\underline{\Fth}) =
\frac{1}{1+e^{-\underline{\Fth}\frac{\partial}{\partial \underline{\lambda}}}}
\left[\alpha(\underline{\lambda};\underline{\Fth})+
 \alpha(-\underline{\lambda};\underline{\Fth})\right],
\ee
where
\be
\underline{\Fth}\frac{\partial}{\partial \underline{\lambda}}\equiv \sum_{\gamma
 \in \Gamma} \Fth_{\gamma}\frac{\partial}{\partial \lambda_{\gamma}}.
\ee
From
\be
\frac{1}{1+e^{-\underline{\Fth}\frac{\partial}{\partial
     \underline{\lambda}}}}=\frac{1}{2}\left(1+\sum_{k=0}^{+\infty} d_k
 \sum_{\underline{p}\geq \underline{0}, |\underline{p}|=2k+1}
 \frac{(2k+1)!}{\underline{p}!}\underline{\Fth}^{\underline{p}}
 \frac{\partial^{2k+1}}{\partial\underline{\lambda}^{\underline{p}}} \right)
\ee
one infers that, for $ \underline{n}$ such that $|\underline{n}|$ is odd
\be
\label{GeneralizeGBSeveralJ}
\Avkappa^{[\underline{n}]}(\underline{\Fth})=\sum_{k=0}^{+\infty} d_k
\sum_{\underline{p}\geq \underline{0}, |\underline{p}|=2k+1}
 \frac{(2k+1)!}{\underline{p}!} \underline{\Fth}^{\underline{p}}\,
 \Avkappa^{[\underline{n}+\underline{p}]} (\underline{\Fth}).
\ee

Again, these relations can be expanded in powers of $\underline{\Fth}$. The corresponding coefficients of the powers in $\Fth$ are the non-linear response coefficients. The relations between the latter non-linear response coefficients are derived by another method  in \cite{AndrieuxGaspard2007JStatMech}.
For instance relations analogous to \eqref{CumRel4} relate various non-linear response coefficients of the cumulants per unit time caused by a variation in thermodynamic forces $\Fth_\alpha$'s to some equilibrium cumulant per unit time, which is symmetric under permutations of the associated cumulative quantities. The latter symmetry is at the origin of the Onsager reciprocity relation in the case of $L_{\alpha\gamma}$ \cite{LebowitzSpohn1999}  (see definition \eqref{defLalphagen}) and of generalized symmetry relations for higher-order response coefficients of cumulants per unit time as noted by Andrieux and Gaspard \cite{AndrieuxGaspard2004,AndrieuxGaspard2007JStatMech}.

\clearpage
\appendix

\section{Property of the  coarse grained dynamics}

\label{CoarseGrainedProperty}

In the present appendix we derive the property \eqref{NumbersEqual} valid over a period of the ergodic  deterministic microscopic dynamics $\Udisc$ when  $\Udisc$ respects both the conservation of $E_\text{dec}$ and the interaction pattern specified when the conservation law \eqref{conserE} was introduced.

Though we believe that a  general development could be pursued,  we
prefer to concentrate on a specific model at this point. 
The system  is made of two large
parts and a small one, which is reduced to  two Ising spins $\sigma_1, \sigma_2=\pm
1$, each one directly in contact with one of the large parts. So a configuration
$C$ can be written as $C=(C_1,\sigma_1,\sigma_2,C_2)$. We assume that, when the small part is isolated, its energy $\En(\sigma_1,\sigma_2)$ does not describe independent spins. Moreover, the microscopic dynamics $\Udisc$ conserves the energy $E_\text{dec}
(C)$ where the interactions between parts is neglected, namely $E_\text{dec}
(C)\equiv E_1(C_1)+E_2(C_2)+\En(\sigma_1,\sigma_2)$. 
 The remnant of interactions
between the parts is embodied in the following restrictions. For any $C$, $\Udisc(C)$
is obtained from $C$ by one of the following operations : 

- (I) a flip of spin $\sigma_1$ together with a change in $C_1$ and a possible change in 
$C_2$ such that $E_1(C_1)+\En(\sigma_1,\sigma_2)$ and $E_2(C_2)$ both remain
constant (i.e. the energy needed to flip the spin $\sigma_1$ entirely comes from or goes to 
large part $1$).

- (II) a flip of spin $\sigma_2$ together with a change in $C_2$ and a possible change in
$C_1$ such that $E_1(C_1)$ and $E_2(C_2)+\En(\sigma_1,\sigma_2)$ both remain
constant (i.e. the energy needed to flip the spin $\sigma_2$ entirely comes from or goes to
large part $2$).

- (III) a change in $C_1$ and/or $C_2$ but no flip of $\sigma_1$ or $\sigma_2$,
such that $E_1(C_1)$ and $E_2(C_2)$ both remain constant, as well as $\En(\sigma_1,\sigma_2)$.

In order to build an  effective mesoscopic dynamics we just keep track of $(E_1,\sigma_1,\sigma_2,E_2)$ as a function of time.  Therefore during  the  time evolution of a given configuration $C$ of the whole system we
concentrate only on  time steps at which a change of type (I) or (II) occurs, namely
when either spin $\sigma_1$ or spin $\sigma_2$ is flipped with known corresponding variations in $E_1$ and $E_2$. We do not follow
precisely the changes (III) that modifies the configurations of  large parts without changing the energy of any part (either $E_1$, $E_2$ or $\En(\sigma_1,\sigma_2)$). 
The possible changes are
\be
\label{typeI}
(E_1, \sigma_1,\sigma_2,  E_2)\rightarrow (E'_1, -\sigma_1, \sigma_2,E_2)\quad\textrm{with}\quad E'_1=E_1+\En(\sigma_1,\sigma_2)-\En(-\sigma_1,\sigma_2)
\quad\textrm{type (I)}\quad
\ee
and 
\be
\label{typeII}
(E_1, \sigma_1,\sigma_2,  E_2)\rightarrow (E_1, \sigma_1,-\sigma_2,E'_2)\quad\textrm{with}\quad E'_2=E_2+\En(\sigma_1,\sigma_2)-\En(\sigma_1,-\sigma_2)
\quad\textrm{type (II)}\quad
\ee

Starting from some initial configuration $(E_1^0,\sigma_1^0,\sigma_2^0,E_2^0)$,
some set of possible $(E_1,\sigma_1,\sigma_2,E_2)$ will be visited during
the time evolution over a period of $\Udisc$, and it is useful to view this set as the vertices of a
graph, whose edges connect two vertices if the system can jump from one to the other by
a single change of type (I) or (II). This graph can be chosen to be unoriented
because a transformation of type (I) or of type (II) is its own inverse.  Then
a trajectory over a period of the underlying
deterministic dynamics corresponds to a closed walk on this graph, which
summarizes the coarse graining due to the macroscopic description of the large
parts.  During a period of the microscopic dynamics the closed walk on  the
graph goes  through each edge a number of times.

The main observation is that, since energy $\En(\sigma_1,\sigma_2)$ does not
describe independent spins, the graph has the topology of a segment. Indeed, the
graph is connected by construction, but each vertex has at most two neighbors.
So the graph is either a segment or a circle. Let us suppose that it is a
circle. Then one can go from $(E_1^0,\sigma_1^0,\sigma_2^0,E_2^0)$ to itself by
visiting each edge exactly once, i.e. by an alternation of moves of type (I) and
(II).  After two steps both spins in the small part are flipped, so the total
length of the circle is a multiple of $4$.  But after $4$ steps a definite
amount of energy has been transferred between the two large parts, namely the
energy in the first large part has changed by
$\En(\sigma_1,\sigma_2)-\En(-\sigma_1,\sigma_2)+\En(-\sigma_1,-\sigma_2)-\En(\sigma_1,-\sigma_2)$
and a trivial computation shows that this cannot vanish unless
$\En(\sigma_1,\sigma_2)$ describes two independent spins, a possibility which
has been discarded. This excludes the circle topology.

Since the graph is a segment, any closed walk on the graph traverses a given
edge the same number of times in one direction and in the other one.  As a
consequence, during a period of the microscopic dynamics $\Udisc$, the motion
induced by $\Udisc$ on the graph is such that the number of transitions of type
(I) $(E_1,\sigma_1,\sigma_2,E_2)\rightarrow (E'_1,-\sigma_1,\sigma_2,E_2)$ is
equal to the number of their inverse transitions
$(E'_1,-\sigma_1,\sigma_2,E_2)\rightarrow(E_1,\sigma_1,\sigma_2,E_2)$, where the
relation between $E'_1$ and $E_1$ is given in \eqref{typeI}.  The same
considerations apply for transition rates associated with the flipping of
$\sigma_2$. The result is summarized in \eqref{NumbersEqual}.

For a more general discrete system, the mesoscopic time evolution can again be characterized by the time steps where the small system variables are flipped while the changes in the large parts are only macroscopically described by their net energies. An analogous graph can be constructed but it
can be much harder to analyze its topology, which is the crucial knowledge needed to
exploit the consequences of ergodicity. These consequences are most stringent
for a tree. There is no reason a priori why the graph should be a tree, but what
this implies, namely that ratios for the transition rates involving the small part and a
large part are given by ratios of energy level degeneracies in the large part, is
physically quite appealing.

\section{The Markovian approximation}
\label{Markovapprox}

This appendix is a short digression on mathematics. The
aim is to briefly recall a trick allowing to replace a fixed sequence of symbols
by a random one with ``analogous'' statistical properties, and then to combine this
trick with a coarse graining procedure.

\subsection{Discrete time stochastic approximation}
\label{DiscreteTMarkov}

Suppose $w \equiv x_1x_2\cdots x_N$ is a given finite sequence
of elements of a finite set $S$. It is convenient to assume that this sequence
is periodic, i.e. that $x_{N+1}\equiv x_1$. For $x\in S$, set $N_x\equiv \#\{i\in
[1,N], x_i=x\}$, i.e. $N_x$ is the number of occurrences of $x$ in the sequence
$w$. There is no loss of generality in assuming that $N_{x}\neq 0$ for
every $x\in S$, should it lead to consider a smaller $S$. For $x,x' \in S$ set
$N_{xx'}\equiv \#\{i\in [1,N], x_i=x,x_{i+1}=x'\}$, i.e. $N_{xx'}$ is the number of
occurrences of the pattern $xx'$ in the sequence $w$. Notice that we accept
that $x=x'$ in this definition. Of course, we could look at the occurrence of more
general patterns. By definition, we have $\sum_{x'\in S} N_{xx'}=\sum_{x'\in
  S}N_{x'x} =N_x$. The result we want to recall is the following.

There is a single time-homogeneous irreducible Markov matrix on $S$ that fulfills the following two requirements. First, 
in the stationary state of the corresponding discrete-time stochastic evolution,
the probability $\Probst(x',i+1;x,i)$ that a sample $\widehat{x}$ takes the
value $x$ at time $i$ and the value $x'$ at time $i+1$\footnote{In the whole paper we use the convention that the
  evolution (here given by the joint probability $\Probst(x',i+1;x,i)$ or the transition matrix $T$) is written from the right
  to the left as in quantum mechanics.} is equal to the
frequency of the pattern $xx'$ in the sequence $w$, namely 
\be
\label{propequiv2}
\Probst(x',i+1;x,i)=\frac{N_{xx'}}{N}.
\ee
Second, the stationary probability that the random variable $\widehat{x}$
takes the value $x$ at any time $i$ is equal to the frequency of $x$ in the
sequence $w$, namely
\be
\label{propequiv1}
\Probst(x)=\frac{N_x}{N}.
\ee
In fact \eqref{propequiv1} is  a consequence of \eqref{propequiv2}, because of
the relations $\Probst(x)=\sum_{x'\in S}\Probst(x',i+1;x,i)$ and
$\sum_{x'\in S} N_{xx'}=N_x$.  To say it in words, there is a unique
time-homogeneous irreducible Markov chain whose stationary statistics for
patterns of length $1$ or $2$ is the same as the corresponding statistics in
$w$.

The proof is elementary. By the Markov property, the Markov matrix element from
$x$ to $x'$, denoted by $T(x' \leftarrow x)$, must satisfy
the relation $\Probst(x',i+1;x,i)= T(x' \leftarrow x)\Probst(x)$, so the only
candidate is 
\be
\label{propequiv3}
T(x' \leftarrow x)=\frac{N_{xx'}}{N_x}.
\ee
Conversely, the corresponding
matrix is obviously a Markov matrix ($\sum_{x'\in S}T(x' \leftarrow x)=1$ since
$\sum_{x'\in S}N_{xx'}=N_x$), and it is irreducible, because, as $w$ contains all
elements of $S$, transitions within $w$ allow to go from every element of $S$ to
every other one. Checking that for this Markov matrix the stationary measure, namely
the solution of $\sum_{x\in S}T(x' \leftarrow x) \Probst(x) =\Probst(x')$, is
given by \eqref{propequiv1} boils down to the identity $\sum_{x\in S}
N_{xx'}=N_{x'}$ recalled above, and then the two-point property \eqref{propequiv2}
follows. This finishes the proof.

In the very specific case where $x_1,x_2,\cdots,x_N$ are all distinct, i.e. $|S|$, the cardinal of $S$, is equal to $N$, $N_x=1$
for all $x\in S$ and $N_{xx'}=\delta_{x',x_{i+1}}$ where $i$ is such that $x_i=x$.
Then the only randomness lies in the choice of the initial distribution, and
each trajectory of the Markov chain reproduces $w$ up to a  translation of  all indices. If $N$ is
large and $|S|$ is $\sim N$, slightly weaker but analogous conclusions survive.
A more interesting case is when $|S| \ll N$ by many orders of magnitude as discussed in next subsection.

This trick has been used for instance to write a random text ``the Shakespeare
way'' by computing the statistics of sequences of two words in one of his
books.

By definition, the Markovian approximation reproduces the statistics of the
original sequence only for length $1$ and length $2$ patterns. It is a delicate
issue to decide whether or not it also does a reasonnable job for other
patterns. For instance, the random Shakespeare book certainly looks
queer. Various physical but heuristic arguments suggest that, for the kind of
sequences $w$ relevant for this work, the Markov approximation is quite
good, but we shall not embark on that.

\subsection{Continuous time approximation}
\label{ContinousTimeApproximation}

By the very same argument, there is a  time-homogeneous irreducible Markov
transition matrix on $S$ (unique up to  a time scale $\tau$)  such that, in the
stationary state of the corresponding continuous-time stochastic evolution, the
expected number of transitions from $x$ to $x'\neq x$ per unit time is
$N_{xx'}/\tau$. The formula one finds for the transition rate is 
\be
\label{Wxxprim} 
W(x'\leftarrow x)= \frac{N_{xx'}}{\tau N_x} \text{ for }x \neq x'.
\ee

The continuous-time approximation becomes more natural in the case when $|S|$,
the cardinal of $S$, is such that $|S|\ll N$, while for all $x$ $N_{xx}\sim N_x$ and for all $x \neq x'$ $N_{xx'} \ll N_x$, which means that transitions are rare and most of the time
$x$ follows $x$ in the sequence $w$.  Again, one can argue that for the kind of
sequences $w$ relevant for this work, this is guaranteed by physics. Then taking
$\tau$ (in some macroscopic time unit) of the order of the largest value of the
$N_{xx'}/N_x$, $x \neq x'$, one ends with a Markov transition matrix with
elements of order unity (in some macroscopic inverse time unit), and $t_i=\tau
i$ can be taken as the physical macroscopic time.

To summarize, in this work, we shall systematically associate to certain
sequences $w$ a continuous-time Markov process and exploit properties of $w$ to
constraint the structure of $W(x' \leftarrow x)$.

\vspace{.3cm}

A natural application of the above ideas is to the case where the sequence $w$ arises from some coarse graining procedure. One starts from a
sequence $\omega=\xi_1\xi_2\cdots \xi_N$ where $\xi_1,\xi_2,\cdots ,\xi_N$
belong to a set $\Sigma$ that is so large that $\omega$ cannot be stored, and
that only some of its features can be kept. Say we partition $\Sigma\equiv \cup_{x
  \in S} \Sigma_x$ where $S$ is of reasonable size.  Then all we keep of
$\omega$ is $w=x_1x_2\cdots x_N$ where, for $i\in [1,N]$, $x_i$ is substituted
for $\xi_i$ when $\xi_i\in \Sigma_{x_i}$. In the applications we
have in mind, $\Sigma$ is an $N$-element set, i.e. all terms in the sequence
$\omega$ are distinct. In that case, even if $\omega$ is constructed in a
perfectly deterministic way, by saying who follows who in the sequence, such a
description is unavailable on $w$, and $w$ may well look quite random, so the
Markov chain approximation is worth a try. In fact, $|S| \ll N$ and we shall
take as a physical input that transitions are rare, so that the (continuous-time)
Markov process is an excellent approximation to the (discrete time) Markov chain.

\section{Microcanonical detailed balance and time reversal invariance in Hamiltonian dynamics}

\label{TimeReversal}

In  view of comparison with the approach developed for discrete variables,   we rederive the microcanonical detailed balance \eqref{ratioWmc} for the probability distribution of some mesoscopic variables, within the framework of statistical ensemble theory, in the case where a microscopic configuration of the full system is described by  continuous variables  and the system has a microscopic deterministic  dynamics whose Hamiltonian $H$, independent of time, is an even function of  momenta. 
Moreover the coordinate of the system in phase space is denoted by $\xi$ (by analogy with the notation in subsubsection \ref{DefinitionMarkovApprox}), and ${x}[\xi]$ is a set of mesoscopic variables defined from the  microscopic coordinate $\xi$  and which are  even functions of  momenta.   
(Similar arguments can be found in derivations which rely on different assumptions in Refs.\cite{Wigner1954,vanKampen1992}.)

From the viewpoint of the statistical ensemble theory, the initial position  of the system in phase space is not known. It is assumed to be  uniformly randomly distributed in the energy shell $E=H[\xi]$, where $E$ is the value of the energy of the full system at some initial time $t_0$. In other words, the initial probability distribution of $\xi$, $\Prob_0(\xi)$, is such that
\be
\label{Prob0def}
d\xi\,\Prob_0(\xi)\equiv \frac{d\xi \,\delta\left(H[\xi]-E\right)}{\int d\xi'\, \delta\left(H[\xi']-E\right)},
\ee
where $d\xi$ is the Liouville measure in phase space. The choice of the uniform probability distribution in the energy shell for the initial position of the system in phase space is motivated by the fact that, at the microscopic level,  it is the measure that is invariant under Hamiltonian dynamics. We recall the derivation of the latte property for the sake of completeness and in  order to introduce notations used below.

The dynamics is invariant by time translation and we simply denote by $f_t(\xi)$ the position of the system in phase space at time $t_0+t$ knowing that it is at position $\xi$ at  time $t_0$.  
(For the same reason, in the following the initial time $t_0$ is set equal to $0$). Then the invariance of the Hamiltonian under the operation $R$ that changes every momentum into its opposite implies that 
\be
\label{fTimeInvariance}
[Rf_tRf_t](\xi)=\xi,
\ee
 namely the microscopic dynamics is invariant  under time reversal. Since the Hamiltonian is a constant of motion  and the  equations of motion conserve the infinitesimal volume in phase space, the probability $d\xi \Prob_0(\xi)$ defined in \eqref{Prob0def} is also conserved under the microscopic evolution, 
\be
\label{Prob0TimeInvariance}
d(f_t[\xi]) \,\Prob_0(f_t[\xi])=d\xi\,\Prob_0(\xi).
\ee

When the initial position of the system in phase space is distributed according to $\Prob_0$, the  probability that at time $t$ the set of mesoscopic variable ${x}$ takes the value ${x}_1$, which is defined as
\be
\label{Probxdef}
\Prob_{\Prob_0}({x}_1, t)\equiv\int d\xi \,\Prob_0(\xi)\,
\delta\left({x}[f_t(\xi)]-{x}_1\right),
\ee
is in fact independent of time (i.e. conserved by the microscopic dynamics). Indeed, let us consider the  change of variable $\xi'=f_t(\xi)$. The conservation of 
$\Prob_0(\xi)d\xi$ by the dynamics \eqref{Prob0TimeInvariance} implies that
$d\xi\,\Prob_0(\xi)=d\xi'\,\Prob_0(\xi')$. Then  the integral in \eqref{Probxdef} reads
$\int d\xi' \,\Prob_0(\xi')\,\delta\left({x}[\xi']-{x}_1\right)$. It is independent of time and, by virtue of the definition \eqref{Prob0def} of the distribution $\Prob_0(\xi)$ in  phase space, it  is in fact equal to  the microcanonical  distribution $\Probmc({x}_1)$,
\be
\label{ProbH}
\Prob_{\Prob_0}({x}_1,t)=\Probmc({x}_1)
\equiv  \frac{\int d\xi \,\delta\left(H[\xi]-E\right) \delta\left({x}[\xi]-{x}_1\right)}{\int d\xi'\, \delta\left(H[\xi']-E\right)}.
\ee

On the other hand, the  probability that the set of mesoscopic variables ${x}$ takes the value  ${x}_1$ at time $t=0$ and the value ${x}_2$ at time $t$, when  the initial position of the system in phase space is distributed according to $\Prob_0(\xi)$, 
 reads
\be
\label{ProbJointH}
\Prob_{\Prob_0}({x}_2,t;{x}_1,0)=\int d\xi \,\Prob_0(\xi)\,
\delta\left({x}[f_t(\xi)]-{x}_2\right) 
\delta\left({x}[\xi]-{x}_1\right).
\ee
Since the  variables denoted by ${x}(\xi)$ are even functions of momenta, we can write  ${x}[f_t(\xi)]={x}[[Rf_t](\xi)]$. 
Let us consider the  change of variable $\xi'=[Rf_t](\xi)$. 
The invariance under time reversal of the microscopic dynamics \eqref{fTimeInvariance} entails that $\xi=[Rf_t](\xi')$, so that 
${x}[\xi]={x}[[Rf_t](\xi')]={x}[f_t(\xi')]$. By virtue of its definition \eqref{Prob0def} $d\xi\,\Prob_0[\xi]$ is invariant under the operation $R$, so
$d\xi'\,\Prob_0(\xi')=d(f_t[\xi]) \,\Prob_0(f_t[\xi])$ and the conservation of $d\xi\,\Prob_0(\xi)$ by the dynamics \eqref{Prob0TimeInvariance} implies that
$d\xi'\,\Prob_0(\xi')=d\xi\,\Prob_0(\xi)$.  Eventually the integral in \eqref{ProbJointH} reads
{$
\int d\xi' \, \Prob_0(\xi')
\delta\left({x}[\xi']-{x}_2\right) 
\delta\left({x}[f_t(\xi')]-{x}_1\right)
$}
and we get 
\be
\label{ProbJointHBis}
 \Prob_{\Prob_0}({x}_2,t;{x}_1,0)=\Prob_{\Prob_0}({x}_1,t;{x}_2,0).
\ee
The joint probability is invariant by exchanging the times at which ${x}_1$ and ${x}_2$ occur.

We now assume that the evolution of  the mesoscopic variable 
${x}$, which is invariant by time translation, can be described by an homogeneous  Markovian stochastic process whose stationary distribution $\Probst({x})$ is the time-independent probability $\Prob_{\Prob_0}({x})$ defined in \eqref{Probxdef} and whose  transition rates, denoted by $W({x}'\leftarrow {x})$, are determined by  the identification 
\be
 \Prob_{\Prob_0}(x', dt;{x},0)\equiv W({x}'\leftarrow {x}) \Prob_{\Prob_0}({x})\times dt,
\ee
where  the time-displaced joint probability for $x'$ and ${x}$ and the probability of the single variable ${x}$ are calculated with the same initial distribution $\Prob_0$ for the microscopic variables $\xi$'s. 
Then,  the time reversal symmetry property \eqref{ProbJointHBis} for the  joint probability and the fact that $ \Prob_{\Prob_0}({x})$ coincides with the microcanonical distribution 
$ \Probmc({x})$ by virtue of \eqref{ProbH}  lead  to the microcanonical detailed balance
\be
 W({x}'\leftarrow {x}) \Probmc({x})= W({x}\leftarrow {x}') \Probmc( {x}').
\ee

\section{Definitions for statistics over histories}

\label{DefHistories}

Consider a history  where the system starts in configuration $\C_0$ at time $t_0=0$ and  ends in configuration $\C_f$ at time $t$ after going through successive configurations $\C_0$, $\C_1$,\ldots, $\C_N=\C_f$. The $N$ instantaneous jumps from one configuration to another occur at $N$ successive intermediate times $T_i$ which are continuous stochastic variables : the system jumps from $\C_{i-1}$ to $\C_i$ at time $T_i$ ($i=1,\ldots,N$) in the time interval $[t_i, t_i+dt_i[$, with $t_0<t_1<t_2<\cdots<t_N<t$. The probability measure for such  a history   is related to the probability density $\Probdist_{\C_f,\C_0}\left[\Hist \right]$ by
\be
\label{defProbdisHist}
d\Prob_{\C_f, \C_0}\left[\Hist \right]
\equiv dt_1\ldots dt_N \Probdist_{\C_f,\C_0}\left[\Hist \right]
\ee
where, for a time-translational invariant process,  
\bea
\Probdist_{\C_f,\C_0}\left[\Hist \right]&=&e^{-(t-t_N)  \Lambda(\C_N)}(\C_N\vert \Trans \vert \C_{N-1}) e^{-(t_N-t_{N-1})\Lambda(\C_{N-1})}
\\
\nonumber
&&\qquad\qquad\qquad\qquad \times
  \cdots
 e^{-(t_2-t_1)\Lambda(\C_1)}(\C_1\vert \Trans \vert \C_0) e^{-(t_1-t_0) \Lambda(\C_0)}
\eea
and $\Lambda(\C)$ is  the total exit rate (also called escape rate) from configuration $\C$,
\be
\Lambda(\C)\equiv\sum_{\C'\not=\C} (\C'\vert \Trans \vert \C).
\ee

\medskip
The average of a functional $F[\Hist]$ over the histories that start in configuration $\C_0$ and end in configuration $\C_f$    is computed as
\be
\label{defavhist}
\Esp{F}_{\C_f,\C_0}=\int d\Prob_{\C_f, \C_0}\left[\Hist \right] F[\Hist]
\ee
with
\be
\int d\Prob_{\C_f, \C_0}\left[\Hist \right] =\sum_{N=0}^{+\infty}\sum_{\C_1}\ldots \sum_{\C_{N-1}} 
\int_{t_0<t_1<\ldots< t_N}
d\Prob_{\C_f, \C_0}\left[\Hist \right].
\ee
Then the average of a functional when the initial distribution of configurations is $\Prob_0$ reads
\be
\label{defEspHist}
\Esp{F}_{\Prob_0}=\sum_{\C_f} \sum_{\C_0}\Prob_0(\C_0)\int d\Prob_{\C_f, \C_0}\left[\Hist \right]  F[\Hist].
\ee

\clearpage

\section{Some remarks on large deviations}

\label{LDRemarks}

Suppose that $X_t$ is some time-dependent random quantity. Typically, what we have in mind is a variation of some physical quantity (energy, matter) exchanged between a reservoir and a system during the interval $[0,t]$. So $X_t$ is expected to scale like $t$ at large times, with an average $\Esp{X_t}/t$ (or even $X_t/t$ almost surely) going to a constant $\Jcumav$ as $t$ goes to $+\infty$.

There are a number of definitions to quantify the probability that $X_t/t$
differs significantly from $\Jcumav$ at some large time. It often happens that
this probability is exponentially small, and a general mathematical theory,
large deviation theory, has emerged to describe this situation.

\subsection{The mathematical definition of large deviations}
\label{MathDefSec}

 The general definition is a bit abstract, dealing with general probability measures depending on $t$. We restrict here to the situation when the probability measure is the distribution of a real random variable $X_t$. If $j_-<j_+$ are two real numbers, observe that 
$\Prob\left(\frac{X_t}{t}\in ]j_-,j_+[\right)\leq \Prob\left(\frac{X_t}{t}\in [j_-,j_+]\right)$ so that
\be
-\frac{1}{t}\ln \Prob\left(\frac{X_t}{t}\in [j_-,j_+]\right)\leq 
-\frac{1}{t}\ln \Prob\left(\frac{X_t}{t}\in ]j_-,j_+[\right).
\ee
For large $t$, the two members may not converge but at least
\be
\liminf_{t\to+\infty}-\frac{1}{t}\ln\Prob\left(\frac{X_t}{t}\in [j_-,j_+]\right)
\leq \limsup_{t\to+\infty}-\frac{1}{t}\ln\Prob\left(\frac{X_t}{t}\in ]j_-,j_+[\right),
\ee
where both sides of the inequality belong to $[0,+\infty]$  with the definitions
$\liminf_{t\to+\infty} g_t\equiv\lim_{t\to+\infty}\inf_{s\geq t}g_s$ and 
$\limsup_{t\to+\infty} g_t=\lim_{t\to+\infty}\sup_{s\geq t}g_s$. A nontrivial lower bound for the left-hand side or upper bound for the right-hand side gives information on the exponential rate of decrease of $\Prob\left(\frac{X_t}{t}\in [j_-,j_+]\right)$ and
$\Prob\left(\frac{X_t}{t}\in ]j_-,j_+[\right)$, the best situation being when
the two exist and coincide. 

One says that the random variable $X_t$ obeys a large deviation principle (LDP)
if there is a lower semi-continuous function $R_X$, 
called the  rate function, such that
\be
\label{defRXa}
\inf_{j\in [j_-,j_+]} R_X(j)
\leq 
\liminf_{t\to+\infty}-\frac{1}{t}\ln \Prob\left(\frac{X_t}{t}\in [j_-,j_+]\right)
\ee
and
\be
\label{defRXb}
\limsup_{t\to+\infty}-\frac{1}{t}\ln \Prob\left(\frac{X_t}{t}\in ]j_-,j_+[\right)
\leq \inf_{j\in ]j_-,j_+[} R_X(j).
\ee
The property that $R_X$ is \textit{lower semicontinuous} means that 
\be
\label{defRXc}
\lim_{\varepsilon \searrow 0} \inf_{j'\in [j-\varepsilon,j+\varepsilon]} R_X(j')= R_X(j)\qquad\textrm{for every $j$}.
\ee
More generally a lower semicontinuous function achieves its minimum on any closed interval.
Notice that using an open interval $]j-\varepsilon,j+\varepsilon[$ instead of
$[j-\varepsilon,j+\varepsilon]$ in  definition \eqref{defRXc} leads to the same notion.
The  definitions \eqref{defRXa} and \eqref{defRXb} involve $\limsup$ and $\liminf$ and make a difference between $]j_-,j_+[$ and 
$ [j_-,j_+]$ to take care of slightly pathological situations. According to the Portmanteau theorem (see Sec. D.2. of 
Ref.\cite{DemboZeitouni1998}) the latter definitions are equivalent to the weak convergence (or  convergence in law) of the measure $-\frac{1}{t}\ln \Prob\left(\frac{X_t}{t}\in I\right)$ to the measure $\inf_{j\in I}R_X(j)$.

By taking $j_+=j+\varepsilon$, $j_-=j-\varepsilon$ and letting $\varepsilon \searrow 0$, we see that if $X_t$ obeys a LDP then the rate function is given by 
\bea
\label{CharacterizationR}
R_X(j)&=&\lim_{\varepsilon \searrow 0} \liminf_{t\to+\infty}-\frac{1}{t}\ln \Prob\left(\frac{X_t}{t}
\in [j-\varepsilon,j+\varepsilon]\right)\\
\nonumber
&=&\lim_{\varepsilon \searrow 0} \limsup_{t\to+\infty}-\frac{1}{t}\ln \Prob\left(\frac{X_t}{t}\in 
]j-\varepsilon,j+\varepsilon[\right).
\eea
Note that we do not claim that the existence of the above limits, equal to some lower semi-continuous function $R_X(j)$, guarantees that $X_t$ satisfies a LDP.

However, if $X_t$ satisfies a LDP with a \textit{continuous} rate function, then $ \inf_{j\in ]j_-,j_+[} R_X(j)=\inf_{j\in [j_-,j_+]} R_X(j)$,
 the $\limsup$ and $\liminf$ become standard limits and the inequalities become equalities :
\be
\label{CharacterizationRContinuous}
\lim_{t\to+\infty}-\frac{1}{t}\ln \Prob\left(\frac{X_t}{t}\in ]j_-,j_+[\right) =
\lim_{t\to+\infty}-\frac{1}{t}\ln \Prob\left(\frac{X_t}{t}\in [j_-,j_+]\right)
=\inf_{j\in [j_-,j_+]} R_X(j).
\ee 
The latter property can be summarized as
\be
\label{CharacterizationRContinuousBis}
\lim_{t\to+\infty} -\frac{1}{t}\ln \Prob\left(\frac{X_t}{t}\in I\right)=\inf_{j \in I} R_X(j)
\ee
where $I$ is some interval. 

Before we turn to other possible characterizations of large deviations, let us explain briefly and heuristically the presence of the $\inf R_X(j)$ in these formulae. If the interval $I$ is split into a finite number of intervals, say $I=\cup_{k=1}^K I_k$, we have
\be
\max_k\Prob\left(\frac{X_t}{t}\in I_k\right)\leq \Prob\left(\frac{X_t}{t}\in I\right)\leq
\sum_k\Prob\left(\frac{X_t}{t}\in I_k\right)\leq K \max_k\Prob\left(\frac{X_t}{t}\in I_k\right).
\ee
We infer that 
\be
-\frac{ \ln K}{t}+\min_k -\frac{1}{t}\ln\Prob\left(\frac{X_t}{t}\in I_k\right) \leq
-\frac{1}{t}\ln\Prob\left(\frac{X_t}{t}\in I\right)\leq
\min_k  -\frac{1}{t}\ln\Prob\left(\frac{X_t}{t}\in I_k\right)
\ee
so that, for $s\geq t$,
\be
-\frac{\ln K}{t}+\min_k -\frac{1}{s}\ln\Prob\left(\frac{X_s}{s}\in I_k\right) \leq
-\frac{1}{s}\ln\Prob\left(\frac{X_s}{s}\in I\right)\leq
\min_k  -\frac{1}{s}\ln\Prob\left(\frac{X_s}{s}\in I_k\right).
\ee
First we take the $\inf$ over $s\geq t$ and interchange freely the $\inf$ and the $\min$ to get 
\be
-\frac{\ln K}{t}+\min_k \inf_{s\geq t} -\frac{1}{s}\ln\Prob\left(\frac{X_s}{s}\in I_k\right) \leq
\inf_{s\geq t} -\frac{1}{s}\ln\Prob\left(\frac{X_s}{s}\in I\right)\leq
\min_k \inf_{s\geq t} -\frac{1}{s}\ln\Prob\left(\frac{X_s}{s}\in I_k\right).
\ee
Letting $t\to +\infty$ yields
\be
\liminf_{t\to+\infty} -\frac{1}{t}\ln\Prob\left(\frac{X_t}{t}\in I\right)=\min_k
\liminf_{t\to+\infty} -\frac{1}{t}\ln\Prob\left(\frac{X_t}{t}\in I_k\right).
\ee
Second we take the $\sup$ over $s\geq t$. This time only one of the inequalities survives the interchange of $\sup$ and $\min$
\be
\sup_{s\geq t} -\frac{1}{s}\ln\Prob\left(\frac{X_s}{s}\in I\right)\leq
\sup_{s\geq t}\min_k -\frac{1}{s}\ln\Prob\left(\frac{X_s}{s}\in I_k\right)
\leq \min_k\sup_{s\geq t} -\frac{1}{s}\ln\Prob\left(\frac{X_s}{s}\in I_k\right)
\ee
and letting $t\to+\infty$ yields
\be
\limsup_{t\to+\infty} -\frac{1}{t}\ln\Prob\left(\frac{X_t}{t}\in I\right)
\leq\min_k
\limsup_{t\to+\infty} -\frac{1}{t}\ln\Prob\left(\frac{X_t}{t}\in I_k\right).
\ee
To summarize 
\bea
\label{subinterval}
\min_k\liminf_{t\to+\infty} -\frac{1}{t}\ln\Prob\left(\frac{X_t}{t}\in I_k\right)
&=&\liminf_{t\to+\infty} -\frac{1}{t}\ln\Prob\left(\frac{X_t}{t}\in I\right)
\\
\nonumber
\limsup_{t\to+\infty} -\frac{1}{t}\ln\Prob\left(\frac{X_t}{t}\in I\right)
&\leq&
\min_k\limsup_{t\to+\infty} -\frac{1}{t}\ln\Prob\left(\frac{X_t}{t}\in I_k\right).
\eea
Informally, this implies indeed that the large deviation estimates for the interval $I$  come solely from some arbitrary small subinterval of $I$ characterized by a minimizing property.

As a conclusion we list some informal rewritings often used in the literature of physicists community. There \eqref{CharacterizationRContinuousBis}  can be found to be rewritten as
\be
\Prob\left(\frac{X_t}{t}\in I\right)\underset{t \to +\infty}{\sim} e^{-t \inf_{j\in I}R_X(j)},
\ee
a notation which neglects the presence of a possibly unbounded $t$-dependent prefactor. The same notation convention is used when writing the consequence \eqref{ExpfXcontinuous} of \eqref{CharacterizationRContinuousBis} for a continuous rate function $R_X(j)$ as 
\be
\label{Rewrite2}
 \Probdist_t(j)\underset{t \to +\infty}{\sim}e^{-t R_X(j)}.
\ee
where $\Probdist_t(j)$ is the  probability density  defined (if it exists) as 
$\Prob\left(\frac{X_t}{t}\in [j,j+dj]\right)= \Probdist_t(j) \, dj$. Sometimes $\Probdist_t(j)$ in \eqref{Rewrite2} is  replaced by the notation 
$\Prob\left(\frac{X_t}{t}=j\right)$, which is quite abusive, since $\Prob\left(\frac{X_t}{t}=j\right)=0$ whenever $\Probdist_t(j)$ exists as a function (in other words $\Prob\left(\frac{X_t}{t}=j\right)=0$ unless $X_t/t$ has a distribution made of $\delta$ peaks). An even  more informal, and possibly misleading, is
\be
\label{Rewrite3}
\Prob\left(\frac{X_t}{t}\in [j,j+dj]\right)\underset{t \to +\infty}{\sim} e^{-tR_X(j)} 
\ee
which holds only under the convention that $\lim_{t\to+\infty} \frac{1}{t}\ln dj=0$ for an infinitesimal $dj$.

\subsection{Alternative definitions}
\label{MathDefAltSec}

There are a number of situations of physical interest when $X_t$ takes values in a (time independent) discrete set $\mathbb{X}$. Typically $\mathbb{X}$ is $\mathbb{Z}$, the set of integers, or $a+\mathbb{Z}$, the set of integers shifted  by $a$, or $a\mathbb{Z}$, the set of integers dilated by $a$.

Though this discreteness by no means prevents from using the general large deviation theory, other reasonable but ad hoc definitions come to mind.

For instance, one may ask whether there is a function $g_X(j)$ from $\mathbb{R}$ to $[0,+\infty]$ such that, for any map $t\to \xi_t$ from $[0,+\infty[$ to $\mathbb{X}$ such that 
$\lim_{t\to+\infty} \frac{\xi_t}{t}=j$, 
\be
\label{defgX}
\lim_{t\to+\infty}-\frac{1}{t}\ln \Prob(X_t=\xi_t)\equiv g_X(j).
\ee

A weaker condition comes from asking whether there is a function $h_X(j)$ 
 from $\mathbb{R}$ to $[0,+\infty]$ such that 
 \be
 \lim_{\substack{t\to +\infty \\tj\in \mathbb{X}}}-\frac{1}{t}\ln \Prob(X_t=tj)\equiv h_X(j).
 \ee

 Of course same care is needed when choosing $\mathbb{X}$, which would better be in some way minimal. For instance, if one takes $\mathbb{X}=\mathbb{Z}$ while $X_t$ takes only even values, $h_X$ or $g_X$ have no chance to exist. However, when $\mathbb{X}$ is chosen appropriately and $\Prob(X_t=x)$ behaves nicely for large $t$ and $x$, it can be expected on physical grounds that all three functions $R_X$, $g_X$ and $h_X$ exist and are the same. Alas, we are not aware of practical and general enough mathematical criteria that guarantee this fact. Of the three, $h_X$ is probably the easiest to tackle with bare hands. However, to deal with $R_X$ one can often rely on powerful theorems of the general theory of large deviations. But whether they exist (and coincide) or not, the three functions are meaningful characteristics of the large deviations of $X_t$. With some imagination, we could  probably invent others.

\clearpage 
\subsection{Inequalities at finite time and fluctuation relation between large deviation functions}
\label{IneqLargeDeviation}

One important feature of large deviation functions in out-of-equilibrium
statistical mechanics is that they satisfy, under appropriate circumstances,
symmetry relations. Usually, these relations are consequences of a, possibly
generalized, time reversal symmetry. Typically, for some $0<m<M<+\infty$ and for
some real $\gamma$ one has 
\be m \Prob(X_t=-x) e^{\gamma x} \leq \Prob(X_t=x)
\leq M\Prob(X_t=-x) e^{\gamma x} 
\ee
 for every $x$. To be fair, this way of
writing things applies, strictly speaking, only when $X_t$ is atomic, i.e. takes
values in a discrete set. A more correct statement is that the laws of $X_t$ and
$-X_t$ are absolutely continuous and the Radon-Nikodym derivative\footnote{One
  says that the Radon-Nikodym derivative of a probability measure $\mu$ with
  respect to a probability measure $\nu$ exists if there is an $\nu$-integrable
  function $f$ such that for every Borel set $B$ one has $\int_B d\mu =\int_B f
  d\nu$. Then $f$ is called the Radon-Nikodym derivative of $\mu$ with respect
  to $\nu$ and one writes $f=d\mu/d\nu$.}  
  \be
\label{inequalityRN}
m e^{\gamma x} \leq \frac{d\Prob_{X_t} (x)}{d\Prob_{-X_t} (x)}\leq M e^{\gamma x},
\ee
of which the previous equation is a special case.

The point we want to make here is that, with any of the three definitions of
large deviation functions given above (when they do exist), the above equation implies a symmetry relation. For $g_X$ or $h_X$ we can use the simplified equation evaluated when $x\in \mathbb{X}$. It implies
\be
-\frac{1}{t}\ln M -\gamma \frac{x}{t}-\frac{1}{t} \ln \Prob(X_t=-x)
\leq-\frac{1}{t} \ln \Prob(X_t=x)\leq
-\frac{1}{t}\ln m -\gamma \frac{x}{t}-\frac{1}{t} \ln \Prob(X_t=-x).
\ee
Assuming the existence of $h_X$ one can take the limit 
$\lim_{\substack{t\to +\infty \\tj\in \mathbb{X}}}$ to get
$h_X(-j)-\gamma j \leq h_X(j) \leq h_X(-j)-\gamma j$, i.e.
\be
 h_X(j)=h_X(-j)-\gamma j,
 \ee
 the announced symmetry relation. If $g_X$ exists, then so does $h_X$ and they are equal, so the symmetry property for $g_X$ (when it exists) is a consequence of the above bounds.

 In the case when $R_X$ exists, the proof of its symmetry based on the inequalities for the 
 Radon-Nikodym derivative is only slightly more elaborate : if $B$ is a Borel subset of $\mathbb{R}$ with $tj_-\leq \inf B$ and $\sup B \leq t j_+$ then, assuming $\gamma\geq 0$ for definiteness, 
\be
\label{bornesProb}
\Prob(-X_t\in B) e^{\gamma tj_-} \leq \int_B d\Prob_{-X_t} (x) e^{\gamma x}
 \leq \Prob(-X_t\in B) e^{\gamma tj_+}
\ee
so, as a consequence of \eqref{inequalityRN},
\be
-\frac{1}{t}\ln M -\gamma j_+-\frac{1}{t} \ln \Prob(X_t\in -B)
\leq-\frac{1}{t} \ln \Prob(X_t\in B)\leq
-\frac{1}{t}\ln m -\gamma j_- -\frac{1}{t} \ln \Prob(X_t\in -B),
\ee 
and taking $B=]t j_-, tj_+[$ and the $\limsup_{t\to+\infty} $  in the second inequality we get
\be
\limsup_{t\to+\infty} -\frac{1}{t}\ln \Prob\left(\frac{X_t}{t}\in ] j_-, j_+[\right)
\leq -\gamma j_-+
\limsup_{t\to+\infty} -\frac{1}{t}\ln \Prob\left(\frac{X_t}{t}\in ]- j_+, -j_-[\right).
\ee
Similarly, taking $B=[t j_-, tj_+]$ and the $\liminf_{t\to+\infty}$ of the first inequality we get
\be
-\gamma j_+ + \liminf_{t\to+\infty} -\frac{1}{t} \ln \Prob\left(\frac{X_t}{t}\in [ -j_+, -j_-]\right)
\leq \liminf_{t\to+\infty} -\frac{1}{t} \ln \Prob\left(\frac{X_t}{t}\in [ j_-, j_+]\right).
\ee
Taking $j_\pm=j\pm \varepsilon$, $j$ fixed, $\varepsilon \searrow 0$ we find
from the characterization \eqref{CharacterizationR} of $R_X$
\be
R_X(j)\leq -\gamma j +R_X(-j)\textrm{ and }-\gamma j +R_X(-j) \leq R_X(j)
\ee
so 
\be
R_X(j)=R_X(-j)-\gamma j,
\ee
 the announced symmetry. So with respect to fluctuation relations, our three
 definitions of large deviations are on the same footing.

We notice that the above relation is usually written in the more symmetric form 
$R_X(j)-R_X(-j)=-\gamma j$, which is however ambiguous when $R_X(j)$ is infinite.
 
\subsection{Cumulated quantities with sub-extensive difference}
\label{JointLargeDeviation}

We would like to stress that some natural and desirable properties of large deviation functions are not automatic with some of our definitions. 

For instance, suppose that $X_t$ and $Y_t$ are random processes with a
sub-extensive difference, i.e.  there is a non-random function $C_t$ such that
$\lim_{t\to+\infty}C_t =0$ and $|X_t-Y_t|\leq  tC_t$. Then it should be expected that the following alternative holds :
\be
\label{alternative}
\begin{cases}
  &\textrm{- $X_t$ and $Y_t$ have the same large deviation function, 
}   \\
  &\textrm{- neither $X_t$ nor $Y_t$ has a large deviation function.}   \end{cases}
\ee 
 We shall show that this holds true within the general theory of large deviations, namely for $R_X$ and $R_Y$. 
 We notice that, when $R_X(j)$ is convex upward, then $-R_X(j)$ can be expressed as in  \eqref{LDfuntionSupInf} and
 the property \eqref{alternative} can also be rederived from the explicit formul\ae\,  \eqref{LDfuntionSupInf}. 
 With our ad-hoc definitions, we can see that even when $X_t$ and $Y_t$ have the required properties, the discrete sets on which $X_t$ and $Y_t$ take their values can be very different. And even if the relation is simple, some problems remain at least with the weakest definition of large deviation functions, namely it may happen that $h_X\neq h_Y$.
 
 To make a proof within the general theory of large deviations, we note that, if $j_-<j_+$, any real function $R$ has the property that 
\be
\lim_{\varepsilon \searrow 0}\inf_{j\in ]j_-+\varepsilon, j_+ -\varepsilon[}R(j)=
\inf_{j\in ]j_-, j_+[ }R(j).
\ee
Moreover, if $R$ is lower semi-continuous then 
\be
\lim_{\varepsilon \searrow 0}\inf_{j\in [j_--\varepsilon, j_+ +\varepsilon]}R(j)=
\inf_{j\in [j_-, j_+] }R(j).
\ee 
Now, let $\varepsilon>0$. For $s$ large enough, $C_s<\varepsilon$ by hypothesis, so that
\be
\Prob\left(\frac{Y_s}{s}\in [j_-,j_+]\right)\leq 
\Prob\left(\frac{X_s}{s}\in [j_--\varepsilon,j_++\varepsilon]\right)
\textrm{ for $s\geq t$} 
\ee
whenever $t$ is such that $C_s\leq \varepsilon$ for $s\geq t$. Then
\be
\inf_{s\geq t} -\frac{1}{s}\ln \Prob\left(\frac{X_s}{s}\in [j_--\varepsilon,j_++\varepsilon]\right)
\leq \inf_{s\geq t} -\frac{1}{s}\ln \Prob\left(\frac{Y_s}{s}\in [j_-,j_+]\right),
\ee
so taking $t\to+\infty$ one gets
\be
\liminf_{t\to+\infty} -\frac{1}{t}\ln \Prob\left(\frac{X_t}{t}\in [j_--\varepsilon,j_++\varepsilon]\right)
\leq \liminf_{t\to+\infty}  -\frac{1}{t}\ln \Prob\left(\frac{Y_t}{t}\in [j_-,j_+]\right).
\ee
Assuming $X_t$ has a large deviation function, the left-hand side is 
$\geq \inf_{j\in [j_--\varepsilon, j_++\varepsilon]}R_X(j)$ by \eqref{defRXa},
so we have proved that for any $\varepsilon >0$ one has 
\be
\label{ineq1}
\inf_{j\in [j_--\varepsilon, j_++\varepsilon]}R_X(j)\leq 
\liminf_{t\to+\infty}  -\frac{1}{t}\ln \Prob\left(\frac{Y_t}{t}\in [j_-,j_+]\right).
\ee
In the same vein, assuming $j_-<j_+$, one has, for $\varepsilon$ small enough ($\varepsilon<\frac{j_+-j_-}{2}$)
\be
\Prob\left(\frac{X_s}{s}\in ]j_-+\varepsilon,j_+-\varepsilon]\right)\leq
\Prob\left(\frac{Y_s}{s}\in ]j_-,j_+[\right)\textrm{ for $s\geq t$} 
\ee
whenever $t$ is such that $C_s\leq \varepsilon$ for $s\geq t$. As above, we infer that
\be
\limsup_{t\to+\infty}  -\frac{1}{t}\ln \Prob\left(\frac{Y_t}{t}\in ]j_-,j_+[\right)\leq
\limsup_{t\to+\infty} -\frac{1}{t}\ln \Prob\left(\frac{X_t}{t}\in ]j_-+\varepsilon,j_+-\varepsilon[\right).
\ee
Assuming $X_t$ has a large deviation function, the right-hand side is 
$\leq \inf_{j\in ]j_-+\varepsilon, j_+-\varepsilon]}R_X(j)$ by \eqref{defRXb},
so for $\varepsilon$ small enough one has 
\be
\label{ineq2}
\limsup_{t\to+\infty}  -\frac{1}{t}\ln \Prob\left(\frac{Y_t}{t}\in ]j_-,j_+[\right)\leq
\inf_{j\in ]j_-+\varepsilon, j_+-\varepsilon[}R_X(j)
\ee
Taking the limit $\varepsilon \searrow 0$ in \eqref{ineq1} and \eqref{ineq2}
(using the lower semi-continuity of $R_X$ in \eqref{ineq1}) we obtain precisely that $Y_t$ satisfies an LDP \eqref{defRXa}--\eqref{defRXb}  with large deviation
function $R_X$. This proves that the alternative \eqref{alternative} holds.

\medskip

We conclude this discussion with a (counter)-example which has the advantage of being extremely simple, but the drawback that it is artificial (in particular, it is not continuous in probability, it even has non-random discontinuity times). 

The example is 
$X_t\equiv \lfloor t\rfloor$ the integer part of $t$, i.e. $\lfloor t\rfloor$ is an integer and 
$\lfloor t\rfloor\leq t< \lfloor t\rfloor+1$, and $Y_t\equiv X_t+1$. As shown below, 

- $X_t$ has a large deviation function $R_X$ and a large deviation function $h_X$ which coincide, namely
\be
\label{exampleRX}
R_X(j)=h_X(j)=\begin{cases}
  0 & \textrm{if $j=1$} \\
  +\infty & \textrm{if $j\not=1$}  
\end{cases}.
\ee

- $Y_t\equiv X_t+1$ has a large deviation function $R_Y$ ($=R_X$ by the previous result) but $h_Y$, while existing, differs from $h_X$, namely 
\be
\label{examplehY}
h_Y(j)=+\infty \textrm{ whatever $j$}.
\ee

- $X_t$ (resp. $Y_t$) has no large deviation function $g_X$ (resp. $g_Y$).
\medskip

Here are proofs for the latter properties. 

$\bullet$ The case of $h_X$ : we take $\mathbb{X}=\mathbb{N}$
\be
 \lim_{\substack{t\to +\infty \\tj\in \mathbb{N}}}-\frac{1}{t}\ln \Prob(X_t=tj)
 = \lim_{\substack{t\to +\infty \\tj\in \mathbb{N}}}-\frac{1}{t}\ln \mathbb{1}_{\lfloor t\rfloor=tj}.
\ee
If $j>1$ $\lfloor t\rfloor\leq t<tj$ so $\mathbb{1}_{\lfloor t\rfloor=tj}=0$. 
If $j<1$ $\lfloor t\rfloor> t-1>tj$ whenever $t>\frac{1}{1-j}$, so 
$\mathbb{1}_{\lfloor t\rfloor=tj}=0$ for $t$ large enough.
If $j=1$ $\mathbb{1}_{\lfloor t\rfloor=t}=1$ for $t\in \mathbb{N}$. This gives the announced formula \eqref{exampleRX} for $h_X$.

\medskip
$\bullet$ The case of $R_X$. 
\\First we consider $\Prob\left(\frac{X_s}{s}\in [j_-,j_+]\right)=
\mathbb{1}_{\lfloor s\rfloor \in [sj_-,sj_+]}
$.

- If $1\in [j_-,j_+]$ and $s=n\in \mathbb{N}$ then $\Prob\left(\frac{X_n}{n}\in
  [j_-,j_+]\right)=1$, so for any $t$ ${\inf_{s\geq
    t}-\frac{1}{s}\ln\Prob\left(\frac{X_s}{s}\in [j_-,j_+]\right)=0}$ and
$\liminf_{t\to+\infty}-\frac{1}{t}\ln\Prob\left(\frac{X_t}{t}\in
  [j_-,j_+]\right)=0$.

- If $j_->1$ then $\lfloor s \rfloor < sj_-$ whenever $s>0$ so for any $t>0$
$\inf_{s>t}-\frac{1}{s}\ln\Prob\left(\frac{X_s}{s}\in [j_-,j_+]\right)=+\infty$
and $\liminf_{t\to+\infty}-\frac{1}{t}\ln\Prob\left(\frac{X_t}{t}\in [j_-,j_+]\right)=+\infty$.

- If $j_+<1$ then $\lfloor s \rfloor > s-1\geq s j_+$ whenever $s\geq \frac{1}{1-j_+}$, so
$\inf_{s>t}-\frac{1}{s}\ln\Prob\left(\frac{X_s}{s}\in [j_-,j_+]\right)=+\infty$ whenever $t\geq \frac{1}{1-j_+}$ and
$\liminf_{t\to+\infty}-\frac{1}{t}\ln\Prob\left(\frac{X_t}{t}\in [j_-,j_+]\right)=+\infty$.
\\
To summarize 
\be\label{eq:cl}
\liminf_{t\to+\infty}-\frac{1}{t}\ln\Prob\left(\frac{X_t}{t}\in [j_-,j_+]\right)=
\begin{cases}
  0&\textrm{if $1\in[j_-,j_+]$}   \\
  +\infty&\textrm{else}  
\end{cases}.
\ee
Second we turn to $\Prob\left(\frac{X_s}{s}\in ]j_-,j_+[\right)=\mathbb{1}_{\lfloor s\rfloor \in ]sj_-,sj_+[}$.

- If $1\notin ]j_-,j_+[$ and $s\not=0$, then $s\notin ]sj_-,sj_+[$  so that for  $s=n\in \mathbb{N}$ we have
$\Prob\left(\frac{X_n}{n}\in ]j_-,j_+[\right)=0$. 
So for any $t$ ${\sup_{s\geq
    t}-\frac{1}{s}\ln\Prob\left(\frac{X_s}{s}\in [j_-,j_+]\right)=+\infty}$ and
$\limsup_{t\to+\infty}-\frac{1}{t}\ln\Prob\left(\frac{X_t}{t}\in
  [j_-,j_+]\right)=+\infty$.

- If $1\in ]j_-,j_+[$ then $\lfloor s\rfloor \in ]sj_-,sj_+[$ as soon as $s\geq \frac{1}{1-j_-}$ so 
$\limsup_{t\to+\infty}-\frac{1}{t}\ln\Prob\left(\frac{X_t}{t}\in ]j_-,j_+[\right)=0$. To summarize
\be \label{eq:op}
\limsup_{t\to+\infty}-\frac{1}{t}\ln\Prob\left(\frac{X_t}{t}\in ]j_-,j_+[\right)=
\begin{cases}
  0&\textrm{if $1\in]j_-,j_+[$}   \\
  +\infty&\textrm{else}  
\end{cases}.
\ee
From \eqref{eq:op}, \eqref{eq:cl} and the definition of an LDP \eqref{defRXa}-\eqref{defRXb} we infer that $X_t$ has a large deviation function 
\be
R_X(j)=
\begin{cases}
  0&\textrm{if $j=1$}   \\
  +\infty&\textrm{else}  
\end{cases}
\ee
and $R_X=h_X$. 

\medskip
$\bullet$ The case of $h_Y$ : we take $\mathbb{X}=\mathbb{N}^\star$
\be
 \lim_{\substack{t\to +\infty \\tj\in \mathbb{N}^\star}}-\frac{1}{t}\ln \Prob(Y_t=tj)
 = \lim_{\substack{t\to +\infty \\tj\in \mathbb{N}^\star}}-\frac{1}{t}\ln \mathbb{1}_{\lfloor t\rfloor+1=tj}.
\ee
If $j>1$ $\lfloor t\rfloor\leq t+1< tj$ whenever $t>\frac{1}{j-1}$ and then 
$\mathbb{1}_{\lfloor t\rfloor+1=tj}=0$. If $j\leq 1$ then 
$\lfloor t\rfloor+1>t\geq tj$ for every $t$ and then 
$\mathbb{1}_{\lfloor t\rfloor+1=tj}=0$. 
So 
$\lim_{\substack{t\to +\infty \\tj\in \mathbb{N}^\star}}-\frac{1}{t}\ln \Prob(Y_t=tj)=+\infty$ for every $j$ and $h_Y(j)=+\infty$ for every $j$ as announced in \eqref{examplehY}.

\medskip
$\bullet$ The case of $g_X$ and $g_Y$. 
\\For $j=1$ we have to consider maps $\xi_t$ such that $\lim_{t\to+\infty}\xi_t/t=1$ and the behavior of $-\frac{1}{t} \ln \Prob(X_t=\xi_t)$ when $t$ goes to $+ \infty$. Among these maps some are such that the corresponding limit is $0=h_X(j=1)$  and some other ones are such that the limit is $+\infty=h_Y(j=1)$ : for instance the map $\xi_t=\lfloor t\rfloor$ is such that $\Prob(X_t=\xi_t)=1$, whereas the map $\xi_t=\lfloor t\rfloor+a$ with $a>0$ is such that $\Prob(X_t=\xi_t)=0$. The same argument can be applied to 
$Y_t\equiv X_t+1$. As a result, neither $g_X$ nor $g_Y$ exist.

\clearpage
\bibliographystyle{unsrt}

\end{document}